\documentclass[10pt, twocolumn, a4paper]{IEEEtran}
\usepackage{mathrsfs}
\usepackage{amsmath}
\usepackage{amsthm}
\usepackage{yhmath}
\usepackage{graphicx}
\usepackage{cite}
\usepackage{amssymb}
\usepackage{tabularx}
\usepackage{bbm}
\usepackage{amsfonts}
\usepackage{bm}
\usepackage{algorithmic,algorithm}
\usepackage{indentfirst}
\usepackage{subfigure}
\usepackage{balance}
\usepackage{epstopdf}
\usepackage{color}
\usepackage[unicode=true,pdfusetitle,
 bookmarks=true,
 breaklinks=false,pdfborder={0 0 1},backref=false,colorlinks=true]
 {hyperref}
\hypersetup{
 linkcolor=blue, citecolor=red}

\newtheorem{remark}{Remark}

\begin{document}
\bibliographystyle{plain}

\title{On Dynamic Time Division Duplex Transmissions for Small Cell Networks}
%\author{Ming Ding$^1$, %
%\thanks{\noindent $^1$Ming Ding is with Data61, Australia. (E-mail: Ming.Ding@nicta.com.au)}%
%David L\'{o}pez-P\'{e}rez$^2$, %
%\thanks{\noindent $^2$David L\'{o}pez P\'{e}rez is with Bell Labs Alcatel-Lucent, Ireland. (E-mail: dr.david.lopez@ieee.org)}%
%Ruiqi Xue$^3$, %
%\thanks{\noindent $^3$Ruiqi Xue and Wen Chen are with Shanghai Jiao Tong University, P. R. China. (E-mail: \{xrq, wenchen\}@sjtu.edu.cn)}%
%Athanasios V. Vasilakos$^4$ and %
%\thanks{\noindent $^4$Athanasios V. Vasilakos is with Kuwait University, Kuwait. (E-mail: th.vasilakos@gmail.com)}%
%Wen Chen$^3$}

\author{\noindent {\normalsize{}Ming Ding, }\textit{\normalsize{}Member, IEEE}{\normalsize{}, David L$\acute{\textrm{o}}$pez-P$\acute{\textrm{e}}$rez,
}\textit{\normalsize{}Member, IEEE}{\normalsize{}, Ruiqi Xue, }\textit{\normalsize{}}
{\normalsize{}\protect \\ Athanasios V. Vasilakos, }\textit{\normalsize{}Senior
Member, IEEE}{\normalsize{}, Wen Chen, }\textit{\normalsize{}Senior
Member, IEEE}%
\thanks{Paper approved by xxx xxx. Manuscript received on xxx xx, 2015;
revised on xxx xx, 2015, xxx xx, 2015, and xxx xx, 2015.}
\thanks{Ming Ding is with Data 61, Australia (e-mail: Ming.Ding@nicta.com.au).}
%\protect \\
\thanks{David L$\acute{\textrm{o}}$pez-P$\acute{\textrm{e}}$rez is with
Bell Labs Alcatel-Lucent, Ireland (email: dr.david.lopez@ieee.org).}
%\protect \\
\thanks{Ruiqi Xue is with Shanghai Jiao Tong University, P. R. China. (e-mail:
xrq@sjtu.edu.cn).}
%\protect \\
\thanks{Athanasios V. Vasilakos is with Department of Computer Science, Electrical and Space Engineering Lule"1¤7University of Technology, Sweden. (e-mail: th.vasilakos@gmail.com).}
%\protect \\
\thanks{Wen Chen is with Shanghai Key Lab of Navigation and Location Based Services, Shanghai Jiao Tong University, and School of Electronic Engineering and Automation, Guilin University of Electronic Technology, China (e-mail: wenchen@sjtu.edu.cn).%
}% <-this % stops a space
\thanks{This work is supported by the National 973 Project \#2012CB316106, the National 863 Project \#2015AA01A710, NSFC \#61328101, and SKL on Mobile Communications \#2013D11.}% <-this % stops a space
%\thanks{Personal use is permitted, but republication/redistribution requires IEEE permission. Please find the final version in IEEE from the link: http://ieeexplore.ieee.org/xpl/articleDetails.jsp?arnumber=xxxxxxx. Digital Object Identifier 10.1109/TVT.2015.xxxxxxx}
}

\maketitle

\begin{abstract}

Motivated by the promising benefits of dynamic Time Division Duplex (TDD),
% at low and medium traffic loads,
in this paper,
we use a unified framework to investigate both the technical issues of applying dynamic TDD in homogeneous small cell networks (HomSCNs),
and the feasibility of introducing dynamic TDD into heterogeneous networks (HetNets).
First, HomSCNs are analyzed,
and a small cell BS scheduler
that dynamically and independently schedules DL and UL subframes is presented,
such that load balancing between the DL and the UL traffic can be achieved.
Moreover, the effectiveness of various inter-link interference mitigation (ILIM) schemes
%such as cell clustering (CC), power control, interference cancellation (IC)
as well as their combinations,
is systematically investigated and compared.
Besides, the interesting possibility of partial interference cancellation (IC) is also explored.
Second, based on the proposed schemes,
the joint operation of dynamic TDD together with cell range expansion (CRE) and almost blank subframe (ABS) in HetNets is studied.
In this regard, scheduling polices in small cells and an algorithm to derive the appropriate macrocell traffic off-load and ABS duty cycle under dynamic TDD operation are proposed.
Moreover, the full IC and the partial IC schemes are investigated for dynamic TDD in HetNets.
The user equipment (UE) packet throughput performance of the proposed/discussed schemes is benchmarked using system-level simulations.
\end{abstract}
% [Ming]: I have completed the Abstract.
%[David]: Good.

% deleted due to page limit

\section{Introduction}

% A brief introduction on the future telecommunication networks
In recent years,
the increase of mobile data traffic has been shown to project an exponential trajectory,
and this trend is expected to continue through the next decade~\cite{Cisco2012}.
In order to meet this formidable traffic demand,
the telecommunication networks have marched beyond the 4th generation (4G) realm~\cite{Book_LTE-A},
and begun to explore new advanced technologies~\cite{Tutor_5G}.
At present, the Third Generation Partnership Project (3GPP) sees exciting activities in the design of Long Term Evolution (LTE) Release~13 networks~\cite{RAN1conf77},
the scopes of which include
advanced interference cancellation receivers~\cite{reviewer_add1},
LTE operations in unlicensed spectrums~\cite{LAA_HetNet2015,LAA_HomoNet2015},
device to device (D2D) communications~\cite{D2D2014,D2D2015},
%three-dimensional (3D) multiple-input multiple-output (MIMO),
enhanced radio resource management~\cite{reviewer_add2,reviewer_add4,reviewer_add5,Lopez2013b,Lopez2014}, etc.
However, the most promising approach to rapidly increase network capacity is network densification through the deployment of small cells in heterogeneous networks (HetNets),
which takes advantage of extensive spatial reuse~\cite{Tutor_5G,Andrews2013,Lopez2011,Lopez2012,SmallCell1,SmallCell2,SmallCell3}.
LTE Release~10 HetNets, i.e., LTE Advanced (LTE-A) HetNets, adopted cell range expansion (CRE) to maximize the benefits of small cells~\cite{Book_LTE-A},~\cite{Lopez2011}.
With CRE, the coverage of a small cell can be artificially increased by instructing UEs to add a positive range expansion bias (REB) to the reference signal receiving power (RSRP) of the small cell.
However, the better spatial reuse and improved uplink (UL) connection offered by CRE comes at the expense of a reduced downlink (DL) signal-to-interference-plus-noise ratio (SINR) for the expanded-region (ER) UEs,
since they no longer connect to the base station (BS) providing the strongest level of signal reception~\cite{Lopez2011}.
% eICIC in HetNets
In order to alleviate this interference problem,
LTE-A HetNets implement time-domain enhanced inter-cell interference coordination (eICIC) by introducing almost blank subframes (ABSs)~\cite{Book_LTE-A},~\cite{Lopez2011}.
In more detail,
in the DL, macrocells schedule ABSs that are subframes in which only common reference signals (CRSs) and the most important cell-specific broadcast information are transmitted,
and small cells typically schedule their ER UEs in those DL subframes overlapping with the macrocell ABSs.
In this way, the inter-tier interference from macrocell BSs (MBSs) to ER UEs can be avoided~\cite{Lopez2011}.
%In~\cite{Lopez2012}, the authors provided an algorithm for calculating REBs and proposed a cooperative scheduling scheme to mitigate both DL and UL inter-cell interference caused by macrocells to ER UEs and macrocell UEs to small cells, respectively.
%The application of ABSs in more complicated scenarios containing three tiers of BSs and low power ABSs (LP-ABSs),
%where macrocells apply a power reduction on macrocell DL subframes instead of blanking,
%were investigated in~\cite{Pedersen2013} and~\cite{Soret2012}, respectively.

% TDD
Besides HetNets, it is also envisaged that future wireless communication networks,
e.g., LTE Release~12$\sim$14 networks,
will embrace time division duplexing (TDD),
which does not require a pair of frequency carriers and holds the possibility of tailoring the amount of DL/UL radio resources to the traffic conditions.
%Moreover, TDD operation facilities smart multi-antenna transmission/reception techniques,
%such as non-codebook based beam-forming and spatial multiplexing,
%due to DL/UL channel reciprocity.
%This avoids the wasting of radio resources in the DL since reference signals (RSs) for channel estimation are not required,
%as well as in the UL since UEs do not need to feed back quantized channel state information~\cite{Larson2013}.
% deleted due to page limit
In the LTE Release~8$\sim$11 networks,
seven TDD configurations~\cite{TS36.213},
each associated with a DL-to-UL subframe ratio in a 10-milisecond transmission frame,
are available for semi-static selection at the network side.
However, the adopted semi-static selection of TDD configuration in LTE Release~8$\sim$11 networks
is not able to adapt DL/UL subframe resources to the fast fluctuations in DL/UL traffic loads.
These fluctuations are exacerbated in small cells due to the low number of connected UEs per small cell and the burstiness of their DL and UL traffic demands.

% Dynamic TDD in HomoNets and the challenges in HetNets
In order to allow small cells to smartly and independently adapt their communication service to the quick variation of DL/UL traffic demands,
a new technology, referred to as dynamic TDD, has drawn much attention in the 3GPP recently~\cite{RAN1conf77}.
%Currently, the 3GPP sees exciting activities in the design of dynamic TDD for the LTE Release~12/13 homogeneous small cell networks (HomSCNs)~\cite{RAN1conf77}.
In dynamic TDD, the configuration of TDD DL-to-UL subframe ratio can be dynamically changed on a per-frame basis,
i.e., once every 10 milliseconds, in each cell or a cluster of cells.
%[Ming]: Technically speaking, I'm afraid TDD transmission direction cannot be changed on a 1\,ms basis in LTE Release~12. When a small cell changes its TDD configuration, the new configuration should be effective for the following frame (i.e., 10\,ms). I have amended the above sentence. Also, I have made relevant revisions throughout the paper with careful thoughts.
%[David]: Agreed.
Dynamic TDD can thus provide a tailored configuration of DL/UL subframe resources for each cell or a cluster of cells
at the expense of allowing inter-link interference,
i.e., the DL transmissions of a cell interfere with the UL ones of a neighbouring cell and vice versa.
Note that although dynamic TDD is a 4G technology,
it serves as the predecessor of the full duplex transmission technology~\cite{FullDup2014},
which has been identified as one of the candidate technologies for the 5th generation (5G) networks.
In a full duplex system,
a BS can transmit to and receive from different UEs simultaneously using the same frequency resource.
Hence, aside from the self-interference issue at the transceiver,
the full duplex transmission shares a common problem with dynamic TDD,
i.e., the inter-link interference.

% Academic research
The application of basic dynamic TDD transmissions in HomSCNs has been investigated in recent works~\cite{Shen2012},~\cite{HomodynTDD_ICC}.
Gains in terms of wide-band (WB) SINR and UE packet throughput (UPT) have been observed.
%mostly in low-to-medium traffic load conditions.
Faster dynamic TDD configuration time scales have also been shown to outperform slower ones.
%However, it is still unclear up to now whether it is feasible to introduce the dynamic TDD transmissions into HetNets,
%because it will complicate the existing CRE and ABS operations,
%and its advantage in the presence of macrocells in terms of UPT has not been confirmed in the existing literature yet~\cite{TR36.828}.
Besides, in~\cite{SG_dynTDD_ICC} the authors present a preliminary analysis based on stochastic geometry for dynamic TDD in HomSCNs,
without the consideration of traffic-adaptive DL/UL schedulers.
However, the introduction of dynamic TDD into HetNets is not straightforward, because it will complicate the existing CRE and ABS operations~\cite{TR36.828}.
An initial study on the feasibility of dynamic TDD in HetNets can be found in~\cite{HetNetdynTDD_ICC}.

% Paper contribution
In this paper, motivated by the promising benefits of dynamic TDD,
we investigate both
the technical issues of applying dynamic TDD in HomSCNs
and the feasibility of introducing dynamic TDD into HetNets.
This paper extends our previous works in~\cite{HomodynTDD_ICC} and~\cite{HetNetdynTDD_ICC} on dynamic TDD by making the following novel contributions:
%the main additional contribution of this paper is two-fold:
\begin{enumerate}
  \item
  Extensive efforts have been done to construct a coherent framework with the same design objectives, modeling assumptions, simulation scenarios and parameters for both HomSCNs and HetNets.
  In particular, an ideal genie-aided link adaptation (LA) mechanism is used in this paper,
  i.e., appropriate modulation and coding schemes are chosen according to the perceived SINRs \emph{after} the DL/UL packets are received.
  Note that some results in our previous work on dynamic TDD in HetNets~\cite{HetNetdynTDD_ICC} were lacking og insights because of the simplistic LA mechanism assumed therein.
  Hence, as a result of the use of a non-ideal link adapter,
  the true performance of dynamic TDD was not fully revealed in~\cite{HomodynTDD_ICC} and~\cite{HetNetdynTDD_ICC},
  especially for dynamic TDD in HetNets~\cite{HetNetdynTDD_ICC}.
  \item
  This paper opens a new avenue of research by analyzing the concept of partial interference cancellation (IC) and its overhead for dynamic TDD.
  Two new partial IC schemes are proposed to mitigate the DL-to-UL interference in dynamic TDD,
  i.e., the BS oriented partial IC scheme and the UE oriented partial IC scheme.
  Results show that the BS oriented partial IC scheme is much more effective than the UE oriented partial IC scheme
  and cancelling a few interferers is usually good enough to mitigate inter-link interference.
  \item
  The dynamic TDD algorithms in this paper have been redefined compared to those in our previous works,~\cite{HomodynTDD_ICC} and~\cite{HetNetdynTDD_ICC},
  such that the algorithms for HomSCNs can be smoothly extended to work in a HetNet scenario.
  This is important for practical implementation since operators can work with the same hardware/software in different scenarios just with minimal upgrades and no drastic changes.
  \item
  In this paper, unlike~\cite{HomodynTDD_ICC}, MIMO transmissions have also been considered for the UL,
  which has an impact on the results due to their larger capacity and thus shorter time for file transmission.
  Moreover, MIMO presents challenges on the appropriated switch between single-stream transmissions and multi-stream transmissions.
  \item
  As a result of the above bulletins, all the experiments have been re-conducted in this paper,
  so that an intriguing comparison between dynamic TDD in HomSCNs and HetNets can be performed to shed new light on dynamic TDD operations.
\end{enumerate}

%Paper structure
The rest of the paper is organized as follows.
In Section~\ref{sec:scenario},
the scenarios
%current scenarios used in the 3GPP
to analyze the dynamic TDD performance for both HomSCNs and HetNets are introduced.
%together with the notation adopted in this paper.
%In Section~\ref{sec:homoDynamicTDD},
%the focus is on dynamic TDD operation in homogeneous small cell networks,
%presenting algorithms to decide the appropriate TDD configuration in each small cell,
%as well as interference mitigation schemes to deal with the inter-link interference.
%In Section~\ref{sec:hetnetDynamicTDD},
%the focus is on dynamic TDD operation in HetNets,
%deriving scheduling policies for small cells that work in conjunction with the presented small cell BS scheduler,
%and allow a joint optimization of dynamic TDD operation together with CRE and ABS in HetNets.
In Sections~\ref{sec:homoDynamicTDD} and~\ref{sec:hetnetDynamicTDD},
the focus is on dynamic TDD operation in HomSCNs and HetNets, respectively.
In Section~\ref{sec:simulator},
our system-level simulator and the 3GPP simulation parameters in our experiments are presented.
In Sections~\ref{sec:homoResults} and~\ref{sec:hetnetResults},
benchmarked network configurations are depicted,
and simulation results for a HomSCN and a HetNet are presented and discussed, respectively.
Finally, a fair performance comparison between dynamic TDD in HomSCNs and HetNets is conducted in~\ref{sec:dynTDD_comp_homo_het},
%to shed new light on dynamic TDD operations
followed by the concluding remarks drawn in Section~\ref{sec:conclusion}.

\section{Network Scenario}
\label{sec:scenario}

During the study of dynamic TDD in the 3GPP~\cite{TR36.828},
a total of eight deployment scenarios were considered for investigation.
The 3GPP prioritised Scenario~3
%and Scenario~4
for further analysis~\cite{RAN1conf77},
and the study of Scenario~6 was left open for further discussion.
%as a logical extension,
%due to the initial doubts on its feasibility.
The definition of Scenarios~3 and 6 is as follows,
%which are summarized in the following:
\begin{itemize}
%\item
%Multiple femtocells deployed on the same carrier frequency, where femtocells can dynamically adjust TDD configurations.
%\item
%Multiple femtocells deployed on the same carrier frequency and multiple macrocells deployed on an adjacent carrier frequency, where all macrocells have the same TDD configuration and femtocells can dynamically adjust TDD configurations.
\item
Scenario~3: Multiple outdoor picocells deployed on the same carrier frequency, where outdoor picocells can dynamically adjust TDD configurations.

%\item
%Scenario~4: Multiple outdoor picocells deployed on the same carrier frequency and multiple macrocells deployed on an adjacent carrier frequency, where all macrocells have the same TDD configuration and outdoor picocells can dynamically adjust TDD configurations.

%\item
%Multiple femtocells and multiple macrocells deployed on the same carrier frequency, where all macrocells have the same TDD configuration and femtocells can adjust TDD configurations.
\item
Scenario~6: Multiple outdoor macrocells and multiple picocells deployed on the same carrier frequency, where all macrocells have the same TDD configuration and outdoor picocells can adjust TDD configurations.
%\item
%Multiple macrocells deployed on the same carrier frequency for one operator and multiple macrocells deployed on an adjacent carrier frequency for another operator, where all victim macrocells deployed on the same carrier have the same TDD configuration and all aggressor macrocells deployed on an adjacent carrier frequency can adjust TDD configurations.
%\item
%Multiple macrocells deployed on the same carrier frequency for one operator, where all macrocells can adjust TDD configurations.
\end{itemize}

\begin{figure}[t]
\begin{minipage}[t]{0.5\linewidth}
\centering
\subfigure[Scenario~3: Homogeneous small cell network.]{
\includegraphics[width=7.5cm]{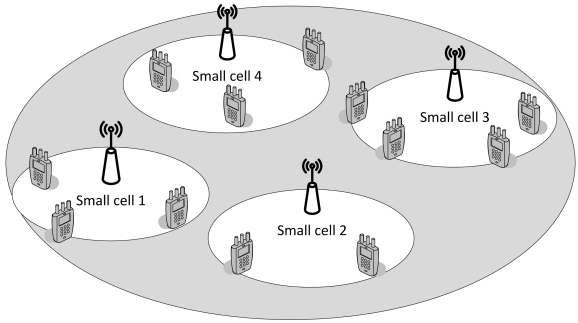}
\label{fig:illust_dynTDDHomo}
}
\end{minipage}
\begin{minipage}[t]{0.5\linewidth}
\subfigure[Scenario~6: Heterogeneous small cell network.]{
\includegraphics[width=8cm]{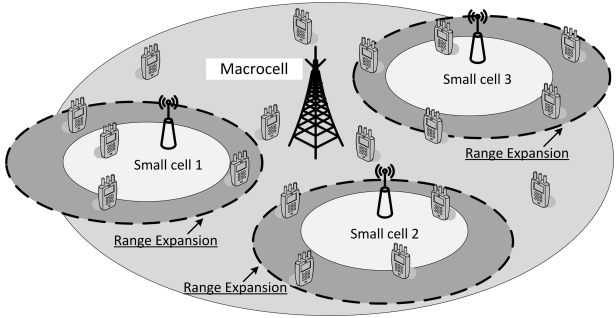}
\label{fig:illust_dynTDDHetNet}
}
\end{minipage}
\caption{Dynamic TDD scenarios.}
\label{fig:illust_dynTDD}
\vspace{-0.4cm}
\end{figure}

%Scenarios~1 and 2 depict multiple femtocells respectively without and with an overlay of macrocells occupying an adjacent carrier frequency.
%Scenarios~3 and 4 are respectively similar to Scenarios 1 and 2,
%but with femtocells substituted with outdoor picocells.
%Furthermore, Scenarios~5 and 6 represent HetNet deployments with macrocells overlaid with femtocells and picocells, respectively.
%Scenarios~7 and 8 consider dynamic TDD transmissions in macrocell-only networks,
%which have a low priority in the 3GPP.
%After preliminary evaluations~\cite{TR36.828},
%it is shown to be technically feasible to apply dynamic TDD schemes for Scenarios~1$\sim$4,
%but it is still unclear whether the same feasibility holds for Scenarios~5$\sim$8,
%especially for the HetNet dynamic TDD operations in Scenarios~5 and 6.
%Therefore, the 3GPP prioritised Scenario~3 and Scenario~4 for further analysis,
%and the study of Scenario~6 was left open as a logical extension.
In this paper, we focus on Scenario~3 and Scenario~6, which are illustrated in Fig.~\ref{fig:illust_dynTDD}.
%Note that Scenario~4 is not covered due to its similarity with Scenario~3.

With regard to notations,
in Fig.~\ref{fig:illust_dynTDD},
the $m$-th MBS, the $n$-th small cell BS (SBS), and the $q$-th UE are denoted as
$b(m), m \in \{1,\dots,M\}$,
$c(n), n \in\{1,\dots,N\}$, and
$u(q), q \in\{1,\dots,Q\}$, respectively.
Moreover, the DL average traffic arriving rate (DATAR), the UL average traffic arriving rate (UATAR), the DL instantaneous data buffer (DIDB) and the UL instantaneous data buffer (UIDB) of UE $u(q)$ are denoted as $\lambda^{\textrm{DL}}(q)$, $\lambda^{\textrm{UL}}(q)$, $\omega^{\textrm{DL}}(q)$ and $\omega^{\textrm{UL}}(q)$, respectively.

In order to determine UE cell association, two measures,
RSRP and WB DL SINR,
have been widely used in practical systems, e.g., LTE-A networks~\cite{Book_LTE-A}.
The RSRP in dBm scale and WB DL SINR in dB scale measured at UE $u(q)$ associated with MBS $b(m)$ are denoted as
${\mu}^\textrm{M}_{m,q}$ and $\gamma^\textrm{M}_{m,q}$, respectively.
The counterpart measures for SBS $c(n)$ are denoted as
${\mu}^\textrm{S}_{n,q}$ and $\gamma^\textrm{S}_{n,q}$, respectively.
% [Ming]: The original notation of the variable RSRP seems cumbersome. I have replaced it with \mu throughout the paper.
%[David]: Agreed.
Based on the best RSRP criterion of UE association,
we assume
\begin{itemize}
\item
%Without CRE in the small cells,
The set of macrocell UEs served by MBS $b(m)$ is denoted by
%$U^{\textrm{M}}_{m}=\left\{u\left(q^\textrm{M}_{m,1}\right),\dots,u\left(q^\textrm{M}_{m,k}\right),\dots,u\left(q^\textrm{M}_{m,K_1(m)}\right)\right\}$,
$U^{\textrm{M}}_{m}=\left\{u\left(q^\textrm{M}_{m,k}\right)\right\}, k\in\left\{1,\dots,K_1(m)\right\}$,
where $Q^{\textrm{M}}_{m}=\left\{q^\textrm{M}_{m,k}\right\}$
%$Q^{\textrm{M}}_{m}=\left\{q^\textrm{M}_{m,1},\dots,q^\textrm{M}_{m,k},\dots,q^\textrm{M}_{m,K_1(m)}\right\}$
is the set of indices of such macrocell UEs and $K_1(m)$ is its cardinality.
Note that the original set of macrocell UEs served by MBS $b(m)$ without the CRE operation is denoted by $U_{m}^{\textrm{M*}}$ and its cardinality is $K_{1}^{*}(m)$.

\item
Without CRE in the small cells,
the set of small cell UEs served by SBS $c(n)$ is denoted by
%$U^{\textrm{S}}_{n}=\left\{u\left(q^\textrm{S}_{n,1}\right),\dots,u\left(q^\textrm{S}_{n,k}\right),\dots,u\left(q^\textrm{S}_{n,K_2(n)}\right)\right\}$,
$U^{\textrm{S}}_{n}=\left\{u\left(q^\textrm{S}_{n,k}\right)\right\}, k\in\left\{1,\dots,K_2(n)\right\}$,
where $Q^{\textrm{S}}_{n}=\left\{q^\textrm{S}_{n,k}\right\}$
%$Q^{\textrm{S}}_{n}=\left\{q^\textrm{S}_{n,1},\dots,q^\textrm{S}_{n,k},\dots,q^\textrm{S}_{n,K_2(n)}\right\}$
is the set of indices of original small cell UEs and $K_2(n)$ is its cardinality.
%The DATAR, UATAR, DIDB and UIDB of small cell UE $u\left(q^\textrm{S}_{n,k}\right)$ are respectively denoted
%as $\lambda^{\textrm{DL}}(q^\textrm{S}_{n,k})$, $\lambda^{\textrm{UL}}(q^\textrm{S}_{n,k})$, $\omega^{\textrm{DL}}(q^\textrm{S}_{n,k})$ and $\omega^{\textrm{UL}}(q^\textrm{S}_{n,k})$.

\item
After the CRE operation,
some macrocell UEs will migrate to small cells leading to traffic off-loading from the macrocell tier to the small cell tier.
Then, the set of off-loaded macrocell UEs
%i.e., ER UEs,
to SBS $c(n)$ is denoted as
%$U^{\textrm{M2S}}_{n}=\left\{u\left(r^\textrm{S}_{n,1}\right),\dots,u\left(r^\textrm{S}_{n,k}\right),\dots,u\left(r^\textrm{S}_{n,K_3(n)}\right)\right\}$,
$U^{\textrm{M2S}}_{n}=\left\{u\left(r^\textrm{S}_{n,k}\right)\right\}, k\in\left\{1,\dots,K_3(n)\right\}$,
where $R^{\textrm{M2S}}_{n}=\left\{r^\textrm{S}_{n,k}\right\}$
%$R^{\textrm{M2S}}_{n}=\left\{r^\textrm{S}_{n,1},\dots,r^\textrm{S}_{n,k},\dots,r^\textrm{S}_{n,K_3(n)}\right\}$
is the set of indices of such ER UEs and $K_3(n)$ is its cardinality.
%The DATAR, UATAR, DIDB and UIDB of ER UE $u\left(r^\textrm{S}_{n,k}\right)$ are respectively denoted as
%$\lambda^{\textrm{DL}}(r^\textrm{S}_{n,k})$, $\lambda^{\textrm{UL}}(r^\textrm{S}_{n,k})$, $\omega^{\textrm{DL}}(r^\textrm{S}_{n,k})$ and $\omega^{\textrm{UL}}(r^\textrm{S}_{n,k})$.
\end{itemize}

\begin{table}
\begin{center}
\caption{Notation of variables}
\vspace{-0.1cm}
\label{Table:var_note}
\scalebox{0.8}{
\begin{tabular}{|l|l|l|l|}
  \hline
  Items & MBS $b(m)$ & SBS $c(n)$ (w/o CRE) & SBS $c(n)$ (w/ CRE)\\ \hline \hline
  Serving UEs & $U^{\textrm{M}}_{m}=\left\{u\left(q^\textrm{M}_{m,k}\right)\right\}$ & $U^{\textrm{S}}_{n}=\left\{u\left(q^\textrm{S}_{n,k}\right)\right\}$ & /\\ \hline
  ER UEs & / & / & $U^{\textrm{M2S}}_{n}=\left\{u\left(r^\textrm{S}_{n,k}\right)\right\}$\\ \hline
  UE indices & $Q^{\textrm{M}}_{m}=\left\{q^\textrm{M}_{m,k}\right\}$ & $Q^{\textrm{S}}_{n}=\left\{q^\textrm{S}_{n,k}\right\}$ & $R^{\textrm{M2S}}_{n}=\left\{r^\textrm{S}_{n,k}\right\}$ \\ \hline
  UE number & $K_1(m)$ & $K_2(n)$ & $K_3(n)$ \\ \hline
  RSRP & ${\mu}^\textrm{M}_{m,q^\textrm{M}_{m,k}}$ & ${\mu}^\textrm{S}_{n,q^\textrm{S}_{n,k}}$ & ${\mu}^\textrm{S}_{n,r^\textrm{S}_{n,k}}$ \\ \hline
  WB DL SINR & $\gamma^\textrm{M}_{m,q^\textrm{M}_{m,k}}$ & $\gamma^\textrm{S}_{n,q^\textrm{S}_{n,k}}$ & $\gamma^\textrm{S}_{n,r^\textrm{S}_{n,k}}$ \\ \hline
  DATAR & $\lambda^{\textrm{DL}}(q^\textrm{M}_{m,k})$ & $\lambda^{\textrm{DL}}(q^\textrm{S}_{n,k})$ & $\lambda^{\textrm{DL}}(r^\textrm{S}_{n,k})$ \\ \hline
  UATAR & $\lambda^{\textrm{UL}}(q^\textrm{M}_{m,k})$ & $\lambda^{\textrm{UL}}(q^\textrm{S}_{n,k})$ & $\lambda^{\textrm{UL}}(r^\textrm{S}_{n,k})$ \\ \hline
  DIDB & $\omega^{\textrm{DL}}(q^\textrm{M}_{m,k})$ & $\omega^{\textrm{DL}}(q^\textrm{S}_{n,k})$ & $\omega^{\textrm{DL}}(r^\textrm{S}_{n,k})$ \\ \hline
  UIDB & $\omega^{\textrm{UL}}(q^\textrm{M}_{m,k})$ & $\omega^{\textrm{UL}}(q^\textrm{S}_{n,k})$ & $\omega^{\textrm{UL}}(r^\textrm{S}_{n,k})$ \\ \hline
\end{tabular}}
\end{center}
\vspace{-0.8cm}
\end{table}

For clarity, the notation of variables related to UE $u(q)$ is summarized in Table~\ref{Table:var_note}.
%[Ming]: Taking advantage of Table~I, I have further simplified the presentation in Section~II, especially for the bulletins related to UE association.
In dynamic TDD, the subframes that can be either DL or UL ones are referred to as dynamic TDD subframes.
For those dynamic TDD subframes converted to DL ones or UL ones,
we refer to them as dynamic DL subframes and dynamic UL subframes, respectively.
%[Ming]: Good suggestion on the definition of "dynamic TDD/DL/UL subframes"! I moved/enhanced the paragraph from Section~III.A to here.
%[David]: Cool.

It is important to note that,
in the following sections,
we propose dynamic TDD schemes based on several coherent optimization objectives,
which are summarized here for the sake of clarity:
\begin{itemize}
  \item \textbf{Objective~1:} To minimize the difference between the DL and the UL \emph{average} traffic demand densities in each small cell.
  \item \textbf{Objective~2:} To minimize the difference between the DL and the UL \emph{instantaneous} traffic demand densities in each small cell.
  \item \textbf{Objective~3:} To minimize the \emph{average} traffic demand density for the macrocell and the small cell tiers.
\end{itemize}

% \cite{HomodynTDD_ICC}
\section{Dynamic TDD Operation in HomSCNs}
\label{sec:homoDynamicTDD}

In Scenario~3,
as illustrated by Fig.~\ref{fig:illust_dynTDDHomo},
multiple outdoor picocells deployed on the same carrier frequency can independently adapt their DL and UL subframe usage to the quick variation of the DL/UL traffic demands.
Two design aspects are fundamental to allow such dynamic TDD operation in each small cell, i.e.,
\begin{itemize}
\item
Algorithms to decide the appropriate TDD configuration.
To be more specific, how many subframes should be scheduled as DL or UL subframes in every $T$ subframes.
\item
Interference mitigation schemes to deal with the new inter-link interference,
i.e., the DL transmissions of small cells interfering with the UL transmissions of neighbouring ones and vice versa.
\end{itemize}

In this section, we present algorithms and schemes to realize these two design aspects.

\subsection{Dynamic DL/UL Subframe Splitting}
\label{sec:pico_split_homonet}

In the following,
we present an algorithm that runs independently in each small cell,
and decides the appropriate TDD configuration for each small cell.
Two cases are distinguished,
whether the small cell has active traffic or not.

%[David]:  The thing that bothers me the most is that in Section III we present first the fast dynamic TDD algorithm based on instantaneous DL/UL traffic, and later as an special case we present the case with no traffic where we compute the solution based on average DL/UL traffic. However, in Section IV, we do the opposite, we present first the algorithm based on average DL/UL traffic because it is required for the offloading and ABS duty cycle calculation, and later we present the fast dynamic TDD algorithm based on instantaneous DL/UL traffic. Do you think it would possible to adopt the same order in both Sections, first present slow algorithm and then fast algorithm or the opposite? I think this would improve a lot the flow, but  it would require to change section III for the first case or section IV for the second case. I think the first case would be better. We can just mention that we consider two cases, without and with traffic and then present the algorithms, first present slow algorithm and then fast algorithm. Let me know what do you think?
%[Ming]: Your suggestion is fantastic! Let me see... I also think the first case is better since reorganizing the algorithm description for HetNets would be very difficult. I have gone through the algorithm description again and made the corresponding revision in the following paragraphs. Please check them.
%[David]: Cool.

% First, the slow algorithm
First, the case in which there is no instantaneous DL or UL traffic at the small cell is considered.
In other words,
the small cell $c\left(n\right)$ is completely idle,
i.e., $\omega^{\textrm{DL}}\left(q_{n,k}^{\textrm{S}}\right)=0$ and $\omega^{\textrm{UL}}\left(q_{n,k}^{\textrm{S}}\right)=0$, $\forall{q}^{\textrm{S}}_{n,k}\in Q_{n}^{\textrm{S}}$.
Then, we  propose that the number of dynamic UL subframes should be set to a statistically optimal value that meets the upcoming traffic and achieves \textbf{Objective~1},
i.e., to minimize the difference between the DL and the UL average traffic demand densities in each small cell,
where the DL(UL) average traffic demand density is defined as
the sum of UEs' DL(UL) average traffic arriving rates over the quantity of the corresponding subframe resources in $T$ subframes.

Formally, the average traffic demand densities in small cell $c\left(n\right)$ in the DL and the UL are respectively defined as
\begin{equation}
	d_{n}^{\textrm{S,DL}}\left(t\right)=\frac{\sum_{k=1}^{K_2\left(n\right)}\lambda^{\textrm{DL}}\left(q_{n,k}^{\textrm{S}}\right)}{T-t},
	\label{eq:d_s_dl}
\vspace{-0.1cm}
\end{equation}
and
\begin{equation}
	d_{n}^{\textrm{S,UL}}\left(t\right)=\frac{\sum_{k=1}^{K_2\left(n\right)}\lambda^{\textrm{UL}}\left(q_{n,k}^{\textrm{S}}\right)}{t},
	\label{eq:d_s_ul}
	\vspace{-0.1cm}
\end{equation}
where the numerator is the sum of the DATARs(UATARs)
$\lambda^{\textrm{DL(UL)}}\left(q_{n,k}^{\textrm{S}}\right)$ of all UEs connected to small cell $c\left(n\right)$,
and the denominator is the number of DL(UL) subframes in every $T$ subframes available to transmit in the DL(UL),
such number denoted as $T-t$ for the DL and $t$ for the UL.
The definitions proposed in~(\ref{eq:d_s_dl}) and~(\ref{eq:d_s_ul}) make sense because the DATAR and the UATAR measure the average traffic influx into the network for the DL and the UL, respectively.

Then, with respect to \textbf{Objective~1},
the statistically optimal number of dynamic TDD UL subframes for small cell $c\left(n\right)$ is selected from
\begin{equation}
t_{n}^{\textrm{STAT\_homo}}=\underset{t=g\left(r\right),r\in\Upsilon^{\textrm{homo}}}{\arg\min}\left\{ \left|d_{n}^{\textrm{S,UL}}\left(t\right) - d_{n}^{\textrm{S,DL}}\left(t\right)\right|\right\},	
	\label{eq:t_STAT_homo}
	\vspace{-0.1cm}
\end{equation}
where $\Upsilon^{\textrm{homo}}$ is the set of all available TDD configurations for the considered HomSCN,
$r$ is one specific TDD configuration,
and $g(r)\in{\left[1,T-1\right]}$ extracts the number of UL subframes in $T$ subframes from TDD configuration $r$.
In general, $t_{n}^{\textrm{STAT\_homo}}$ indicates a reasonable stand-by state,
which tunes each small cell to be prepared for the upcoming traffic.

% deleted due to page limit
It is important to note that:
\begin{itemize}
\item
$g\left(r\right)$ may not be limited to integer values
since in practical systems certain special subframes consist of DL symbols, UL symbols and a transition interval between the DL and the UL symbols~\cite{TS36.213}.
The proportion of these three parts depends on the specific TDD configuration $r$.
\item
In order to keep the DL/UL control/reference signal channels always open for the TDD system to function properly,
we assume that $\Upsilon^{\textrm{homo}}$ only contains reasonable TDD configurations with $g(r)\in{\left[1,T-1\right]}$.
%the minimum and maximum values of $t$ being no less than one and no larger than $T-1$, respectively,
In other words, there are at least one DL and one UL subframe available in every $T$ subframes.
Note that all the 3GPP TDD configurations satisfy the above constraint~\cite{TS36.213}.
%the minimum value of $t^{\textrm{INST\_homo}}_n$ is set to one, even if there is no instantaneous UL traffic demand.
\end{itemize}

%[David]: Should we mention that the minimum value of DL subframes should also be 1 to allow CRS estimation?
%[Ming]: Sure. I have amended the sentence. Please check it.
%[David]: Looks good.

%[David]: We do not need to force this in the optimization but in the set of available TDD configurations, and I guess the set of available TDD configurations already consider this. No need to re-run experiments.
%[Ming]: You are right. In our studies, we only considered meaningful TDD configurations in $\Upsilon^{\textrm{homo}}$. So no need to re-conduct our simulations.
%[David]: Looks good.

% Second, the fast algorithm
In the following, the case in which there is some active DL and/or UL traffic at the small cell is considered.
In this case, the optimization objective is changed to \textbf{Objective~2},
i.e., to minimize the difference between the DL and the UL instantaneous traffic demand densities in each small cell,
where the DL(UL) instantaneous traffic demand density is defined as
the sum of UEs' DIDBs(UIDBs) over the quantity of the corresponding subframe resources in $T$ subframes.
This optimization objective ensures that load balancing between the DL and the UL transmissions can be dynamically achieved,
since both the DIDB and the UIDB are instantaneous information characterizing the immediate network loads.

Formally and similar to ${d}_{n}^{\textrm{DL}}\left(t\right)$ and ${d}_{n}^{\textrm{UL}}\left(t\right)$,
the instantaneous traffic demand densities of small cell $c\left(n\right)$ in the DL and the UL are respectively defined as
\begin{equation}
	\tilde{d}_{n}^{\textrm{S,DL}}\left(t\right)=\frac{\sum_{k=1}^{K_2\left(n\right)}\omega^{\textrm{DL}}\left(q_{n,k}^{\textrm{S}}\right)}{T-t}\,,
	\label{eq:d_s_dl_tilde}
	\vspace{-0.1cm}
\end{equation}
and
\begin{equation}
	\tilde{d}_{n}^{\textrm{S,UL}}\left(t\right)=\frac{\sum_{k=1}^{K_2\left(n\right)}\omega^{\textrm{UL}}\left(q_{n,k}^{\textrm{S}}\right)}{t}\,,
	\label{eq:d_s_ul_tilde}
	\vspace{-0.1cm}
\end{equation}
where the numerator is the sum of the DIDBs(UIDBs)
$\omega^{\textrm{DL(UL)}}\left(q_{n,k}^{\textrm{S}}\right)$ of all UEs connected to small cell $c\left(n\right)$.

Then and similarly as in~(\ref{eq:t_STAT_homo}),
with respect to \textbf{Objective~2},
the instantaneous optimal number of dynamic TDD UL subframes in $T$ subframes for small cell $c\left(n\right)$ is selected from
\begin{equation}
t_{n}^{\textrm{INST\_homo}}=\underset{t=g\left(r\right),r\in\Upsilon^{\textrm{homo}}}{\arg\min}\left\{\left| \tilde{d}_{n}^{\textrm{S,UL}}\left(t\right) - \tilde{d}_{n}^{\textrm{S,DL}}\left(t\right)\right|\right\}.
\label{eq:t_INST_homo}
\vspace{-0.1cm}
\end{equation}

In Algorithm~\ref{algo_pico_dyn_split_homo},
we summarize the proposed method to compute the dynamic TDD DL/UL subframe splitting for a given small cell according to its traffic condition in a HomSCN.

\begin{algorithm}[t]
\caption{\small{Selection of the optimal number of instantaneous dynamic TDD UL subframes in a small cell, i.e., $t^{\textrm{INST\_homo}}_n$}, for a HomSCN}
\label{algo_pico_dyn_split_homo}
\begin{algorithmic}
\small{
%\STATE .
\STATE Compute $d_{n}^{\textrm{S,DL}}\left(t\right)$, $d_{n}^{\textrm{S,UL}}\left(t\right)$, $\tilde{d}_{n}^{\textrm{DL}}\left(t\right)$ and $\tilde{d}_{n}^{\textrm{UL}}\left(t\right)$, using (\ref{eq:d_s_dl}) and (\ref{eq:d_s_ul}), (\ref{eq:d_s_dl_tilde}) and (\ref{eq:d_s_ul_tilde}), respectively.

\STATE Select $t^{\textrm{INST\_homo}}_n$ using the following procedure.

%\IF{$(\exists\omega^{\textrm{UL}}(q^\textrm{S}_{n,k})\neq 0,k \in K_2(n)$ $\textbf{and}$
%$\forall\omega^{\textrm{DL}}(q^\textrm{S}_{n,k})= 0,k \in K_2(n)$}
%\STATE $t^{\textrm{INST\_homo}}_n = T$

%[David]: This would be T-1 if we make the minimum value to DL subframes equal to 1.See next comment
%[Ming]: Actually, it may not be very accurate to specify t^{\textrm{INST\_homo}}_n to T or T-1. It depends on the available TDD configurations in $\Upsilon^{\textrm{homo}}$. For example, if the maximum t in $\Upsilon^{\textrm{homo}}$ is T-2, then t^{\textrm{INST\_homo}}_n should be T-2. I suppose that the branches $t^{\textrm{INST\_homo}}_n = T and $t^{\textrm{INST\_homo}}_n = 1 are not necessary here because eq.\ref{eq:t_INST_homo} covers these two branches. I guess your intention is to reuse Algorithm~\ref{algo_pico_dyn_split_homo} in the algorithm description for HetNets and I fully agree with you. Here I simplify Algorithm~\ref{algo_pico_dyn_split_homo} and later I will present our assumption on the $\Upsilon$ for HetNets so that Algorithm~\ref{algo_pico_dyn_split_homo} can still be reused.

%\ELSIF {$\forall\omega^{\textrm{UL}}(q^S_{n,k})= 0,k \in K_2(n)$ $\textbf{and}$
%$\exists\omega^{\textrm{DL}}(q^\textrm{S}_{n,k})\neq 0,k \in K_2(n)$}
%\STATE $t^{\textrm{INST\_homo}}_n = 1$

\IF {$\omega^{\textrm{UL}}\left(q^\textrm{S}_{n,k}\right)= 0,\forall{q}^\textrm{S}_{n,k} \in Q^\textrm{S}_{n}$ $\textbf{and}$
$\omega^{\textrm{DL}}\left(q^\textrm{S}_{n,k}\right)= 0,\forall{q}^\textrm{S}_{n,k} \in Q^\textrm{S}_{n}$}
\STATE Obtain $t^{\textrm{INST\_homo}}_n = t^{\textrm{STAT\_homo}}_n$ using (\ref{eq:t_STAT_homo}).

\ELSE
\STATE Obtain $t^{\textrm{INST\_homo}}_n$ using (\ref{eq:t_INST_homo}).

\ENDIF
}

\end{algorithmic}

\end{algorithm}

\subsection{Inter-link Interference Mitigation Schemes}
\label{sec:ILIMS}

It can be expected that the dynamic TDD DL/UL subframe splitting described in Section~\ref{sec:pico_split_homonet} enables traffic-adaptive scheduling,
i.e., more UL subframes will be diverted to DL transmissions when the DL traffic demand density in a small cell is higher than the UL one and vice versa.
However, dynamic TDD DL/UL subframe splitting gives rise to a new type of interference,
which is the inter-link interference between DL and UL transmissions resulted from non-uniform TDD subframe configurations among adjacent cells.
Such kind of inter-link interference is particularly severe in the DL-to-UL case
because \emph{i)} a BS-to-BS path loss is normally much smaller than a UE-to-BS one and  \emph{ii)} the DL signal from a high-power BS may easily overwhelm a low-power UE's UL signal intended for another BS.

Various inter-link interference mitigation (ILIM) schemes can be applied to address this DL-to-UL interference problem,
such as cell clustering (CC)~\cite{TR36.828}, DL power reduction (DLPR)~\cite{SHARP_DLPC_dynTDD}, UL power boosting (ULPB)~\cite{SHARP_ULPC_dynTDD}, interference cancellation (IC)~\cite{HomodynTDD_ICC}, as well as their combinations.
For brevity, the DLPR scheme will not be considered hereafter, due to its known poor performance,
i.e., the DL performance is heavily scarified in exchange of decreasing the DL-to-UL interference and improving the UL performance~\cite{HomodynTDD_ICC}.
More advanced techniques such as the machine-learning techniques~\cite{Machine_Learning1},~\cite{Machine_Learning2} can also be applied in dynamic TDD to tackle the DL-to-UL interference problem.
For example, the machine-learning techniques could be invoked at BSs to determine the right frequency and power allocation in view of buffer status and inter-link interference conditions.
However, the potential performance gains come at the cost of overhead and complexity.
We will consider such advanced ILIM schemes in our future work with emphasis on performance improvement and convergence issues.

In the following, we discuss the CC scheme, the ULPB scheme and the IC scheme,
whose performance will be compared in later sections.
Note that these ILIM schemes can be classified into two strategies to cope with the DL-to-UL interference,
i.e., \emph{(i)} to weaken the DL interference or \emph{(ii)} to strengthen the UL signal.
The CC scheme and the IC scheme belong to the first strategy,
while the ULPB scheme represents the second strategy.

\subsubsection{Cell Clustering}

The CC scheme semi-statically organizes the small cells into cell clusters based on metrics such as coupling loss $\rm{PL}^{\textrm{CC}}$,
i.e., the path loss between SBSs~\cite{TR36.828}.
Then, the dynamic TDD configuration is conducted on a per-cluster basis,
rather than on a per-cell basis.
In other words, the TDD configuration of all the small cells in a cell cluster is the same,
thus the inter-link interference is eliminated within the cell cluster.
In this case, negotiation and coordination of TDD configurations within cell clusters are required through inter-cell communications over backhaul links or the air interface.
A simple method to perform dynamic TDD DL/UL subframe splitting for a given cell cluster is to sum the cell specific
$\tilde{d}_{n}^{\textrm{S,DL}}\left(t\right)$ and $\tilde{d}_{n}^{\textrm{S,UL}}\left(t\right)$
as well as  $d_{n}^{\textrm{S,DL}}\left(t\right)$ and $d_{n}^{\textrm{S,UL}}\left(t\right)$ over the small cells in such cell cluster,
and proceed accordingly with Algorithm~\ref{algo_pico_dyn_split_homo} for each cell cluster.
Note that a more dynamic CC scheme considering joint optimization of DL/UL scheduling among multiple small cells might be possible.
However, it is out of the scope of our paper.
Here, we only consider the semi-static CC scheme that allows distributed operations among small cell clusters~\cite{TR36.828}.

\subsubsection{Power Control}

%The power control strategy includes the DLPR~\cite{SHARP_DLPC_dynTDD} and ULPB~\cite{SHARP_ULPC_dynTDD}  schemes.

The power control strategy considered here is based on ULPB~\cite{SHARP_ULPC_dynTDD}.
%The DLPR scheme decreases the amount of transmit power allocated by the BS to the UEs
%compared with a traditional allocation.
%The DL performance is thus scarified in exchange of decreasing the DL-to-UL interference and improving the UL performance.
The ULPB scheme increases the amount of transmit power used at the UEs
compared to the traditional fractional path loss compensation power control~\cite{TS36.213}.
This UL power boost helps to combat the DL-to-UL interference coming from neighbouring small cells.
%The implementation of these power control schemes is relatively simple,
%i.e., a fixed power offset $\Delta P^{\textrm{DL}}$ and $\Delta P^{\textrm{UL}}$ can be configured on top of the DL and UL power level, respectively.
The implementation of the ULPB scheme is relatively simple,
e.g., a fixed power offset $\Delta P^{\textrm{UL}}$ can be configured on top of the UL power level.

%[David]: I have deleted DLPR
%[Ming]: Agreed.

\subsubsection{Interference Cancellation}

In this paper, the IC scheme refers to the DL-to-UL IC and not to the UL-to-DL IC
because it is technically more feasible to assume that BSs are capable of exchanging information and cancelling inter-link interference coming from neighbouring BSs.
In contrast, the assumption of UEs performing UL-to-DL IC with regard to other peer UEs would seem to be too farfetched and thus impractical
(it is unlikely that UEs can exchange information).
In theory, the IC scheme should provide the best ILIM for the UL compared to the CC and the ULPB schemes,
but requires good backhaul connections for the exchange of inter-cell information on DL transmission assumptions,
such as resource allocation, modulation and coding scheme,
%configuration of demodulation reference signals,
etc.
Besides, strong signal processing units are required in the BSs to detect, reconstruct and cancel the DL interference for UL.

To reduce the complexity and cost of the IC scheme,
partial IC schemes can be further considered.
To be more specific, in the following,
we propose a UE-oriented IC (UOIC) scheme and a BS-oriented IC (BOIC) scheme.

In the UOIC scheme,
only cell-edge UEs will be granted the use of IC to mitigate the DL-to-UL interference.
Here, cell-edge UEs can be identified as those UEs,
which have at least one RSRP associated with a neighbouring BS that is larger than the RSRP associated with the serving BS by a bias of $x_1$\,dB.
Formally, for a UE $u\left(q^\textrm{S}_{n,k}\right)$, $q^\textrm{S}_{n,k}\in{Q^{\textrm{S}}_{n}}$,
it is a cell-edge UE entitled for IC if the following condition is valid.
%such criterion of determining whether it is a cell-edge UE can be expressed as
\begin{equation}
\exists~{\mu}^\textrm{S}_{m,q^\textrm{S}_{n,k}}>{\mu}^\textrm{S}_{n,q^\textrm{S}_{n,k}}-x_1, ~~m\neq{n}\,.
\label{eq:edge_UE_criterion}
\vspace{-0.1cm}
\end{equation}

In the BOIC scheme,
only DL interference from neighbouring BSs,
whose path losses to the serving BS are less than $x_2$\,dB are cancelled.
Formally, for an SBS $c(n)$, a neighbouring SBS $c(m)$ satisfying the following condition will be treated in the IC process.
\begin{equation}
PL^{\textrm{S2S}}_{m,n}<x_2, ~~m\neq{n}\,,
\label{eq:domi_BS_criterion}
\vspace{-0.1cm}
\end{equation}
where $PL^{\textrm{S2S}}_{m,n}$ is the path loss from SBS $c(m)$ to SBS $c(n)$ in dB scale.

Note that in both partial IC schemes,
the selected UE set and the selected BS set for the IC operations are cell-specific.
%[David]: This looks good.
%[Ming]: Thanks!

\section{Small Cell Dynamic TDD Operation in HetNets}
\label{sec:hetnetDynamicTDD}

In Scenario~6 illustrated by Fig.~\ref{fig:illust_dynTDDHetNet},
it is assumed that multiple outdoor macrocells and multiple picocells are deployed on the same carrier frequency,
and that all macrocells have the same TDD configuration while outdoor picocells can adjust their TDD configurations.
% deleted due to page limit
This is a logical assumption since macrocell traffic dynamics are usually averaged out due to the fairly large number of macrocell UEs per macrocell site.
Moreover, with a quasi-static configuration of DL/UL subframe splitting,
the detrimental DL-to-UL interference in the macrocell tier can be avoided.
In contrast, the traffic behaviour is completely different in the small cell tier,
mostly because of the low number of connected UEs per small cell and the burstiness of their DL and UL traffic demands.
This leads to drastic DL/UL traffic fluctuations,
which are particularly suitable for dynamic TDD operations.
Here, we propose that the macrocell tier uses a quasi-static configuration of DL/UL subframe splitting,
which matches its statistical DL/UL traffic ratio,
and consider dynamic TDD only for the small cell tier.

Moreover, in a HetNet, dynamic TDD operation at small cells cannot ignore CRE and ABS operations that have already been adopted at the macrocells,
and these technologies need to be designed together.
Hence, the following design aspects have to be considered:
\begin{itemize}
\item
Scheduling policy in small cells,
i.e., what is the behaviour of small cells in macrocell DL, UL and ABS subframes.
\item
UE cell association after CRE and optimal macrocell ABS duty cycle.
\item
Dynamic TDD scheduling at small cells.
\end{itemize}

In the following subsections, we examine these issues one by one in detail.

\subsection{Scheduling Policy in Small Cells}

For small cell UEs, any UL transmission attempt to SBSs will find itself in an extremely adverse situation in the subframes aligned with macrocell DL subframes,
since DL signals emitted from MBSs are of high power and thus can easily jam small cell UEs' UL signals.
Macrocell DL to small cell UL IC techniques based on full or partial prior information of macrocell DL transmissions may solve this problem.
However, the involved complexity in this kind of inter-tier IC is extremely high,
considering the dominant role of the DL interference coming from macrocells and the heavy traffic flow in macrocells.
Thus, it may not be wise to abuse the IC technique to cancel the DL-to-UL interference from the macrocell tier to the small cell tier.
Thus, we propose that small cells only conduct DL transmissions in the subframes aligned with macrocell DL subframes.

As for the subframes aligned with macrocell UL and ABS subframes,
since the interference suffered by SBSs and small cell UEs will probably be low
because strong interfering macrocell UEs are very likely to have been off-loaded to small cells as ER UEs,
we propose that small cells can perform dynamic TDD when macrocells transmit UL or ABS subframes.

As a result of these scheduling policies,
not all subframes in the small cell tier are dynamic TDD subframes,
and the number of dynamic TDD subframes is denoted as $f^{\textrm{S,dynTDD}}$.

Having decided which subframe type should be scheduled at each time at small cells,
it is important to define which small cell UEs should be scheduled in the subframes overlapping with macrocell DL subframes and in the dynamic TDD subframes.
A widely adopted assumption in LTE-A DL HetNets is that
DL packets of ER UEs should be scheduled with a high priority in subframes overlapping with the macrocell ABSs,
and that they should not be scheduled in subframes overlapping with the macrocell DL subframes
due to the strong inter-tier interference~\cite{R1-132304}.
Taking into account the previous scheduling policy and extending these ideas to the HetNet dynamic TDD scenario,
we propose the following:
\begin{enumerate}
\item
Small cell DL packets of ER UEs,
i.e., $U_{n}^{\textrm{M2S}}$,
are transmitted in small cell dynamic TDD DL subframes.
\item
Small cell DL packets of non-ER UEs,
i.e., $U_{n}^{\textrm{S}}$,
are transmitted in subframes overlapping with the macrocell DL subframes.
If the small cell dynamic TDD DL subframes are not occupied,
DL packets of non-ER UEs can also be carried by these subframes.
\item
Small cell UL packets of all connected UEs,
i.e., $U_{n}^{\textrm{S}}\bigcup U_{n}^{\textrm{M2S}}$,
are transmitted in small cell dynamic TDD UL subframes.
\end{enumerate}

\subsection{UE Cell Association and Macrocell ABS Duty Cycle}
\label{sec:Aopt_UE_association}

In light of the CRE and ABS operations,
and given the proposed scheduling policy in small cells,
next important questions to be answered are:
To which small cell should each off-loaded macrocell UE go?
Which is the optimal ABS duty cycle for the macrocell tier?
In order to answer these questions,
in this paper,
a new semi-dynamic algorithm is proposed to jointly determine UE cell association and macrocell ABS duty cycle,
with the consideration of dynamic TDD operation at small cells.
%This new algorithm is aimed at achieving load balancing between the macrocell and the small cell tiers
%and offload macrocell UEs that may suffer from low capacity to the small cells
%where they may benefit from higher performance.

The proposed semi-dynamic scheme considers a subframe splitting algorithm that is consistent with that presented in Section~\ref{sec:pico_split_homonet},
targeted at providing load balancing between the DL and the UL average traffic demand densities, i.e., \textbf{Objective~1}.
Considering the multiple cell tiers in HetNets,
the proposed semi-dynamic scheme also tries to find the optimal macrocell ABS duty cycle,
which achieves \textbf{Objective~3},
i.e., to minimize the average cell traffic demand density for the macrocell and the small cell tiers,
proving load balancing between tiers.
%In addition to jointly determine UE cell association and macrocell ABS duty cycle,
%the resulting average subframe splitting can also be engaged to deal with the special use case that
%occurs when a small cell $c\left(n\right)$ is completely idle with no DL or UL traffic,
%and the optimal number of dynamic DL and UL subframes should be set to a statistically optimal value that matches the average upcoming traffic.
%A more dynamic algorithm to deal with the instantaneous subframe splitting at the small cells in a HetNet scenario will be proposed in Section~\ref{sec:pico_split_hetnet}.
%[Ming]: The above paragraph is a very nice stitching :)
%[David]: Thanks.

The proposed algorithm to jointly determine UE cell association and macrocell ABS duty cycle is summarized in Algorithm~\ref{algo_Aopt_UE_association_hetnet},
where $A$ is the number of ABSs given up by the macrocells every $T$ subframes with $A \in \{0,1,\dots,T-1\}$,
$\alpha^{\textrm{M,DL}}$  and $\alpha^{\textrm{M,UL}}$ are the ratios of DL-to-total subframes and UL-to-total subframes for macrocells, respectively,
with $\alpha^{\textrm{M,DL}}+\alpha^{\textrm{M,UL}}=1$,
and round$\{x\}$ is an operator that maps $x$ to its closest integer.
Moreover, and similar to the DL/UL average traffic demand densities defined in (\ref{eq:d_s_dl}) and (\ref{eq:d_s_ul}) for small cell $c\left(n\right)$,
the average traffic demand densities for macrocell $b\left(m\right)$ in the DL and the UL are respectively defined as
\begin{equation}
	d^{{\textrm{M,DL}}}_{m} = \frac{\sum_{k=1}^{K_1(m)}\lambda^{\textrm{DL}}(q^{\textrm{M}}_{m,k})}{f^{{\textrm{M,DL}}}},
	\label{eq:d_m_dl}
	\vspace{-0.1cm}
\end{equation}
and
\begin{equation}
	d^{{\textrm{M,UL}}}_{m} = \frac{\sum_{k=1}^{K_1(m)}\lambda^{\textrm{UL}}(q^{\textrm{M}}_{m,k})}{f^{{\textrm{M,UL}}}},
	\label{eq:d_m_ul}
	\vspace{-0.1cm}
\end{equation}
where the numerator is the sum of DATARs/UATARs $\lambda^{\textrm{DL(UL)}}\left(q^\textrm{M}_{m,k}\right)$ of all UEs connected to macrocell $b(m)$,
and the denominator is the number of DL(UL) subframes in every $T$ subframes available to transmit it,
such number denoted as $f^{{ \textrm{M,DL}}}$(${f^{{ \textrm{M,UL}}}}$).

It is important to note that due to the static TDD configuration in the macrocell tier,
${f^{{\textrm{M,DL}}}}$ and ${f^{{\textrm{M,UL}}}}$ take network-wide values for all macrocells,
and that ${f^{{\textrm{M,DL}}}}+{f^{{\textrm{M,UL}}}}+A=T$.

\begin{algorithm}[t]
\caption{\small{Joint selection of UE cell association and macrocell ABS duty cycle in a HetNet}}
\label{algo_Aopt_UE_association_hetnet}
\begin{algorithmic}
\small{
\FOR {$A=0:T-1$}
\STATE Compute $f^{\textrm{M,DL}}=\textrm{round}\left\{(T-A)\times \alpha^{\textrm{M,DL}}\right\}$, $f^{\textrm{M,UL}}=T-A-f^{\textrm{M,DL}}$, and $f^{\textrm{S,dynTDD}}=f^{\textrm{M,UL}}+A$.

\FOR {$m=1:M$}
\STATE Initialization: $U_{m}^{\textrm{M}}=U_{m}^{\textrm{M*}}$, $K_{1}\left(m\right)=K_{1}^{*}\left(m\right)$.
\STATE Obtain $\bar{U}^{\textrm{M}}_{m}$ by sorting $u\left(q^\textrm{M}_{m,k}\right)$ according to the ascending order of $\gamma_{m,k}^\textrm{M}$.
\FOR {$j=1:K_1(m)$}
\STATE Regarding the candidate ER UE $u\left(q^\textrm{M}_{m,\pi(j)}\right)$, calculate $d^{\prime\textrm{M,DL}}_{m}$ and $d^{\prime\textrm{M,UL}}_{m}$ using (\ref{eq:d_m_dl_prime}) and (\ref{eq:d_m_ul_prime}).\\
Obtain $C(q^\textrm{M}_{m,\pi(j)})$ by sorting all small cells according to the descending order of  ${\mu}^\textrm{S}_{n,q^\textrm{M}_{m,\pi(j)}}$.

\FOR {$l=1:N$}
\STATE Compute $d^{\textrm{S,DL,M\_DL\_sf}}_{\zeta(l)}$ using (\ref{eq:d_s_dl_M_DL_sf}).\\
Obtain $t^{\textrm{STAT\_het}}_{\zeta(l)}$, $d^{\textrm{S,DL,dynTDD\_sf}}_{\zeta(l)}\left(t^{\textrm{STAT\_het}}_{\zeta(l)}\right)$ and $d^{\textrm{S,UL,dynTDD\_sf}}_{\zeta(l)}\left(t^{\textrm{STAT\_het}}_{\zeta(l)}\right)$ for the candidate outsourcing small cell  $c\left(\zeta(l)\right)$ using Algorithm~\ref{algo_pico_stat_split_hetnet}.\\
Update $d^{\textrm{S,DL}}_{\zeta(l)}$ and $d^{\textrm{S,UL}}_{\zeta(l)}$ using (\ref{eq:d_s_DL_Het}) and (\ref{eq:d_s_UL_Het}).

\IF {$d^{\textrm{S,DL}}_{\zeta(l)}<d^{\prime\textrm{M,DL}}_{m}$ $\textbf{and}$
$d^{\textrm{S,UL}}_{\zeta(l)}<d^{\prime\textrm{M,UL}}_{m}$ $\textbf{and}$ ${\mu}^\textrm{M}_{m,q^\textrm{M}_{m,\pi(j)}}-{\mu}^\textrm{S}_{\zeta(l),q^\textrm{M}_{m,\pi(j)}}<y$}
\STATE UE $u\left(q^\textrm{M}_{m,\pi(j)}\right)$ is outsourced to $c(\zeta(l))$.\\
Update the UE cell association as\\
$K_3(\zeta(l))=K_3(\zeta(l))+1$;\\
$R^{\textrm{M2S}}_{\zeta(l)} = R^{\textrm{M2S}}_{\zeta(l)}+\left\{u\left(q^\textrm{M}_{m,\pi(j)}\right)\right\}$;\\
$K_1(m) = K_1(m)-1$;\\
$U^M_{m} = U^\textrm{M}_{m}-\left\{u\left(q^\textrm{M}_{m,\pi(j)}\right)\right\}$.\\
Record the average traffic demand density of macrocell $b(m)$ as $d^\textrm{M}_m(A)=\frac{d^{\textrm{M,DL}}_m+d^{\textrm{M,UL}}_m}{2}$.\\
Obtain the average traffic demand density for small cell $c(\zeta(l))$ as $d^\textrm{S}_{\zeta(l)}(A)=\frac{d^{\textrm{S,DL}}_{\zeta(l)}+d^{\textrm{S,UL}}_{\zeta(l)}}{2}$.\\
$\textbf{break}$;
\ENDIF~~~~\scriptsize{\texttt{\{judgement of a successful outsourcing\}}}
\ENDFOR~~~~~~~~~~~~~~~\scriptsize{\texttt{\{loop of candidate small cells\}}}
\ENDFOR~~~~~~~~~~~~~~~~~~~~~~~\scriptsize{\texttt{\{loop of candidate ER UEs\}}}
\ENDFOR~~~~~~~~~~~~~~~~~~~~~~~~~~~~~~~~~\scriptsize{\texttt{\{loop of macrocells\}}}
\ENDFOR~~~~~~\scriptsize{\texttt{\{loop of candidate macrocell ABS duty cycles\}}}\\
\normalsize
Choose the appropriate macrocell ABS duty cycle using (\ref{eq:Aopt}). And UE cell association is eventually determined based on $A^{\textrm{opt}}$.
%\mathop{{max}}_{\mathbf{Z},\mathbf{Y}}
}
\end{algorithmic}
\end{algorithm}

Due to the limited solution space of $A$,
%solutions can be computed efficiently,
Algorithm~\ref{algo_Aopt_UE_association_hetnet} performs an exhaustive search on $A$ and its objective is to find the optimal $A^{\textrm{opt}}$,
which achieves \textbf{Objective~3},
i.e., to minimize the average cell traffic demand density for the macrocell and the small cell tiers.
% deleted due to page limit
Note that in practice,
different operators may have different objectives and could select different optimization targets,
but in those there is always a trade-off between the macrocell and small cell UPTs~\cite{Lopez2013},
i.e., increasing macrocell UPT reduces small cell UPT and vice versa.
Intuitively, $A^{\textrm{opt}}$ tends to be larger if some operator wants to put more emphasis on the performance of the small cell tier
and vice versa.

The procedure of Algorithm~\ref{algo_Aopt_UE_association_hetnet} is explained as follows,
where for each possible $A$ the following operations are performed.

%1)
For each macrocell $b(m)$,
all connected UEs in $U_{m}^{{\textrm{M}}}$ are sorted according to their ascending order of wideband SINR $\gamma^{\rm{M}}_{m,k}$,
and the following sorted set is obtained
$\bar{U}^{{\textrm{M}}}_{m}=\left\{u\left(q^{\textrm{M}}_{m,\pi(1)}\right),\dots,u\left(q^{\textrm{M}}_{m,\pi(j)}\right),\dots,u\left(q^{\textrm{M}}_{m,\pi(K_1(m))}\right)\right\}$.
The first UE in the sorted set is the first candidate UE to be off-loaded to a small cell,
and candidate UEs are examined sequentially.

%2)
For an examined candidate UE $u\left(q^{\textrm{M}}_{m,\pi(j)}\right)$ to be off-loaded,
the average DL and UL traffic demand densities for macrocell $b\left(m\right)$ in (\ref{eq:d_m_dl}) and (\ref{eq:d_m_ul}) are updated as follows:
\begin{equation}
	d^{\prime{\textrm{M,DL}}}_{m} = d^{{\textrm{M,DL}}}_{m}- \frac{\lambda^{{\textrm{DL}}}\left(q^{\textrm{M}}_{m,\pi(j)}\right)}{f^{{\textrm{M,DL}}}},
	\label{eq:d_m_dl_prime}
	\vspace{-0.1cm}
\end{equation}
and
\begin{equation}
	d^{\prime{\textrm{M,UL}}}_{m} = d^{{\textrm{M,UL}}}_{m}- \frac{\lambda^{{\textrm{UL}}}\left(q^{\textrm{M}}_{m,\pi(j)}\right)}{f^{{\textrm{M,UL}}}}.
	\label{eq:d_m_ul_prime}
	\vspace{-0.1cm}
\end{equation}

%3)
Then, in order to determine the new serving cell of candidate UE $u\left(q^{\textrm{M}}_{m,\pi(j)}\right)$,
all small cells $c(n)$ are sorted according to their descending order of RSRP, i.e., ${\mu}^{\textrm{S}}_{n,q^{{\textrm{M}}}_{m,\pi(j)}}$.
The sorted small cell set is UE-specific,
and is denoted as $C\left(q^{\textrm{M}}_{m,\pi(j)}\right)=\left\{c(\zeta(1)),\dots,c(\zeta(l)),\dots,c(\zeta(N))\right\}$.
Because of its highest signal strength,
the first small cell in the sorted set is the first candidate small cell to host the candidate UE,
and candidate small cells are examined sequentially.

%4)
For each candidate small cell $c(\zeta(l))$,
its average DL traffic demand density in the subframes overlapping with macrocell DL subframes,
denoted by $d^{\textrm{S,DL,M\_DL\_sf}}_{\zeta(l)}$ is defined as
\begin{equation}
	d^{\textrm{S,DL,M\_DL\_sf}}_{\zeta(l)} = \frac{\sum_{k=1}^{K_2(\zeta(l))}\lambda^{\textrm{DL}}\left(q^\textrm{S}_{\zeta(l),k}\right)}{f^{\textrm{M,DL}}},
	\label{eq:d_s_dl_M_DL_sf}
	\vspace{-0.1cm}
\end{equation}
where the numerator is the sum of DATARs $\lambda^{\textrm{DL}}\left(q^\textrm{S}_{\zeta(l),k}\right)$ of all non-ER UEs in small cell $c(\zeta(l))$.
The proposed definition is predicated on the fact that according to our scheduling policy,
small cell DL packets of non-ER UEs should be typically transmitted in subframes overlapping with the macrocell DL subframes,
which number is $f^{\textrm{M,DL}}$.

%5)
Once the average DL traffic demand density in the subframes overlapping with macrocell DL subframes has been calculated,
the algorithm looks for the statistically optimal splitting of dynamic TDD subframes in the DL and the UL for the candidate small cell $c(\zeta(l))$.
For future use in Section~\ref{sec:pico_split_hetnet},
the presentation of the proposed statistically optimal splitting of dynamic TDD DL/UL subframes in the DL and the UL for a small cell
is isolated from Algorithm~\ref{algo_Aopt_UE_association_hetnet} and presented in Algorithm~\ref{algo_pico_stat_split_hetnet}.
In this case, and following the same approach as in Section~\ref{sec:pico_split_homonet},
we propose that the statistically optimal number of dynamic TDD UL subframes for the candidate small cell $c(\zeta(l))$
%denoted as $t^{\textrm{STAT\_het}}_{\zeta(l)}$,
should be derived with \textbf{Objective~1},
i.e., to minimize the difference between the average DL and UL traffic demand densities.
In this way, a balanced DL/UL UPT performance in such small cell can be achieved.

\begin{algorithm}[t]
\caption{\small{Selection of the optimal number of average dynamic TDD UL subframes, i.e., $t^{\textrm{STAT\_het}}_{\zeta(l)}$} in a HetNet}
\label{algo_pico_stat_split_hetnet}
\begin{algorithmic}
\small{
\FOR {each $t=g\left(r\right),r\in\Upsilon^{\textrm{het}}$}
\STATE Compute $d^{\textrm{S,DL,dynTDD\_sf}}_{\zeta(l)}(t)$ and $d^{\textrm{S,UL,dynTDD\_sf}}_{\zeta(l)}(t)$ using (\ref{eq:d_s_DL_dynTDD_sf}) and (\ref{eq:d_s_UL_dynTDD_sf}), respectively.
\ENDFOR
\\Select $t^{\textrm{STAT\_het}}_{\zeta(l)}$ using (\ref{eq:t_stat_hetnet}).
}
\end{algorithmic}
\end{algorithm}

In Algorithm~\ref{algo_pico_stat_split_hetnet},
$\Upsilon^{\textrm{het}}$ is the set of all available TDD configurations for the considered HetNet.
For a candidate number of dynamic TDD UL subframes $t$,
based on our proposed scheduling policy,
ER UE DL traffic and all UE UL traffic should be served by dynamic TDD subframes aligned with macrocell UL and ABS subframes.
Considering the candidate ER UE $u\left(q^{\textrm{M}}_{m,\pi(j)}\right)$,
the average DL and UL traffic demand density in dynamic TDD subframes for
the candidate host small cell $c(\zeta(l))$ can be respectively computed as
\begin{equation}
	d^{\textrm{S,DL,dynTDD\_sf}}_{\zeta(l)}(t)=\frac{\sum_{k=1}^{K_3(\zeta(l))} \lambda^{\textrm{DL}}\left(r^\textrm{S}_{\zeta(l),k}\right)+\lambda^{\textrm{DL}}\left(q^{\textrm{M}}_{m,\pi(j)}\right)}{f^{\textrm{S,dynTDD}}-t},
	\label{eq:d_s_DL_dynTDD_sf}
	\vspace{-0.1cm}
\end{equation}
and
\begin{equation}
\begin{split}
d&^{\textrm{S,UL,dynTDD\_sf}}_{\zeta(l)}(t)=\\
&\frac{1}{t}\left[\sum_{k=1}^{K_2(\zeta(l))}\hspace{-6pt}\lambda^{\textrm{UL}}\left(q^{\textrm{S}}_{\zeta(l),k}\right)\hspace{-2pt}+\hspace{-8pt}
\sum_{k=1}^{K_3(\zeta(l))}\hspace{-6pt}\lambda^{\textrm{UL}}\left(r^\textrm{S}_{\zeta(l),k}\right)\hspace{-2pt}+\hspace{-2pt}\lambda^{\textrm{UL}}\left(q^{\textrm{M}}_{m,\pi(j)}\right)\right].
\end{split}
\label{eq:d_s_UL_dynTDD_sf}
\vspace{-0.1cm}
%d^{\textrm{S,UL,dynTDD\_sf}}_{\zeta(l)}(t)=
%\frac{1}{t}\left[\sum_{k=1}^{K_2(\zeta(l))}\hspace{-6pt}\lambda^{\textrm{UL}}\left(q^{\textrm{S}}_{\zeta(l),k}\right)\hspace{-2pt}+\hspace{-8pt}
%\sum_{k=1}^{K_3(\zeta(l))}\hspace{-6pt}\lambda^{\textrm{UL}}\left(r^\textrm{S}_{\zeta(l),k}\right)\hspace{-2pt}+\hspace{-2pt}\lambda^{\textrm{UL}}\left(q^{\textrm{M}}_{m,\pi(j)}\right)\right].
%\label{eq:d_s_UL_dynTDD_sf}
%\vspace{-0.1cm}
\end{equation}

Then, based on such computations and similar to (\ref{eq:t_STAT_homo}) considering \textbf{Objective~1},
the statistically optimal number of dynamic TDD UL subframes for small cell $c(\zeta(l))$ becomes
\begin{equation}
\begin{split}
t&^{\textrm{STAT\_het}}_{\zeta(l)}=\\
&\underset{t=g\left(r\right),r\in\Upsilon^{\textrm{het}}}{\arg\min}
\left\{\left| d^{\textrm{S,UL,dynTDD\_sf}}_{\zeta(l)}(t) - d^{\textrm{S,DL,dynTDD\_sf}}_{\zeta(l)}(t)\right|\right\}.
\end{split}
\label{eq:t_stat_hetnet}
\vspace{-0.1cm}
%t^{\textrm{STAT\_het}}_{\zeta(l)}=
%\underset{t=g\left(r\right),r\in\Upsilon^{\textrm{het}}}{\arg\min}
%\left\{\left| d^{\textrm{S,UL,dynTDD\_sf}}_{\zeta(l)}(t) - d^{\textrm{S,DL,dynTDD\_sf}}_{\zeta(l)}(t)\right|\right\}.
%\label{eq:t_stat_hetnet}
%\vspace{-0.1cm}
\end{equation}

%6)
Having obtained $t^{\textrm{STAT\_het}}_{\zeta(l)}$,
we propose that
the average DL traffic demand density for the candidate small cell $c(\zeta(l))$,
used in the following step,
should be the larger one of the average DL traffic demand density associated with ER UEs and with non-ER UEs,
which is expressed as
\begin{equation}
	d^{\textrm{S,DL}}_{\zeta(l)} = \max\left\{d^{\textrm{S,DL,dynTDD\_sf}}_{\zeta(l)}\left(t^{\textrm{STAT\_het}}_{\zeta(l)}\right),d^{\textrm{S,DL,M\_DL\_sf}}_{\zeta(l)}\right\},
	\label{eq:d_s_DL_Het}
	\vspace{-0.1cm}
\end{equation}
while the average UL traffic demand density for the candidate small cell $c(\zeta(l))$ is
\begin{equation}
	d^{{\textrm{S,UL}}}_{\zeta(l)} = d^{{\textrm{S,UL,dynTDD\_sf}}}_{\zeta(l)}\left(t^{\textrm{STAT\_het}}_{\zeta(l)}\right).
	\label{eq:d_s_UL_Het}
	\vspace{-0.1cm}
\end{equation}

%7)
Now, before executing the offloading of candidate UE $u\left(q^{\textrm{M}}_{m,\pi(j)}\right)$,
we propose that two constraints should be checked.
First, the average traffic demand density of the candidate small cell after offloading should not be larger than that of the source macrocell to avoid small cells taking upon too much burden and becoming new traffic bottlenecks.
This is a necessary condition in the load balanced state,
and is mathematically formulated as
$d^{\textrm{S,DL}}_{\zeta(l)}<d^{\prime\textrm{M,DL}}_{m}$
and
$d^{\textrm{S,UL}}_{\zeta(l)}<d^{\prime\textrm{M,UL}}_{m}$.
Second, the link quality between the candidate ER UE and the candidate small cell should be good enough,
i.e., ${\mu}^\textrm{M}_{m,q^\textrm{M}_{m,k}}-{\mu}^\textrm{S}_{\zeta(l),q^\textrm{M}_{m,k}}<y$ ,
where $y$ is the REB parameter in dB scale for the CRE operation.
Intuitively, the proposed two constraints require that a candidate macrocell UE should be offloaded to a small cell that is neither overloaded nor far away from the concerned macrocell UE.
Otherwise, the off-loading will not be performed.

%8)
Once these constraints are met,
the candidate UE $u\left(q^{\textrm{M}}_{m,\pi(j)}\right)$ is offloaded to the candidate small cell $c(\zeta(l))$,
and all related parameters are updated as described in Algorithm~\ref{algo_Aopt_UE_association_hetnet}.
The average traffic demand density of the offloaded macrocell
$b(m)$ is updated as $d^\textrm{M}_m(A)=\frac{d^{\textrm{M,DL}}_m+d^{\textrm{M,UL}}_m}{2}$,
and that of the candidate small cell $c(\zeta(l))$ is updated as $d^\textrm{S}_{\zeta(l)}(A)=\frac{d^{\textrm{S,DL}}_{\zeta(l)}+d^{\textrm{S,UL}}_{\zeta(l)}}{2}$.

Finally, after iterating over all macrocells,
all candidate UEs and all candidate small cells,
we select the macrocell ABS duty cycle $A^{\textrm{opt}}$ using (\ref{eq:Aopt}) with respect to \textbf{Objective~3},
i.e., to minimize the average traffic demand density for the macrocell and the small cell tiers.
\begin{equation}
    A^{\textrm{opt}}=\mathop{{\arg\min}}\limits_A \left\{\frac{1}{M+N}\left[\sum\limits_{m=1}^M d^\textrm{M}_m(A)+\sum\limits_{n=1}^N d^\textrm{S}_n(A)\right]\right\}.
	\label{eq:Aopt}
	\vspace{-0.1cm}
\end{equation}
The final UE cell association is established according to the selected $A^{\textrm{opt}}$.

In the proposed Algorithm~\ref{algo_Aopt_UE_association_hetnet},
two parameters need to be chosen for its operation.
The first parameter is $T$, which can be set to 10 according to the 3GPP specifications~\cite{TS36.213},
because each transmission frame consists of 10 subframes in the current LTE networks.
The other parameter is $y$, which is the REB parameter in dB.
As suggested in some previous work on CRE~\cite{R1-132304}, a reasonable value of $y$ can be $y=9$\,dB.

\subsection{Discussion on the Convergence and the Complexity of Algorithm~\ref{algo_Aopt_UE_association_hetnet}}
\label{sec:discussion_Algorithm2}

Before we delve deeper into the problem of DL/UL subframe splitting in the small cell tier,
it is beneficial to have a full assessment on the convergence and the complexity of the proposed Algorithm~\ref{algo_Aopt_UE_association_hetnet},
which jointly optimizes UE cell association and macrocell ABS duty cycle.
Note that Algorithm~\ref{algo_Aopt_UE_association_hetnet} is a one-shot exhaustive searching algorithm
%to jointly optimize the UE association and the macrocell duty cycle
with no iterative steps, and thus convergence is not an issue for Algorithm~\ref{algo_Aopt_UE_association_hetnet}.

The complexity of Algorithm~\ref{algo_Aopt_UE_association_hetnet},
on the other hand,
could be a serious issue that may prevent its implementation in practice.
In more detail,
the complexity of Algorithm~\ref{algo_Aopt_UE_association_hetnet} is in the order of $TN\sum_{m=1}^{M}K_{1}^{*}(m)$
because $T$ candidate values of $A$ and $N$ candidate outsourcing small cells need to be tested for \emph{each and every} macrocell UE.
One way to reduce the complexity of the algorithm without compromising its performance is to adopt a macrocell-UE-specific number of candidate outsourcing small cells based on the value of $y$,
which should be much smaller than $N$,
because the small cells that are too far away from the considered macrocell UE do not need to go through the off-loading test due to poor signal strength.
Another way to reduce the complexity is to perform Algorithm~\ref{algo_Aopt_UE_association_hetnet} inside a macrocell cluster,
the size of which can be adjusted based on the implementation feasibility.

Having said that,
the real challenge to implement Algorithm~\ref{algo_Aopt_UE_association_hetnet} comes from the \emph{time-variant} network,
where UEs can come and go,
and thus the UE cell association and the macrocell ABS duty cycle need to be updated on the fly.
In more detail,
it is generally feasible to execute Algorithm~\ref{algo_Aopt_UE_association_hetnet} only once for a \emph{time-invariant} network scenario.
%for a specific network scenario that is merely \emph{a time-invariant snapshot} of the \emph{time-variant} network,
%all the UEs are pre-determined and they do not vary in the time window of our investigation.
%Then,
%the proposed Algorithm~\ref{algo_Aopt_UE_association_hetnet} needs to be executed for only once,
%and hence it is feasible to do so since the time window of our investigation is at least tens of seconds or even minutes~\cite{TR36.828}.
However, when the network becomes \emph{time-variant} due to UE mobility and bursty traffic, etc.,
we need to frequently recall Algorithm~\ref{algo_Aopt_UE_association_hetnet},
which is not practical due to its high complexity.
Note that it is not necessary to consider fast time-variant networks caused by high UE mobility in the framework of HetNets dynamic TDD,
since UEs with high mobility will be connected to the macrocell tier only,
thus avoiding handover failure issues~\cite{Book_LTE-A}.
Here,
the considered time-variant network changes in the order of seconds or hundreds of milliseconds,
since a UE with a speed of 10\,km/h will only move about 2.78\,m in one second,
and it may take seconds or tens of seconds for a UE to finish reading a web page before requesting a new DL/UL transmission~\cite{TR36.814}.
Even so,
it is still infeasible to conduct the entire Algorithm~\ref{algo_Aopt_UE_association_hetnet} every time when a UE arrives at a cell or a UE leaves a cell.
%In the following, we summarize the key points of algorithm design for time-variant networks.
Therefore,
we need to design new algorithms for the time-variant networks,
and use Algorithm~\ref{algo_Aopt_UE_association_hetnet} in the initialization stage only.
Based on the best RSRP criterion of UE association discussed in Section~\ref{sec:scenario},
we propose to classify the events of network changing into four cases:
% due to UE arrival and UE leaving
\begin{itemize}
  \item Case~1:
  A new macrocell UE $u\left(z\right)$ arrives at macrocell $b\left(m_{0}\right)$.
  Then we have $U_{m_{0}}^{\textrm{M*}}=U_{m_{0}}^{\textrm{M*}}\cup u\left(z\right)$ and $K_{1}^{*}\left(m_{0}\right)=K_{1}^{*}\left(m_{0}\right)+1$.
  There are two alternatives for algorithm design.
  \begin{itemize}
    	\item "1¤7   	Alt.~1:
    	For macrocell $b\left(m_{0}\right)$,
    	%we can update $U_{m_{0}}^{\textrm{M*}}$ and $K_{1}^{*}\left(m_{0}\right)$ as $U_{m_{0}}^{\textrm{M*}}=U_{m_{0}}^{\textrm{M*}}\cup u\left(z\right)$ and $K_{1}^{*}\left(m_{0}\right)=K_{1}^{*}\left(m_{0}\right)+1$ , respectively. And then
    	we perform Algorithm~\ref{algo_Aopt_UE_association_hetnet} for macrocell $b\left(m_{0}\right)$ only,
    	the complexity of which is in the order of $TNK_{1}^{*}\left(m_{0}\right)$.
        %Note that Alt. 1 is a simple adaptation of Algorithm~\ref{algo_Aopt_UE_association_hetnet} for time-variant networks.
    	
	\item "1¤7Alt.~2:
    	For UE $u\left(z\right)$,
    	we can design a new algorithm, denoted by Algorithm~\ref{algo_Aopt_UE_association_hetnet}-A,
	to check the $N$ candidate outsourcing small cells and decide whether UE $u\left(z\right)$ should stay in macrocell $b\left(m_{0}\right)$, or it should be off-loaded to a small cell $c\left(n_{0}\right)$.
    	The complexity of Algorithm~\ref{algo_Aopt_UE_association_hetnet}-A is in the order of $TN$.
        %, which is lower than that of Alt. 1.
        %Unlike Alt. 1, Alt. 2 does not change the existing UE association, but it may suffer from some performance loss, which is for further study.
  \end{itemize}

  \item Case~2:
  A macrocell UE $u\left(q_{m_{0},k_{0}}^{\textrm{M}}\right)$ leaves from macrocell $b\left(m_{0}\right)$.
  Then, we have $U_{m_{0}}^{\textrm{M*}}=U_{m_{0}}^{\textrm{M*}}\setminus u\left(q_{m_{0},k_{0}}^{\textrm{M}}\right)$ and $K_{1}^{*}\left(m_{0}\right)=K_{1}^{*}\left(m_{0}\right)-1$.
  Also, there are two alternatives for algorithm design.
  \begin{itemize}
        	\item "1¤7Alt.~1:
	For macrocell $b\left(m_{0}\right)$,
        %we can update $U_{m_{0}}^{\textrm{M*}}$ and $K_{1}^{*}\left(m_{0}\right)$ as $U_{m_{0}}^{\textrm{M*}}=U_{m_{0}}^{\textrm{M*}}\setminus u\left(q_{m_{0},k_{0}}^{\textrm{M}}\right)$ and $K_{1}^{*}\left(m_{0}\right)=K_{1}^{*}\left(m_{0}\right)-1$, respectively. And then
        we perform Algorithm~\ref{algo_Aopt_UE_association_hetnet} for macrocell $b\left(m_{0}\right)$ only, the complexity of which is in the order of $TNK_{1}^{*}\left(m_{0}\right)$.
            %Note that this is a simple adaptation of Algorithm~\ref{algo_Aopt_UE_association_hetnet} for time-variant networks.

        \item "1¤7Alt.~2:
        Since a UE leaves from macrocell $b\left(m_{0}\right)$, the traffic load of macrocell $b\left(m_{0}\right)$ should be reduced.
        Therefore, we should design a new algorithm, denoted by Algorithm~\ref{algo_Aopt_UE_association_hetnet}-B,
        to examine the $K_{1}^{*}\left(m_{0}\right)-K_{1}\left(m_{0}\right)$ UEs that have been outsourced to small cells and check whether some of them should come back to macrocell $b\left(m_{0}\right)$.
        The complexity of Algorithm~\ref{algo_Aopt_UE_association_hetnet}-B is in the order of $T\left[K_{1}^{*}\left(m_{0}\right)-K_{1}\left(m_{0}\right)\right]$.
      \end{itemize}

  \item Case~3:
  A new small cell UE $u\left(z\right)$ arrives at small cell $c\left(n_{0}\right)$.
  Note that such UE cannot be an ER UE because we consider the best RSRP criterion of UE association and all the potential ER UEs should go through Case 1 first.
  Due to the arrival of UE $u\left(z\right)$, the traffic load of small cell $c\left(n_{0}\right)$ should be increased.
  Therefore, we should design a new algorithm, denoted by Algorithm~\ref{algo_Aopt_UE_association_hetnet}-C,
  to check the $K_{3}\left(n_{0}\right)$ UEs that have been outsourced to small cell $c\left(n_{0}\right)$ and check whether some of them should come back to their original macrocells.
  The complexity of Algorithm~\ref{algo_Aopt_UE_association_hetnet}-C is in the order of $TK_{3}\left(n_{0}\right)$.

 \item Case~4:
 A small cell UE $u\left(q_{n_{0},k_{0}}^{\textrm{S}}\right)$ leaves from small cell $c\left(n_{0}\right)$.
 Note that such UE can be an ER UE or a non-ER UE.
 Either way, the traffic load of small cell $c\left(n_{0}\right)$ should be reduced.
 Therefore, we should design a new algorithm, denoted by Algorithm~\ref{algo_Aopt_UE_association_hetnet}-D,
 to examine all the macrocell UEs, the number of which is $\sum_{m=1}^{M}K_{1}\left(m\right)$, and check whether some of them are eligible to be outsourced by small cell $c\left(n_{0}\right)$.
 The complexity of Algorithm~\ref{algo_Aopt_UE_association_hetnet}-D is in the order of $T\sum_{m=1}^{M}K_{1}\left(m\right)$.
\end{itemize}

In this paper, we would like to focus on \emph{time-invariant} networks,
both in algorithm design and simulation,
to show the full potential of dynamic TDD in HetNets.
In our future work, we will study Case~1$\sim$4 as well as Algorithm~\ref{algo_Aopt_UE_association_hetnet}-A$\sim$D for \emph{time-variant} networks.

%Besides, the arrival or the leave of a macrocell UE only impacts the calculation associated with the interested macrocell UE, and the corresponding complexity is in the order of $TN$, which is relatively small. Therefore, we suppose that the complexity of Algorithm~\ref{algo_Aopt_UE_association_hetnet} is not prohibitively high, largely because its operation only relies on the semi-static values of $\lambda^{\textrm{DL}}(q)$ and $\lambda^{\textrm{UL}}(q)$, not on the dynamic values of $\omega^{\textrm{DL}}(q)$ and $\omega^{\textrm{UL}}(q)$.

%As a result, Algorithm 2 will run based on smooth statistics, which is updated much slower than the coherence time of channel state information. Nevertheless, finding an alternative low-complexity algorithm is desirable and it will be our future work.

%The purpose of presenting Algorithm~\ref{algo_Aopt_UE_association_hetnet} is to demonstrate the gains of dynamic TDD in HetNets. As will be shown in later sections, Algorithm~\ref{algo_Aopt_UE_association_hetnet} serves its purpose very well.

\subsection{Dynamic DL/UL Subframe Splitting in the Small Cell Tier}
\label{sec:pico_split_hetnet}

Following the dynamic DL/UL subframe splitting algorithm (see Algorithm~\ref{algo_pico_dyn_split_homo}) proposed for the HomSCNs,
we also propose a dynamic algorithm to compute
the instantaneous small cell dynamic TDD DL/UL subframe splitting for a given small cell according to its instantaneous traffic conditions in a HetNet.
%Following the design of Algorithm~\ref{algo_pico_stat_split_hetnet} with its main target was to find ABS duty cycle and UE association,
Similar to Algorithm~\ref{algo_pico_dyn_split_homo},
the proposed algorithm is performed every $T$ subframes and is based on the criterion of \textbf{Objective~2},
i.e., to minimize the difference between the instantaneous DL and UL traffic demand densities in each small cell.
Considering our previous discussion in Section~\ref{sec:Aopt_UE_association},
the instantaneous DL and UL traffic demand densities of $c(n)$ for given number of dynamic UL subframes $t$ are defined in a similar way as in (\ref{eq:d_s_DL_dynTDD_sf}) and (\ref{eq:d_s_UL_dynTDD_sf}) with $\omega$ instead of $\lambda$,
i.e.,
\begin{equation}
\tilde{d}^{\textrm{S,DL,dynTDD\_sf}}_{n}(t)=\frac{\sum_{k=1}^{K_3(n)} \omega^{\textrm{DL}}\left(r^\textrm{S}_{n,k}\right)}{f^{\textrm{S,dynTDD}}-t},
	\label{eq:d_tilde_s_DL_dynTDD_sf}
	\vspace{-0.1cm}
\end{equation}
and
\begin{equation}
\begin{split}
\tilde{d}&^{\textrm{S,UL,dynTDD\_sf}}_{n}(t)=\\
&\frac{1}{t}\left[\sum_{k=1}^{K_2(n)}\omega^{\textrm{UL}}\left(q^{\textrm{S}}_{n,k}\right)\hspace{-0pt}+\hspace{-4pt}
\sum_{k=1}^{K_3(n)}\omega^{\textrm{UL}}\left(r^\textrm{S}_{n,k}\right)\right].
\end{split}
\label{eq:d_tilde_s_UL_dynTDD_sf}
\vspace{-0.1cm}
%\tilde{d}^{\textrm{S,UL,dynTDD\_sf}}_{n}(t)=
%\frac{1}{t}\left[\sum_{k=1}^{K_2(n)}\omega^{\textrm{UL}}\left(q^{\textrm{S}}_{n,k}\right)\hspace{-0pt}+\hspace{-4pt}
%\sum_{k=1}^{K_3(n)}\omega^{\textrm{UL}}\left(r^\textrm{S}_{n,k}\right)\right].
%\label{eq:d_tilde_s_UL_dynTDD_sf}
%\vspace{-0.1cm}
\end{equation}

Then, similar to (\ref{eq:t_stat_hetnet}) considering \textbf{Objective~2},
the optimal number of instantaneous dynamic TDD UL subframes for small cell $c(n)$ can be selected as
\begin{equation}
\begin{split}
t&^{\textrm{INST\_het}}_{n}=\\
&\underset{t=g\left(r\right),r\in\Upsilon^{\textrm{het}}}{\arg\min}\left\{\left| \tilde{d}^{\textrm{S,UL,dynTDD\_sf}}_{n}(t) - \tilde{d}^{\textrm{S,DL,dynTDD\_sf}}_{n}(t)\right|\right\}.
\end{split}
\label{eq:t_dyn_hetnet}
%\vspace{-0.1cm}
%t^{\textrm{INST\_het}}_{n}=
%\underset{t=g\left(r\right),r\in\Upsilon^{\textrm{het}}}{\arg\min}\left\{\left| \tilde{d}^{\textrm{S,UL,dynTDD\_sf}}_{n}(t) - \tilde{d}^{\textrm{S,DL,dynTDD\_sf}}_{n}(t)\right|\right\}.
%\label{eq:t_dyn_hetnet}
%\vspace{-0.1cm}
\end{equation}

The proposed algorithm to split the dynamic TDD DL/UL subframes for small cell $c(n)$ in a HetNet is summarized in Algorithm~\ref{algo_pico_dyn_split_hetnet}.
Note that Algorithm~\ref{algo_pico_dyn_split_hetnet} is built on the same principle as that of  Algorithm~\ref{algo_pico_dyn_split_homo}
so that our design of dynamic TDD for small cells is coherent for both HomSCNs and HetNets.
Similar to the consideration on the range of $t$ for Algorithm~\ref{algo_pico_dyn_split_homo},
here we also impose constraints on $t$
so that the DL/UL control/reference signal channels are always available for the small cell TDD system to function properly.
Since $f^{\rm M, DL}\geq1$ (the macrocell DL should never be completely deactivated),
which indicates the availability of DL subframes for the small cell tier in every $T$ subframes,
we assume that $\Upsilon^{\textrm{het}}$ contains TDD configurations with $g(r)\in{\left[1,f^{\rm S,dynTDD}\right]}$.
%the minimum and maximum values of $t$ being no less than one and no larger than $f^{\rm S,dynTDD}$, respectively.
Moreover, as indicated in Algorithm~\ref{algo_pico_dyn_split_homo},
when a small cell is completely idle with neither DL nor UL traffic demand,
we propose that $t^{\textrm{INST\_het}}_n$ should be set to $t^{\textrm{STAT\_het}}_n$
so that the DL/UL subframe splitting in the small cell matches its statistical traffic pattern.

\begin{algorithm}[t]
\caption{\small{Selection of the optimal number of instantaneous dynamic TDD UL subframes in a small cell, i.e., $t^{\textrm{INST\_het}}_n$}, for a HetNet}
\label{algo_pico_dyn_split_hetnet}
\begin{algorithmic}
\small{
\STATE Obtain $f^{\rm S,dynTDD} = f^{\rm M, UL} + A^{\rm opt}$ via Algorithm~\ref{algo_Aopt_UE_association_hetnet}.
\STATE Compute $\tilde{d}^{\textrm{S,DL,dynTDD\_sf}}_{n}(t)$ and $\tilde{d}^{\textrm{S,UL,dynTDD\_sf}}_{n}(t)$ using (\ref{eq:d_tilde_s_DL_dynTDD_sf}) and (\ref{eq:d_tilde_s_UL_dynTDD_sf}), respectively.
\STATE Select $t^{\textrm{INST\_het}}_n$ using the following procedure.

\IF {$\omega^{\textrm{UL}}(q^\textrm{S}_{n,k})= 0,\forall{q}^\textrm{S}_{n,k} \in Q^\textrm{S}_{n}$ $\textbf{and}$
$\omega^{\textrm{UL}}(r^\textrm{S}_{n,k})= 0,\forall{r}^\textrm{S}_{n,k} \in R^\textrm{M2S}_{n}$ $\textbf{and}$
$\omega^{\textrm{DL}}(r^\textrm{S}_{n,k})= 0,\forall{r}^\textrm{S}_{n,k} \in R^\textrm{M2S}_{n}$}
\STATE Obtain $t^{\textrm{INST\_het}}_n = t^{\textrm{STAT\_het}}_n$, which is computed using Algorithm~\ref{algo_pico_stat_split_hetnet} with $u\left(q^{\textrm{M}}_{m,\pi(j)}\right)=\emptyset$.
\ELSE
\STATE Obtain $t^{\textrm{INST\_het}}_n$ using (\ref{eq:t_dyn_hetnet}).
\ENDIF
}
\end{algorithmic}
\end{algorithm}

\section{System-Level Simulation}
\label{sec:simulator}

In order to verify the effectiveness of the proposed dynamic TDD schemes,
system-level simulations are used.
As indicated in Section~\ref{sec:scenario},
we concentrate our analysis on the 3GPP dynamic TDD Scenario~3 and Scenario~6,
illustrated in Fig.~\ref{fig:illust_dynTDDHomo} and Fig.~\ref{fig:illust_dynTDDHetNet}, respectively.
Detailed information on our system-level simulator used for this analysis can be found in~\cite{simulator}.
The full list of system parameters and traffic modelling methodology can be found in~\cite{TR36.828} and~\cite{TR36.814}, respectively.
Some key parameters in our simulations are presented in Table~\ref{Table:Sim_para}.

\begin{table}
\begin{center}
\caption{Key simulation parameters}
\vspace{-0.1cm}
\label{Table:Sim_para}
\scalebox{0.8}{
\begin{tabular}{|l|l|}
  \hline
  Parameters & Assumptions\\ \hline \hline
  Scenario & Scenario~3 or Scenario~6~\cite{TR36.828} \\ \hline
  Network layout & 7 cell sites, 3 macrocells per cell site, wrap-around \\ \hline
  Inter-site distance & 500\,m~\cite{TR36.814} \\ \hline
  \# small cells per macrocell & 4 (84 small cells in total)~\cite{TR36.828} \\ \hline
  Small cell deployment & Random deployment, 40\,m radius of coverage~\cite{TR36.828} \\ \hline
  \# UEs per macrocell & 0 (Scenario~3), 10 (Scenario~6)~\cite{TR36.828} \\ \hline
  \# UEs per small cell & 10 (Scenario~3), 5 (Scenario~6)~\cite{TR36.828} \\ \hline
  System bandwidth & 10\,MHz~\cite{TR36.814} \\ \hline
  UE deployment & Uniform and random deployment in cell coverage \\ \hline
  \# macro/small cell antenna & 4 (for both transmission and reception) \\ \hline
  \# UE antenna & 2 (for both transmission and reception) \\ \hline
  Receiver type & Basic MMSE Rx for both the DL and the UL~\cite{TR36.828} \\ \hline
  Codebook for PMI feedback & LTE~Release 11 codebook with WB rank adaptation \\ \hline
%  CSI feedback periodicity & 50 ms \\ \hline
%  CSI feedback delay & 10 ms \\ \hline
%  Max macrocell Tx power & 46\,dBm~\cite{TR36.828} \\ \hline
%  Max small cell Tx power & 24\,dBm~\cite{TR36.828} \\ \hline
%  Max UE Tx power & 23\,dBm~\cite{TR36.828} \\ \hline
  UE scheduling in each cell & Proportional fairness (PF) \\ \hline
  Packet scheduling for each UE & Round Robin (RR) \\ \hline
  Modulation \& coding schemes & QPSK, 16QAM, 64QAM, 256 QAM \\ \hline
  Ideal genie-aided LA & Target BLER being 0.1 for both the DL and the UL \\ \hline
  IC capability & For DL: none\\ & For UL: with or without perfect DL-to-UL IC \\ \hline
  Non-data overhead & 3 out of 14 OFDM symbols per subframe \\ \hline
  HARQ modelling & Retransmission in the first available subframe \\ \hline
  Small-scale fading channel & Explicitly modelled (EPA channel~\cite{EPAchannel})\\ \hline
%[David]: it seems that the models used in the homo and hetnet papers are not the same EPA channel model
%[Ming]: We re-conducted the simulations for both papers and the same multi-path fading model was adopted. Please see our final versions for those two ICC papers.
%[David]: Very good.

\end{tabular}}
\end{center}
\vspace{-0.8cm}
\end{table}
%[David]: Do all parameters hold for the homo and hetnet scenarios?
%[Ming]: Fortunately, YES!!!:) Note that previously we discussed that the traffic load parameters should be different for homonet (84 pico with 10 UEs per pico + 0 macro) and hetnet (84 pico with 5 UEs per pico + 21 macro with 10 UEs per macro). I have added some explanation in the last paragraph of this subsection. Also please note that in our ICC paper on homonet, the UE antenna configuration is "1 for Tx and 2 for Rx", which is different from that in our hetnet paper. As promised, I have finished the simulations for the configuration of "2 for both Tx and Rx". So in this paper, let's present new results for the homonet and try to draw some coherent conclusions for both the homonet and the hetnet. Cheers!:)
%[David]: Very  good.

In our simulations, the traffic model is assumed to be Poisson distributed with
$\lambda^{\textrm{DL}}\left(u(q)\right)$ taking a uniform value for all UEs~\cite{TR36.828}.
Different values of $\lambda^{\textrm{DL}}\left(u(q)\right)$ correspond to different traffic load conditions,
i.e., low, medium, and high traffic loads.
Besides, $\lambda^{\textrm{UL}}\left(u(q)\right)$ is assumed to be half of $\lambda^{\textrm{DL}}\left(u(q)\right)$,
i.e., $\lambda^{\textrm{UL}}\left(u(q)\right)=\frac{1}{2}\lambda^{\textrm{DL}}\left(u(q)\right)$~\cite{TR36.828}.
The packet size is 0.5$\,$Mbytes.
Packets are independently generated for the DL and the UL in each small cell,
and they are randomly assigned to small cell UEs.
Finally, we assume that $T=10$~\cite{TS36.213}.

Due to the inherently different topology of HomSCNs and HetNets as well as the CRE and eICIC operations in HetNets,
it is generally very difficult to accurately compare the performance of two networks respectively associated with Scenario~3 and Scenario~6.
Nevertheless, in the following sections we will try to draw some useful conclusions regarding the comparison of dynamic TDD in HomSCNs and that in HetNets.
To that end,
in our simulations, as suggested in~\cite{TR36.828},
we deploy 10 UEs per small cell in Scenario~3,
while we deploy 10 UEs per macrocell as well as 5 UEs per small cell in Scenario~6.
So the simulated Scenario~3 network is slightly more crowded with UEs than the simulated Scenario~6 network.
As a result, for Scenario~3,
the values of $\lambda^{\textrm{DL}}\left(u(q)\right)$ are set to \{0.05, 0.25, 0.45\} packets per UE per second to represent the low, medium and high traffic loads, respectively.
In contrast, for Scenario~6 to achieve a similar load,
the values of $\lambda^{\textrm{DL}}\left(u(q)\right)$ are slightly increased to \{0.1, 0.3, 0.5\} packets per UE per second due to its relatively lower UE density.
Note that here we assume $\lambda^{\textrm{DL}}\left(u(q)\right)$ is independent of the UE index $q$,
because we want to focus on a case with the same $\lambda^{\textrm{DL}}\left(u(q)\right)$ for all UEs.
This facilitates the extraction of conclusions on the functioning of dynamic TDD and interference mitigation techniques that are not biased by the traffic model.
However, it should be clarified that no restriction is imposed on the values of $\lambda^{\textrm{DL}}\left(u(q)\right)$ in the proposed algorithms,
which ensures their feasibility in general cases.
Also note that the aggregate traffic load for each cell should be the product of $\lambda^{\textrm{DL}}\left(u(q)\right)$ and the number of served UEs,
which roughly injects more than 4 packets into each small per second in case of high traffic load,
i.e., $\lambda^{\textrm{DL}}\left(u(q)\right)=0.45$.

With regard to key performance indicators,
UPT is adopted in this paper.
According to~\cite{TR36.814},
UPT is defined as the ratio of successfully transmitted bits over the time consumed to transmit the said data bits,
where the consumed time starts when the DL/UL packet arrives at the UE DL/UL buffer and ends when the last bit of the DL/UL packet is correctly decoded.

% deleted due to page limit
It is important to note that an ideal genie-aided LA mechanism is adopted for both HomSCNs and HetNets in this study.
In more detail, appropriate modulation and coding schemes are chosen according to the perceived SINRs \emph{after} the DL/UL transmissions.
We make such assumption due to the following reasons:
\begin{itemize}
  \item
  Some results in our previous work on dynamic TDD~\cite{HomodynTDD_ICC}~\cite{HetNetdynTDD_ICC} were lacking insights and seemed counter-intuitively small
  because a simple LA mechanism was assumed therein,
  and hence the true value of dynamic TDD was not fully revealed,
  as a result of using such a non-ideal practical link adapter.
  \item
  To make a fair performance comparison between dynamic TDD in HomSCNs and in HetNets,
  a common LA algorithm should be assumed and the ideal genie-aided LA mechanism is a good choice
  since it provides the performance upper bounds for the considered networks.
  \item As can be well imagined,
  the fluctuation of interference in dynamic TDD transmissions should be significantly larger than that in static TDD ones.
  How to harness such interference fluctuation and perform a good LA function in practical networks are far from trivial,
  and are out of the scope of this paper.
  Therefore, considering an ideal genie-aided LA mechanism becomes the logical choice.
\end{itemize}

% deleted due to page limit
It is also important to note that in both our algorithm design and our simulation evaluation,
we adopt some ideal assumptions such as the ideal genie-aided LA mechanism,
the perfect inter-cell IC function (if considered),
the perfect knowledge of $\omega^{\textrm{DL}}(q)$ and $\omega^{\textrm{UL}}(q)$ for the instantaneous splitting of dynamic TDD subframes, etc.
Our intention is to conduct a performance evaluation to show the \emph{potentials} of dynamic TDD in current and future networks,
and it will be our future work to consider more practical assumptions in our study.
For example, although it is feasible for a UE to report its UL buffer size to its serving BS in the LTE networks,
some mismatch between the reported buffer size and the actual one still exists due to the quantization error and the feedback error~\cite{Book_LTE-A}. S
uch errors should be considered properly in a more detailed study.

\section{HomSCN Results}
\label{sec:homoResults}

In this section, we present numerical results to compare the performance of the existing static TDD scheme in LTE Release~11
with that of dynamic TDD transmissions in LTE Release~12
and an enhanced version with full flexibility of dynamic TDD configuration,
which probably falls into the scope of LTE future releases.
We also investigate the performance gains of dynamic TDD with the basic ILIM schemes presented in Section~\ref{sec:ILIMS} and their combinations.
The study is performed for $\lambda^{\textrm{DL}}\left(u(q)\right)=\{0.05, 0.25, 0.45\}$,
as explained in Section~\ref{sec:simulator}.

For LTE future releases,
apart from the existing 7~TDD configurations defined for $\Upsilon^{\textrm{homo}}$ in LTE Release~12,
another 3~TDD configurations favouring the UL transmissions with DL/UL subframe ratios being 1/9, 2/8, and 3/7, respectively, are added into $\Upsilon^{\textrm{homo}}$.
It should be noted that the DL/UL subframe ratio in LTE Release~12 cannot go below 2/3~\cite{TS36.213},
while in the hypothetical LTE future release network,
the ratio now ranges freely from 1/9 to 9/1,
and hence the system can achieve full flexibility of dynamic TDD configuration.
The purpose of investigating a hypothetical $\Upsilon^{\textrm{homo}}$ of LTE future releases is to check the performance limit.

Considering the ILIM schemes addressed in Section~\ref{sec:ILIMS},
the corresponding parameters are explained in the following.
For the CC scheme,
the coupling loss threshold $PL^{\textrm{CC}}$ for small cells within a cell cluster is chosen as 90$\,$dB~\cite{TR36.828}.
%For the DLPR and the ULPB schemes,
%$\Delta P^{\textrm{DL}}$ and $\Delta P^{\textrm{UL}}$ are set to -20\,dB~\cite{SHARP_DLPC_dynTDD} and 10\,dB~\cite{SHARP_ULPC_dynTDD}, respectively.
For the ULPB scheme,
$\Delta P^{\textrm{UL}}$ is set to 10\,dB~\cite{SHARP_ULPC_dynTDD}.
For the UOIC scheme,
the parameter $x_1$ is set to $x_1=9$\,dB so that about half of the UEs are labelled as cell-edge UEs and treated in IC in our simulations,
reducing the complexity by approximately 50\,\% compared with the full IC scheme.
Besides, for the BOIC scheme,
the parameter $x_2$  is chosen as $x_2=120$\,dB,
leading to around 2.3 BSs treated in IC on average in our simulations.
As a result, the complexity of the BOIC scheme is slashed to approximately 2.3/83$\approx$2.77\,\% compared with the full IC scheme (83 neighbouring BSs in our simulations).
Note that an additional parametric study for the proposed UOIC and BOIC schemes could be useful.
However, the basic conclusion should be obvious: different $x_1$ and $x_2$ parameters can achieve difference balances between complexity and performance,
i.e., the proposed UOIC(BOIC) scheme will become the full IC scheme with the highest complexity when $x_1$($x_2$) approaches infinity,
and the proposed UOIC(BOIC) scheme will degenerate to the non-IC scheme with the lowest complexity when $x_1$($x_2$) approaches zero.
In order to keep our discussion concise and concentrate on the complexity reduction of the proposed partial IC schemes,
we omit the parametric study and directly show the efficiency of the proposed UOIC/BOIC scheme using the parameters that achieve comparable performance with the full IC scheme.

\subsection{Performance of DL/UL UPTs with Basic ILIM}

In this subsection,
we investigate the performance of DL/UL UPTs for dynamic TDD with various basic ILIM schemes:
\begin{itemize}
\item Scheme~1: LTE Release~12 baseline static TDD with TDD configuration~3
as in~\cite{TS36.213},
where the DL/UL subframe ratio is 7:3.
Note that the assumed TDD DL/UL subframe splitting optimally matches the ratio of $\lambda^{\textrm{DL}}\left(u(q)\right)$ over $\lambda^{\textrm{UL}}\left(u(q)\right)$ when $T=10$.
\item Scheme~2: LTE Release~12 dynamic TDD ($T_{0}$) with no ILIM.
\item Scheme~3: LTE Release~12 dynamic TDD ($T_{1}$) with no ILIM.
\item Scheme~4: Scheme 3 with CC.
%\item Scheme~5: Scheme 3 with DLPR.
\item Scheme~5: Scheme 3 with ULPB.
\item Scheme~6: Scheme 3 with full IC.
\item Scheme~6(a): Scheme 3 with UOIC.
\item Scheme~6(b): Scheme 3 with BOIC.
\item Scheme 7: Hypothetical LTE future release dynamic TDD ($T_{1}$) with full IC.
\end{itemize}

Here, the periodicities of dynamic TDD reconfiguration are $T_{0}=200\,\textrm{ms}$ and $T_{1}=10\,\textrm{ms}$ for comparison purposes.
Schemes~6(a) and~6(b) are the proposed partial IC schemes previously discussed in Section~\ref{sec:ILIMS}.

Table~\ref{tab:Homo_relative_gain_results_basicILIM} shows the relative performance gains of dynamic TDD with basic ILIM
compared with the static TDD scheme (Scheme~1) in terms of 95-, 50-, and 5-percentile DL/UL UPTs, respectively.
The absolute results for Scheme~1 are also provided in Table~\ref{tab:Homo_relative_gain_results_basicILIM}
so that the absolute results for other schemes can be easily derived.

%\begin{figure}
% \centering
% \subfigure[95-percentile DL and UL UPTs (basic ILIM)]{
%  \includegraphics[width=8.5cm]{figures/Homo_95ile_UPT_basic}
%   \label{fig:Homo_95ile_UPT_basicILIM}
%   }
% \subfigure[50-percentile DL and UL UPTs (basic ILIM)]{
%  \includegraphics[width=8.5cm]{figures/Homo_50ile_UPT_basic}
%   \label{fig:Homo_50ile_UPT_basicILIM}
%   }
% \subfigure[5-percentile DL and UL UPTs (basic ILIM)]{
%  \includegraphics[width=8.5cm]{figures/Homo_5ile_UPT_basic}
%   \label{fig:Homo_5ile_UPT_basicILIM}
%   }
% \caption[DL and UL UPTs (basic ILIM)]{DL and UL UPTs (basic ILIM)}
% \label{fig:Homo_UPT_basicILIM}
%\end{figure}

%[David]: For the sake of space, I am wondering whether we should delete the figures and just keep the pics, adding the absolution values in addition to the gains. I think this would be the best approach.
%[Ming]: I guess you mean "keep the tabs". I agree with you. Compared with the tables containing all the numerical results, the figures seem to be redundant and occupy 2+ pages. Let's remove them from this version and forward versions. No worries about the time I've spent to put these figures together. The figures have been and would be excellent illustrative materials during our internal or external discussions. Feel free to use them in the future :)
%[David]: Agreed.

\noindent
\begin{center}
\begin{table*}
\caption{Relative performance gains of DL and UL UPTs (HomSCN, basic ILIM)}
\vspace{-0.1cm}
\label{tab:Homo_relative_gain_results_basicILIM}

\centering{}{\footnotesize }
\scalebox{0.95}{
\begin{tabular}{|c|r|r|r|r|r|r|r|r|r|}
\hline
{\footnotesize 95-percentile UPTs} & {\footnotesize $\begin{array}{c}
\textrm{Sch.~1 (Mbps)}
\end{array}$ } & {\footnotesize $\begin{array}{c}
\textrm{Sch.~2}
\end{array}$} & {\footnotesize $\begin{array}{c}
\textrm{Sch.~3}
\end{array}$ } & {\footnotesize $\begin{array}{c}
\textrm{Sch.~4}
\end{array}$} & {\footnotesize $\begin{array}{c}
\textrm{Sch.~5}
\end{array}$ } & {\footnotesize $\begin{array}{c}
\textrm{Sch.~6}
\end{array}$ } & {\footnotesize $\begin{array}{c}
%\textrm{Sch.~7}
%\end{array}$ } & {\footnotesize $\begin{array}{c}
\textrm{Sch.~6(a)}
\end{array}$ } & {\footnotesize $\begin{array}{c}
\textrm{Sch.~6(b)}
\end{array}$ } & {\footnotesize $\begin{array}{c}
\textrm{Sch.~7}
\end{array}$}\tabularnewline
\hline
\hline
{\footnotesize DL $\lambda^{\textrm{DL}}\left(u(q)\right)=0.05$} & {\footnotesize 62.50 (-)} & {\footnotesize 4.92\%} & {\footnotesize 25.49\%} & {\footnotesize 25.49\%} & {\footnotesize 25.49\%} & {\footnotesize 25.49\%} & {\footnotesize 25.49\%} & {\footnotesize 25.49\%} & {\footnotesize 25.49\%}\tabularnewline
\hline
{\footnotesize DL $\lambda^{\textrm{DL}}\left(u(q)\right)=0.25$} & {\footnotesize 57.97 (-)} & {\footnotesize 1.47\%} & {\footnotesize 25.45\%} & {\footnotesize 23.21\%} & {\footnotesize 27.78\%} & {\footnotesize 25.45\%} & {\footnotesize 25.45\%} & {\footnotesize 25.45\%} & {\footnotesize 27.78\%}\tabularnewline
\hline
{\footnotesize DL $\lambda^{\textrm{DL}}\left(u(q)\right)=0.45$} & {\footnotesize 47.06 (-)} & {\footnotesize -4.49\%} & {\footnotesize 23.19\%} & {\footnotesize 11.70\%} & {\footnotesize 34.92\%} & {\footnotesize 28.40\%} & {\footnotesize 23.19\%} & {\footnotesize 25.00\%} & {\footnotesize 32.81\%}\tabularnewline
\hline
{\footnotesize UL $\lambda^{\textrm{UL}}\left(u(q)\right)=0.05/2$} & {\footnotesize 13.75 (-)} & {\footnotesize 78.53\%} & {\footnotesize 100.69\%} & {\footnotesize 102.08\%} & {\footnotesize 246.43\%} & {\footnotesize 106.38\%} & {\footnotesize 102.08\%} & {\footnotesize 103.50\%} & {\footnotesize 212.90\%}\tabularnewline
\hline
{\footnotesize UL $\lambda^{\textrm{UL}}\left(u(q)\right)=0.25/2$} & {\footnotesize 12.66 (-)} & {\footnotesize 70.81\%} & {\footnotesize 88.10\%} & {\footnotesize 89.22\%} & {\footnotesize 232.63\%} & {\footnotesize 106.54\%} & {\footnotesize 97.50\%} & {\footnotesize 100.00\%} & {\footnotesize 219.19\%}\tabularnewline
\hline
{\footnotesize UL $\lambda^{\textrm{UL}}\left(u(q)\right)=0.45/2$} & {\footnotesize 10.84 (-)} & {\footnotesize 57.69\%} & {\footnotesize 80.88\%} & {\footnotesize 74.06\%} & {\footnotesize 229.46\%} & {\footnotesize 108.42\%} & {\footnotesize 88.27\%} & {\footnotesize 93.19\%} & {\footnotesize 229.46\%}\tabularnewline
\hline
\hline
{\footnotesize 50-percentile UPTs} & {\footnotesize $\begin{array}{c}
\textrm{Sch.~1 (Mbps)}
\end{array}$ } & {\footnotesize $\begin{array}{c}
\textrm{Sch.~2}
\end{array}$} & {\footnotesize $\begin{array}{c}
\textrm{Sch.~3}
\end{array}$ } & {\footnotesize $\begin{array}{c}
\textrm{Sch.~4}
\end{array}$} & {\footnotesize $\begin{array}{c}
\textrm{Sch.~5}
\end{array}$ } & {\footnotesize $\begin{array}{c}
\textrm{Sch.~6}
\end{array}$ } & {\footnotesize $\begin{array}{c}
%\textrm{Sch.~7}
%\end{array}$ } & {\footnotesize $\begin{array}{c}
\textrm{Sch.~6(a)}
\end{array}$ } & {\footnotesize $\begin{array}{c}
\textrm{Sch.~6(b)}
\end{array}$ } & {\footnotesize $\begin{array}{c}
\textrm{Sch.~7}
\end{array}$}\tabularnewline
\hline
\hline
{\footnotesize DL $\lambda^{\textrm{DL}}\left(u(q)\right)=0.05$} & {\footnotesize 38.83 (-)} & {\footnotesize -7.21\%} & {\footnotesize 24.10\%} & {\footnotesize 19.77\%} & {\footnotesize 27.16\%} &
{\footnotesize 24.10\%} & {\footnotesize 23.03\%} & {\footnotesize 22.62\%} & {\footnotesize 24.10\%}\tabularnewline
\hline
{\footnotesize DL $\lambda^{\textrm{DL}}\left(u(q)\right)=0.25$} & {\footnotesize 23.12 (-)} & {\footnotesize -4.42\%} & {\footnotesize 16.11\%} & {\footnotesize -0.29\%} & {\footnotesize 30.08\%} & {\footnotesize 22.26\%} & {\footnotesize 20.56\%} & {\footnotesize 20.98\%} & {\footnotesize 21.00\%}\tabularnewline
\hline
{\footnotesize DL $\lambda^{\textrm{DL}}\left(u(q)\right)=0.45$} & {\footnotesize 7.18 (-)} & {\footnotesize 14.61\%} & {\footnotesize 21.24\%} & {\footnotesize -1.07\%} & {\footnotesize 52.39\%} & {\footnotesize 48.14\%} & {\footnotesize 28.34\%} & {\footnotesize 41.01\%} & {\footnotesize 53.02\%}\tabularnewline
\hline
{\footnotesize UL $\lambda^{\textrm{UL}}\left(u(q)\right)=0.05/2$} & {\footnotesize 11.46 (-)} & {\footnotesize 63.85\%} & {\footnotesize 92.82\%} & {\footnotesize 92.82\%} & {\footnotesize 245.54\%} & {\footnotesize 98.86\%} & {\footnotesize 96.07\%} & {\footnotesize 94.97\%} & {\footnotesize 195.76\%}\tabularnewline
\hline
{\footnotesize UL $\lambda^{\textrm{UL}}\left(u(q)\right)=0.25/2$} & {\footnotesize 6.98 (-)} & {\footnotesize 74.43\%} & {\footnotesize 91.64\%} & {\footnotesize 94.90\%} & {\footnotesize 239.05\%} & {\footnotesize 143.83\%} & {\footnotesize 119.54\%} & {\footnotesize 131.98\%} & {\footnotesize 258.13\%}\tabularnewline
\hline
{\footnotesize UL $\lambda^{\textrm{UL}}\left(u(q)\right)=0.45/2$} & {\footnotesize 3.18 (-)} & {\footnotesize 61.70\%} & {\footnotesize 69.77\%} & {\footnotesize 77.68\%} & {\footnotesize 191.88\%} & {\footnotesize 156.73\%} & {\footnotesize 95.65\%} & {\footnotesize 135.36\%} & {\footnotesize 209.09\%}\tabularnewline
\hline
\hline
{\footnotesize 5-percentile UPTs} & {\footnotesize $\begin{array}{c}
\textrm{Sch.~1 (Mbps)}
\end{array}$ } & {\footnotesize $\begin{array}{c}
\textrm{Sch.~2}
\end{array}$} & {\footnotesize $\begin{array}{c}
\textrm{Sch.~3}
\end{array}$ } & {\footnotesize $\begin{array}{c}
\textrm{Sch.~4}
\end{array}$} & {\footnotesize $\begin{array}{c}
\textrm{Sch.~5}
\end{array}$ } & {\footnotesize $\begin{array}{c}
\textrm{Sch.~6}
\end{array}$ } & {\footnotesize $\begin{array}{c}
%\textrm{Sch.~7}
%\end{array}$ } & {\footnotesize $\begin{array}{c}
\textrm{Sch.~6(a)}
\end{array}$ } & {\footnotesize $\begin{array}{c}
\textrm{Sch.~6(b)}
\end{array}$ } & {\footnotesize $\begin{array}{c}
\textrm{Sch.~7}
\end{array}$}\tabularnewline
\hline
\hline
{\footnotesize DL $\lambda^{\textrm{DL}}\left(u(q)\right)=0.05$} & {\footnotesize 17.32 (-)} & {\footnotesize 1.03\%} & {\footnotesize 21.80\%} & {\footnotesize 13.99\%} & {\footnotesize 26.92\%} & {\footnotesize 22.87\%} & {\footnotesize 20.06\%} & {\footnotesize 19.93\%} & {\footnotesize 20.27\%}\tabularnewline
\hline
{\footnotesize DL $\lambda^{\textrm{DL}}\left(u(q)\right)=0.25$} & {\footnotesize 2.76 (-)} & {\footnotesize 65.84\%} & {\footnotesize 65.83\%} & {\footnotesize 21.91\%} & {\footnotesize 91.71\%} & {\footnotesize 113.02\%} & {\footnotesize 113.64\%} & {\footnotesize 98.14\%} & {\footnotesize 114.52\%}\tabularnewline
\hline
{\footnotesize DL $\lambda^{\textrm{DL}}\left(u(q)\right)=0.45$} & {\footnotesize 0.50 (-)} & {\footnotesize 35.11\%} & {\footnotesize 26.84\%} & {\footnotesize 6.16\%} & {\footnotesize 69.19\%} & {\footnotesize 90.02\%} & {\footnotesize 70.66\%} & {\footnotesize 90.03\%} & {\footnotesize 99.06\%}\tabularnewline
\hline
{\footnotesize UL $\lambda^{\textrm{UL}}\left(u(q)\right)=0.05/2$} & {\footnotesize 6.26 (-)} & {\footnotesize 90.03\%} & {\footnotesize 125.42\%} & {\footnotesize 116.44\%} & {\footnotesize 275.59\%} & {\footnotesize 133.67\%} & {\footnotesize 129.68\%} & {\footnotesize 124.82\%} & {\footnotesize 224.94\%}\tabularnewline
\hline
{\footnotesize UL $\lambda^{\textrm{UL}}\left(u(q)\right)=0.25/2$} & {\footnotesize 2.31 (-)} & {\footnotesize 84.91\%} & {\footnotesize 120.20\%} & {\footnotesize 141.03\%} & {\footnotesize 225.14\%} & {\footnotesize 212.25\%} & {\footnotesize 181.79\%} & {\footnotesize 202.44\%} & {\footnotesize 285.11\%}\tabularnewline
\hline
{\footnotesize UL $\lambda^{\textrm{UL}}\left(u(q)\right)=0.45/2$} & {\footnotesize 0.71 (-)} & {\footnotesize 6.19\%} & {\footnotesize 3.60\%} & {\footnotesize 42.26\%} & {\footnotesize 48.28\%} & {\footnotesize 114.41\%} & {\footnotesize 65.98\%} & {\footnotesize 103.73\%} & {\footnotesize 125.52\%}\tabularnewline
\hline
\end{tabular}
}
\vspace{-0.2cm}
\end{table*}
\end{center}

%Sanity check: When the traffic load is low, e.g., $\lambda^{\textrm{DL}}\left(u(q)\right)=0.05$,
%the performance gains of Scheme~7 in terms of the 95-percentile DL UPT and the 95-percentile UL UPT should be around
%2/7$\approx$28.6\,\% (7 DL subframes increasing to a maximum of 9 DL subframes per 10 subframes) and
%3/3=100\,\% (3 UL subframes increasing to a maximum of 6 UL subframes per 10 subframes), respectively.
%This is because that the 95-percentile DL and UL UPTs are contributed by cell-interior UEs,
%which are insusceptible to small inter-cell interference particularly when the traffic load is low and the IC function is engaged to mitigate the inter-link interference.
%Besides, when the traffic load is low, the coupling of DL scheduling and UL scheduling is quite weak.
%Hence, the UPT gains are mainly determined by the amount of additional transmission subframes in the DL or in the UL.
%Similarly, when the traffic load is low, e.g., $\lambda^{\textrm{DL}}\left(u(q)\right)=0.05$,
%the performance gains of Scheme~8 in terms of the 95-percentile UL UPT should be around 6/3=200\,\% (3 UL subframes increasing to a maximum of 9 UL subframes per 10 subframes), respectively.

% Straightforward
Compared with the baseline static TDD scheme (Scheme~1),
the straightforward dynamic TDD scheme with $T_{1}$ (Scheme~3) shows solid gains in most performance categories.
However, it shows no gain in terms of the 5-percentile UL UPT when the traffic load is high,
i.e.., $\lambda^{\textrm{DL}}\left(u(q)\right)=0.45$.
This is due to the lack of ILIM to mitigate the DL-to-UL interference.
%At high traffic loads, there is also little room to divert DL subframes to UL or vice versa.
Moreover, a faster dynamic TDD configuration time scale (Scheme~3) is shown to outperform a slower one (Scheme~2) in almost every performance category,
as previously reported in~\cite{Shen2012} and~\cite{HomodynTDD_ICC}.

% CC
In order to improve performance in terms of the 5-percentile UL UPT,
the CC scheme (Scheme~4) can be adopted.
Note that the efficiency of the CC scheme degrades when the traffic grows,
since the flexibility of dynamic TDD is reduced as
all the small cells in a cluster adapt their TDD configuration according to the aggregated traffic in the cluster rather than to their individual traffic conditions.
Still, CC brings a considerable improvement of 42.26$\,$\% in the 5-percentile UL UPT when the traffic load is high,
i.e., $\lambda^{\textrm{DL}}\left(u(q)\right)=0.45$
at the expense of 10$\sim$20$\,$\% sacrifice in DL UPTs
compared with the straightforward dynamic TDD scheme (Scheme~3).

%DLPR
%The DLPR scheme with 10dB power reduction (Scheme~5) shows no or marginal performance gain in all DL and UL UPTs compared with the CC scheme (Scheme~4) when $\lambda^{\textrm{DL}}\left(u(q)\right)=0.05$. When the traffic load is relatively high, e.g., $\lambda^{\textrm{DL}}\left(u(q)\right)=0.45$,
%the DLPR scheme (Scheme~5) is shown to be far less effective than the CC scheme (Scheme~4) in mitigating the DL-to-UL interference,
%showing a significant loss in terms of the 5-percentile UL UPT.
%This is because that the CC scheme (Scheme~4) can completely remove some strong interference from adjacent BSs,
%while the DLPR scheme (Scheme~5) only reduces the interference to some extent.

% ULPB
The ULPB scheme (Scheme~5) is also quite useful to boost the UL UPTs by 225.14\,\%$\sim$275.59\,\% when the traffic load is low to medium,
i.e., $\lambda^{\textrm{DL}}\left(u(q)\right)\leq0.25$,
indicating that the UL network is generally power limited.
However, when the traffic load is high
i.e., $\lambda^{\textrm{DL}}\left(u(q)\right)=0.45$,
the performance gain in terms of the 5-percentile UL UPT, albeit considerable, decreases by 48.28\,\%,
since the power headroom of a cell-edge UE tends to be quickly drained up and increasing UL power leads to more serious UL interference.
Overall, ULPB follows a similar trend as CC.

% IC
The IC schemes (Schemes~6 \&~7) are shown to bring substantial gains in every performance category compared with the baseline static TDD scheme (Schemes~1),
for all the considered traffic loads.
In particular, among the considered ILIM schemes,
the IC schemes (Schemes~6 \&~7) provide the largest performance gain of 114.41\,\%$\sim$125.52\,\% in terms of the 5-percentile UL UPT
with no loss in the DL UPTs when the traffic load is relatively high,
e.g., $\lambda^{\textrm{DL}}\left(u(q)\right)=0.45$.

As for the proposed partial IC schemes,
it is interesting to find that the BOIC scheme (Schemes~6(b)) achieves similar results with small losses in every performance category compared with the full IC scheme (Schemes~6).
This is because in current networks where small cells are not ultra-densely deployed,
only a few BSs are the dominant interferers.
Thus, cancelling the DL-to-UL interference from those BSs is already good enough to achieve satisfactory performances~\cite{TS36.213}.
%To be more specific, in practical small cell networks,
%the existence of line-of-sight (LoS) transmission has a significant impact on the BS-to-BS path loss.
%For example, in~\cite{TS36.213},
%compared with the BS-to-BS path loss with LoS,
%the BS-to-BS path loss with non-LoS (NLoS) is severely penalized by about 70\,dB as well as a double loss exponent.
%Since the probability of BS-to-BS LoS decreases drastically with the increase of inter-BS distance due to building obstruction~\cite{TR36.828},
%the number of neighbouring BSs with LoS and thus low path losses to a given BS is quite limited on average.
%
In contrast, the UOIC scheme (Schemes~6(a)) turns out to be much less effective than the BOIC scheme (Schemes~6(b)),
especially in improving UL UPTs when the traffic load is medium to high.
This is because in realistic scenarios
cell-center UEs are also vulnerable to dominant DL-to-UL interference in dynamic TDD,
since BS-to-BS path loss could be orders of magnitude smaller than UE-to-BS path loss~\cite{TS36.213}.
%For example, as indicated in~\cite{TS36.213},
%compared with the dominant BS-to-BS path loss with LoS,
%the UE-to-BS path loss with NLoS is severely penalized by about 50\,dB as well as a large path loss exponent of 3.75.
As a result, even a cell-center UE with good link quality cannot combat such a large difference in signal reception levels.
Thus, we conclude that if partial IC should be used to reduce the complexity/cost of full IC,
the BOIC scheme is a much more preferable choice than the UOIC scheme.

% deleted due to page limit
Another important note is that compared with our previous work on dynamic TDD in HomSCNs~\cite{HomodynTDD_ICC},
the performance gains of dynamic TDD are considerably larger in this study.
In particular, unlike that in~\cite{HomodynTDD_ICC},
the straightforward dynamic TDD scheme (Scheme~3) is shown to be able to work on its own with positive gains in all performance categories over the baseline Scheme~1.
This is because an ideal genie-aided LA mechanism is used,
as discussed in Section~\ref{sec:simulator},
so that the full potential of dynamic TDD in HomSCNs can be exposed.
This shows the importance of LA and the need for designing a practical LA algorithm in dynamic TDD networks,
which will be part of our future work.

% deleted due to page limit
As a summary, dynamic TDD provides substantial UTP gains compared to the static TDD,
the gains depending on the quality of the considered ILIM scheme.
The (partial) IC schemes have been shown to provide the most significant gains at the expense of a higher complexity,
and such performance gain in terms of the 5-percentile UL UPT becomes much more obvious when the traffic load is medium to heavy,
where the DL-to-UL interference occurs frequently.

\subsection{Performance of DL/UL UPTs with Combined ILIM}

In the following, the following combined  ILIM schemes are considered:
\begin{itemize}
\item Scheme~8: Combined Schemes~4 and 5.
\item Scheme~9: Combined Schemes~4 and 6.
%\item Scheme~11: Combined Schemes~4 and 8.
\item Scheme~10: Combined Schemes~5 and 6.
\item Scheme~10(b): Combined Schemes~5 and 6(b).
%\item Scheme~13: Combined Schemes~6 and 8.
\item Scheme~11: Combined Schemes~4, 5 and 6.
\item Scheme~12: Combined Schemes~5 and 7.
%\item Scheme~15: Combined Schemes~4, 6 and 8.
\end{itemize}

Table~\ref{tab:Homo_relative_gain_results_combILIM} shows the relative performance gains of dynamic TDD with combined ILIM
compared with the static TDD scheme (Scheme~1) in terms of 95-, 50-, and 5-percentile DL/UL UPTs.
Note that the absolute results for Scheme~1 are also provided in Table~\ref{tab:Homo_relative_gain_results_combILIM}
so that the absolute results for other schemes can be easily obtained.

%[David]: For the sake of space, I am wondering whether we should delete the figures and just keep the pics, adding the absolution values in addition to the gains. I think this would be the best approach.
%[Ming]: I agree with you. Compared with the tables containing all the numerical results, the figures seem to be redundant and occupy 2+ pages. Let's remove them from this version and forward versions. No worries about the time I've spent to put these figures together. The figures have been and would be excellent illustrative materials during our internal or external discussions. Feel free to use them in the future :)

%\begin{figure}
% \centering
% \subfigure[95-percentile DL and UL UPTs (combined ILIM)]{
%  \includegraphics[width=8.5cm]{figures/Homo_95ile_UPT_combn}
%   \label{fig:Homo_95ile_UPT_combILIM}
%   }
% \subfigure[50-percentile DL and UL UPTs (combined ILIM)]{
%  \includegraphics[width=8.5cm]{figures/Homo_50ile_UPT_combn}
%   \label{fig:Homo_50ile_UPT_combILIM}
%   }
% \subfigure[5-percentile DL and UL UPTs (combined ILIM)]{
%  \includegraphics[width=8.5cm]{figures/Homo_5ile_UPT_combn}
%   \label{fig:Homo_5ile_UPT_combILIM}
%   }
% \caption[DL and UL UPTs (combined ILIM)]{DL and UL UPTs (combined ILIM)}
% \label{fig:Homo_UPT_combILIM}
%\end{figure}

\noindent
\begin{center}
\begin{table*}
\caption{Relative performance gains of DL and UL UPTs (HomSCN, combined ILIM)}
\vspace{-0.1cm}
\label{tab:Homo_relative_gain_results_combILIM}

\centering{}{\footnotesize }
\scalebox{1.0}{
\begin{tabular}{|c|r|r|r|r|r|r|r|}
\hline
{\footnotesize 95-percentile UPTs} & {\footnotesize $\begin{array}{c}
\textrm{Sch.~1 (Mbps)}
\end{array}$ } & {\footnotesize $\begin{array}{c}
\textrm{Sch.~8}
\end{array}$} & {\footnotesize $\begin{array}{c}
\textrm{Sch.~9}
\end{array}$ } & {\footnotesize $\begin{array}{c}
\textrm{Sch.~10}
\end{array}$} & {\footnotesize $\begin{array}{c}
\textrm{Sch.~10(b)}
\end{array}$ } & {\footnotesize $\begin{array}{c}
\textrm{Sch.~11}
\end{array}$ } & {\footnotesize $\begin{array}{c}
\textrm{Sch.~12}
\end{array}$}\tabularnewline
\hline
\hline
{\footnotesize DL $\lambda^{\textrm{DL}}\left(u(q)\right)=0.05$} & {\footnotesize 62.50 (-)} & {\footnotesize 25.49\%} & {\footnotesize 25.49\%} & {\footnotesize 25.49\%} & {\footnotesize 25.49\%} & {\footnotesize 25.49\%} & {\footnotesize 25.49\%}\tabularnewline
\hline
{\footnotesize DL $\lambda^{\textrm{DL}}\left(u(q)\right)=0.25$} & {\footnotesize 57.97 (-)} & {\footnotesize 25.45\%} & {\footnotesize 23.21\%} & {\footnotesize 27.78\%} & {\footnotesize 27.78\%} & {\footnotesize 25.45\%} & {\footnotesize 27.78\%}\tabularnewline
\hline
{\footnotesize DL $\lambda^{\textrm{DL}}\left(u(q)\right)=0.45$} & {\footnotesize 47.06 (-)} & {\footnotesize 25.00\%} & {\footnotesize 13.33\%} & {\footnotesize 34.92\%} & {\footnotesize 34.92\%} & {\footnotesize 25.00\%} & {\footnotesize 37.10\%}\tabularnewline
\hline
{\footnotesize UL $\lambda^{\textrm{UL}}\left(u(q)\right)=0.05/2$} & {\footnotesize 13.75 (-)} & {\footnotesize 246.43\%} & {\footnotesize 104.93\%} & {\footnotesize 250.60\%} & {\footnotesize 246.43\%} & {\footnotesize 250.60\%} & {\footnotesize 429.09\%}\tabularnewline
\hline
{\footnotesize UL $\lambda^{\textrm{UL}}\left(u(q)\right)=0.25/2$} & {\footnotesize 12.66 (-)} & {\footnotesize 243.48\%} & {\footnotesize 100.00\%} & {\footnotesize 263.22\%} & {\footnotesize 255.06\%} & {\footnotesize 255.06\%} & {\footnotesize 454.39\%}\tabularnewline
\hline
{\footnotesize UL $\lambda^{\textrm{UL}}\left(u(q)\right)=0.45/2$} & {\footnotesize 10.84 (-)} & {\footnotesize 241.67\%} & {\footnotesize 91.19\%} & {\footnotesize 288.42\%} & {\footnotesize 272.73\%} & {\footnotesize 258.25\%} & {\footnotesize 495.16\%}\tabularnewline
\hline
\hline
{\footnotesize 50-percentile UPTs} & {\footnotesize $\begin{array}{c}
\textrm{Sch.~1 (Mbps)}
\end{array}$ } & {\footnotesize $\begin{array}{c}
\textrm{Sch.~8}
\end{array}$} & {\footnotesize $\begin{array}{c}
\textrm{Sch.~9}
\end{array}$ } & {\footnotesize $\begin{array}{c}
\textrm{Sch.~10}
\end{array}$} & {\footnotesize $\begin{array}{c}
\textrm{Sch.~10(b)}
\end{array}$ } & {\footnotesize $\begin{array}{c}
\textrm{Sch.~11}
\end{array}$ } & {\footnotesize $\begin{array}{c}
\textrm{Sch.~12}
\end{array}$}\tabularnewline
\hline
\hline
{\footnotesize DL $\lambda^{\textrm{DL}}\left(u(q)\right)=0.05$} & {\footnotesize 38.83 (-)} & {\footnotesize 24.10\%} & {\footnotesize 19.77\%} & {\footnotesize 27.16\%} & {\footnotesize 24.10\%} & {\footnotesize 22.62\%} & {\footnotesize 28.75\%}\tabularnewline
\hline
{\footnotesize DL $\lambda^{\textrm{DL}}\left(u(q)\right)=0.25$} & {\footnotesize 23.12 (-)} & {\footnotesize 20.14\%} & {\footnotesize 1.17\%} & {\footnotesize 36.45\%} & {\footnotesize 36.22\%} & {\footnotesize 18.49\%} & {\footnotesize 35.69\%}\tabularnewline
\hline
{\footnotesize DL $\lambda^{\textrm{DL}}\left(u(q)\right)=0.45$} & {\footnotesize 7.18 (-)} & {\footnotesize 40.83\%} & {\footnotesize 4.14\%} & {\footnotesize 85.67\%} & {\footnotesize 83.83\%} & {\footnotesize 45.81\%} & {\footnotesize 84.44\%}\tabularnewline
\hline
{\footnotesize UL $\lambda^{\textrm{UL}}\left(u(q)\right)=0.05/2$} & {\footnotesize 11.46 (-)} & {\footnotesize 252.53\%} & {\footnotesize 96.07\%} & {\footnotesize 259.79\%} & {\footnotesize 252.53\%} & {\footnotesize 254.01\%} & {\footnotesize 428.79\%}\tabularnewline
\hline
{\footnotesize UL $\lambda^{\textrm{UL}}\left(u(q)\right)=0.25/2$} & {\footnotesize 6.98 (-)} & {\footnotesize 276.97\%} & {\footnotesize 104.64\%} & {\footnotesize 381.51\%} & {\footnotesize 369.67\%} & {\footnotesize 292.47\%} & {\footnotesize 653.95\%}\tabularnewline
\hline
{\footnotesize UL $\lambda^{\textrm{UL}}\left(u(q)\right)=0.45/2$} & {\footnotesize 3.18 (-)} & {\footnotesize 250.42\%} & {\footnotesize 92.94\%} & {\footnotesize 437.61\%} & {\footnotesize 419.71\%} & {\footnotesize 267.84\%} & {\footnotesize 580.00\%}\tabularnewline
\hline
\hline
{\footnotesize 5-percentile UPTs} & {\footnotesize $\begin{array}{c}
\textrm{Sch.~1 (Mbps)}
\end{array}$ } & {\footnotesize $\begin{array}{c}
\textrm{Sch.~8}
\end{array}$} & {\footnotesize $\begin{array}{c}
\textrm{Sch.~9}
\end{array}$ } & {\footnotesize $\begin{array}{c}
\textrm{Sch.~10}
\end{array}$} & {\footnotesize $\begin{array}{c}
\textrm{Sch.~10(b)}
\end{array}$ } & {\footnotesize $\begin{array}{c}
\textrm{Sch.~11}
\end{array}$ } & {\footnotesize $\begin{array}{c}
\textrm{Sch.~12}
\end{array}$}\tabularnewline
\hline
\hline
{\footnotesize DL $\lambda^{\textrm{DL}}\left(u(q)\right)=0.05$} & {\footnotesize 17.32 (-)} & {\footnotesize 23.53\%} & {\footnotesize 16.29\%} & {\footnotesize 29.05\%} & {\footnotesize 25.34\%} & {\footnotesize 23.76\%} & {\footnotesize 26.07\%}\tabularnewline
\hline
{\footnotesize DL $\lambda^{\textrm{DL}}\left(u(q)\right)=0.25$} & {\footnotesize 2.76 (-)} & {\footnotesize 79.97\%} & {\footnotesize 52.70\%} & {\footnotesize 136.06\%} & {\footnotesize 126.26\%} & {\footnotesize 59.93\%} & {\footnotesize 151.90\%}\tabularnewline
\hline
{\footnotesize DL $\lambda^{\textrm{DL}}\left(u(q)\right)=0.45$} & {\footnotesize 0.50 (-)} & {\footnotesize 62.68\%} & {\footnotesize 17.99\%} & {\footnotesize 153.39\%} & {\footnotesize 137.10\%} & {\footnotesize 56.98\%} & {\footnotesize 171.25\%}\tabularnewline
\hline
{\footnotesize UL $\lambda^{\textrm{UL}}\left(u(q)\right)=0.05/2$} & {\footnotesize 6.26 (-)} & {\footnotesize 291.72\%} & {\footnotesize 117.18\%} & {\footnotesize 331.42\%} & {\footnotesize 317.32\%} & {\footnotesize 291.72\%} & {\footnotesize 480.45\%}\tabularnewline
\hline
{\footnotesize UL $\lambda^{\textrm{UL}}\left(u(q)\right)=0.25/2$} & {\footnotesize 2.31 (-)} & {\footnotesize 353.66\%} & {\footnotesize 162.18\%} & {\footnotesize 539.48\%} & {\footnotesize 518.94\%} & {\footnotesize 369.65\%} & {\footnotesize 597.27\%}\tabularnewline
\hline
{\footnotesize UL $\lambda^{\textrm{UL}}\left(u(q)\right)=0.45/2$} & {\footnotesize 0.71 (-)} & {\footnotesize 162.55\%} & {\footnotesize 53.74\%} & {\footnotesize 303.12\%} & {\footnotesize 287.03\%} & {\footnotesize 183.24\%} & {\footnotesize 366.93\%}\tabularnewline
\hline
\end{tabular}
}
\vspace{-0.6cm}
\end{table*}
\end{center}

%[Ming]: Adding another table with absolute results would consume a lot of space. Hence, I propose that we only add the absolute results for Scheme~1. What do you think of it?
%[David]: Yes, I think it is the best solution.

As can be observed from Table~\ref{tab:Homo_relative_gain_results_combILIM},
the combined CC and ULPB scheme (Scheme~8) is strictly superior to the combined CC and IC scheme (Scheme~9).
%and the combined CC and IC scheme with the hypothetical LTE future release dynamic TDD configurations (Scheme~11).
This is because the CC scheme and IC scheme are somehow redundant,
i.e., the CC scheme already eliminates dominant interfering small cells for the UL by coordination,
rendering the IC process less effective.
%In contrast, in the combined CC and ULPB scheme (Scheme~10),
%on top of the interference mitigation provided by CC,
%UEs are granted to use a larger power that leads to a better performance.
When the traffic load is relatively heavy,
e.g., $\lambda^{\textrm{DL}}\left(u(q)\right)=0.45$,
Scheme~8 greatly outperforms the static TDD scheme (Scheme~1) by 25.00\,\%$\sim$62.68\,\% and 162.55\,\%$\sim$250.42\,\%
%around 25$\sim$60$\,$\% and 160$\sim$250$\,$\%
in terms of the DL and the UL UPTs, respectively.
%Scheme~11 outperforms Scheme~10 in terms of UL UPTs because of the full flexibility of dynamic TDD configurations.
%(Note by Ming: I intend to remove Scheme~11 to keep our discussion more concise.)
%[David]: We can go ahead removing Scheme~11 but them I feel we should remove all cases using Scheme~8, thus I have remove Scheme~13 and Scheme~15 too.
%[Ming]: Please have a look at the current selected schemes. The thought of removing scheme~7 and scheme~12 also occurred to me. The only reason I'd like to keep them is because they give the performance limits of the network with full flexibility of dynamic TDD configurations. Other than that, they are not very meaningful in our paper. What do you think?

The combined ULPB and IC scheme (Schemes~10) is the most powerful combination,
which substantially increases the UL performance due to the larger transmit power at UEs and the IC capabilities at BSs.
Some of the tremendous performance gain in the UL is also shown to be transferred to the DL by means of the traffic-adaptive dynamic TDD scheduling.
To be more specific,
since the performance in the UL is enhanced,
some UL subframes can be transformed into DL subframes,
thus improving the DL performance.
When the traffic load is medium to high,
e.g., $\lambda^{\textrm{DL}}\left(u(q)\right) \geq 0.25$,
Scheme~10 is shown to significantly outperform the static TDD scheme (Scheme~1) by
27.78\,\%$\sim$153.39\,\% and 263.22\,\%$\sim$539.48\,\%
%approximately 25$\sim$150$\,$\% and 260$\sim$540$\,$\%
in terms of the DL and the UL UPTs, respectively.
In order to reduce the complexity of IC,
the combination of the ULPB and the BOIC schemes (Scheme~10(b)) is proposed here.
As can be seen from Table~\ref{tab:Homo_relative_gain_results_combILIM},
Schemes~10(b) achieves a similar UPT performance compared with Schemes~10,
but with a much lower complexity of the IC operations.
%[David]: I have removed Scheme~13.
%[Ming]: Agreed.

Finally, the combination of all three ILIM schemes (Scheme~11) only gives similar performance as that of the combination of CC and ULPB (Scheme~8),
which does not justify the employment of IC on top of the joint operation of CC and ULPB.
This is again because the CC scheme and  IC scheme are somehow redundant.
%Only with the hypothetical LTE future release dynamic TDD configurations activated,
%the full combination scheme (Scheme~15) can outperform Scheme~9 in the UL.
%This is again because the CC scheme is not very compatible with the IC scheme. (Note by Ming: I didn't do Scheme~15 this time to keep our discussion concise.)
%[David]: I have removed Scheme~15.
%[Ming]: Agreed.
Moreover, the combined ULPB and IC scheme with full flexibility of dynamic TDD configuration (Schemes~12) is investigated to show the performance upper bound.
As can be seen from Table~\ref{tab:Homo_relative_gain_results_combILIM},
Scheme~12 significantly outperforms the static TDD scheme (Scheme~1) by
25.49\,\%$\sim$171.25\,\% and 366.93\,\%$\sim$653.95\,\%
%around 25$\sim$170$\,$\% and 360$\sim$650$\,$\%
in terms of the DL and the UL UPTs, respectively.

% deleted due to page limit
To sum up, if it is preferable to find an easy-to-implement scheme with reasonable performance gains,
Scheme~8 should be called upon.
But if complexity issue is a minor concern,
Scheme~10(b) should be engaged to realize the full potential of dynamic TDD.

\section{HetNet Results}
\label{sec:hetnetResults}

In this section,
we present numerical results to benchmark the performance of
static/dynamic TDD in HetNets.

%Similar as in the previous section,
%since $\lambda^{\textrm{DL}}\left(u(q)\right):\lambda^{\textrm{UL}}\left(u(q)\right)=2:1$,
%We assume that in dynamic TDD the probabilities of observed DL and UL subframes are $\alpha^{\textrm{M,DL}}=\frac 2 3$ and $\alpha^{\textrm{M,UL}}=\frac 1 3$, respectively.
We assume that the range expansion bias is $y=9$\,dB,
as suggested in some previous work on CRE~\cite{R1-132304}.
Moreover, the study is performed for $\lambda^{\textrm{DL}}\left(u(q)\right)=\{0.1, 0.3, 0.5\}$,
as explained in Section~\ref{sec:simulator}.
For the considered HetNet,
after running Algorithm~\ref{algo_Aopt_UE_association_hetnet},
we found that
$f^{\textrm{M,DL}}=5$,
$f^{\textrm{M,UL}}=3$,
$A^{\textrm{opt}}=2$,
and that approximately 1/3 of macrocell UEs are off-loaded to small cells.
Thus, 5\,subframes in every 10\,subframes are used as dynamic TDD subframes in small cells,
i.e., $f^{\textrm{S,dynTDD}}=5$.
%In the following,
%the performance in terms of 95-, 50- and 5-percentile UPTs for the DL and the UL in such HetNet scenario is investigated,

In this light,
the following schemes are considered for benchmarking,
\begin{itemize}
\item
Scheme~A~(Static TDD scheme without CRE and ABS):
LTE Release~12 TDD configuration 3 for both macrocells and small cells (DL/UL subframe ratio = 7:3~\cite{TS36.213}).
%Note that the assumed TDD DL/UL subframe splitting optimally matches the ratio of DL/UL traffic arriving rates when $T=10$.
\item
Scheme~B~(Straightforward dynamic TDD scheme without CRE and ABS):
macrocell (DL/UL subframe ratio = 7:3),
small cell (dynamic TDD without CRE and ABS).
Note that Algorithm~\ref{algo_Aopt_UE_association_hetnet} is used to determine the dynamic TDD DL/UL subframe splitting for the small cells.
%with the constraint in LTE Release~12 that the lowest and highest DL/UL subframe ratios are 2/3 and 9/1, respectively~\cite{TS36.213}.
\item
Scheme~C~(Static TDD scheme with CRE and ABS):
macrocell (DL/ABS/UL subframe ratio = 5:2:3),
small cell (DL/UL subframe ratio = 7:3).
Note that the scheduling policy in~\cite{R1-132304} is adopted where
DL packets of ER UEs should be scheduled with a high priority in subframes overlapping with the macrocell ABSs
and that they should not be scheduled in subframes overlapping with the macrocell DL subframes.
\item
Scheme~D~(Proposed scheme without IC):
macrocell (DL/ABS/UL subframe ratio = 5:2:3),
small cell (DL/dynamic TDD subframe ratio = 5:5),
dynamic TDD reconfiguration per 10 ms,
and no IC.
\item
Scheme~E:~Scheme~D plus small cell DL to macrocell UL IC.
\item
Scheme~F:~Scheme~E plus small cell DL to small cell UL full IC,
i.e., Scheme~6 in Section~\ref{sec:homoResults}.
\item
Scheme~F(b):~Scheme~E plus the BS oriented small cell DL to small cell UL partial IC,
i.e., Scheme~6(b) in Section~\ref{sec:homoResults}.
\end{itemize}

%Fig.~\ref{fig:Het_UPT_all} Fig.~\ref{fig:Het_UPT_macro} and Fig.~\ref{fig:Het_UPT_pico}
Table~\ref{tab:Het_relative_gain_results} shows the relative performance gains of the considered schemes
compared with the baseline static TDD scheme (Scheme~A) in terms of 95-, 50-, and 5-percentile DL/UL UPTs.
Note that the absolute results for Scheme~A are also provided in Table~\ref{tab:Het_relative_gain_results}
so that the absolute results for other schemes can be easily obtained.
Moreover, in Table~\ref{tab:Het_relative_gain_results},
apart from the overall performances,
the UPT results are broken down to show the contributions from the macrocell and the small cell tiers, respectively.

\noindent
\begin{center}
\begin{table*}
\caption{Relative performance gains of DL and UL UPTs (HetNets)}
\vspace{-0.1cm}
\label{tab:Het_relative_gain_results}
\centering{}{\footnotesize }
\scalebox{1.0}{
\begin{tabular}{|c|r|r|r|r|r|r|r|}
\hline
{\footnotesize 95-percentile UPTs (overall)} & {\footnotesize $\begin{array}{c}
\textrm{Sch.~A (Mbps)}
\end{array}$ } & {\footnotesize $\begin{array}{c}
\textrm{Sch.~B}
\end{array}$} & {\footnotesize $\begin{array}{c}
\textrm{Sch.~C}
\end{array}$ } & {\footnotesize $\begin{array}{c}
\textrm{Sch.~D}
\end{array}$} & {\footnotesize $\begin{array}{c}
\textrm{Sch.~E}
\end{array}$ } & {\footnotesize $\begin{array}{c}
\textrm{Sch.~F}
\end{array}$ } & {\footnotesize $\begin{array}{c}
\textrm{Sch.~F(b)}
\end{array}$}\tabularnewline
\hline
\hline
{\footnotesize DL $\lambda^{\textrm{DL}}\left(u(q)\right)=0.1$} & {\footnotesize 42.55 (-)} & {\footnotesize 11.90\%} & {\footnotesize 9.30\%} & {\footnotesize 38.24\%} & {\footnotesize 40.30\%} & {\footnotesize 38.24\%} & {\footnotesize 38.24\%}\tabularnewline
\hline
{\footnotesize DL $\lambda^{\textrm{DL}}\left(u(q)\right)=0.3$} & {\footnotesize 15.81 (-)} & {\footnotesize 54.75\%} & {\footnotesize 91.88\%} & {\footnotesize 155.56\%} & {\footnotesize 155.56\%} & {\footnotesize 158.16\%} & {\footnotesize 158.16\%}\tabularnewline
\hline
{\footnotesize DL $\lambda^{\textrm{DL}}\left(u(q)\right)=0.5$} & {\footnotesize 11.20 (-)} & {\footnotesize 63.76\%} & {\footnotesize 97.40\%} & {\footnotesize 166.42\%} & {\footnotesize 162.50\%} & {\footnotesize 166.42\%} & {\footnotesize 164.44\%}\tabularnewline
\hline
{\footnotesize UL $\lambda^{\textrm{UL}}\left(u(q)\right)=0.1/2$} & {\footnotesize 13.58 (-)} & {\footnotesize -25.99\%} & {\footnotesize 17.81\%} & {\footnotesize 79.59\%} & {\footnotesize 79.92\%} & {\footnotesize 83.28\%} & {\footnotesize 82.94\%}\tabularnewline
\hline
{\footnotesize UL $\lambda^{\textrm{UL}}\left(u(q)\right)=0.3/2$} & {\footnotesize 12.16 (-)} & {\footnotesize -30.00\%} & {\footnotesize 14.46\%} & {\footnotesize 71.35\%} & {\footnotesize 72.25\%} & {\footnotesize 95.83\%} & {\footnotesize 84.83\%}\tabularnewline
\hline
{\footnotesize UL $\lambda^{\textrm{UL}}\left(u(q)\right)=0.5/2$} & {\footnotesize 11.50 (-)} & {\footnotesize -32.21\%} & {\footnotesize 10.05\%} & {\footnotesize 50.54\%} & {\footnotesize 47.63\%} & {\footnotesize 75.63\%} & {\footnotesize 65.28\%}\tabularnewline
\hline
\hline
{\footnotesize 50-percentile UPTs (overall)} & {\footnotesize $\begin{array}{c}
\textrm{Sch.~A (Mbps)}
\end{array}$ } & {\footnotesize $\begin{array}{c}
\textrm{Sch.~B}
\end{array}$} & {\footnotesize $\begin{array}{c}
\textrm{Sch.~C}
\end{array}$ } & {\footnotesize $\begin{array}{c}
\textrm{Sch.~D}
\end{array}$} & {\footnotesize $\begin{array}{c}
\textrm{Sch.~E}
\end{array}$ } & {\footnotesize $\begin{array}{c}
\textrm{Sch.~F}
\end{array}$ } & {\footnotesize $\begin{array}{c}
\textrm{Sch.~F(b)}
\end{array}$}\tabularnewline
\hline
\hline
{\footnotesize DL $\lambda^{\textrm{DL}}\left(u(q)\right)=0.1$} & {\footnotesize 17.62 (-)} & {\footnotesize 21.39\%} & {\footnotesize 18.23\%} & {\footnotesize 36.75\%} & {\footnotesize 36.75\%} & {\footnotesize 36.75\%} & {\footnotesize 36.75\%}\tabularnewline
\hline
{\footnotesize DL $\lambda^{\textrm{DL}}\left(u(q)\right)=0.3$} & {\footnotesize 2.19 (-)} & {\footnotesize 41.86\%} & {\footnotesize 285.02\%} & {\footnotesize 432.07\%} & {\footnotesize 425.94\%} & {\footnotesize 435.19\%} & {\footnotesize 422.92\%}\tabularnewline
\hline
{\footnotesize DL $\lambda^{\textrm{DL}}\left(u(q)\right)=0.5$} & {\footnotesize 0.88 (-)} & {\footnotesize 27.83\%} & {\footnotesize 299.10\%} & {\footnotesize 476.37\%} & {\footnotesize 468.79\%} & {\footnotesize 488.31\%} & {\footnotesize 478.20\%}\tabularnewline
\hline
{\footnotesize UL $\lambda^{\textrm{UL}}\left(u(q)\right)=0.1/2$} & {\footnotesize 9.96 (-)} & {\footnotesize -36.77\%} & {\footnotesize 7.07\%} & {\footnotesize 45.47\%} & {\footnotesize 56.84\%} & {\footnotesize 59.33\%} & {\footnotesize 57.05\%}\tabularnewline
\hline
{\footnotesize UL $\lambda^{\textrm{UL}}\left(u(q)\right)=0.3/2$} & {\footnotesize 3.20 (-)} & {\footnotesize -84.26\%} & {\footnotesize 121.28\%} & {\footnotesize 176.60\%} & {\footnotesize 188.07\%} & {\footnotesize 215.71\%} & {\footnotesize 201.26\%}\tabularnewline
\hline
{\footnotesize UL $\lambda^{\textrm{UL}}\left(u(q)\right)=0.5/2$} & {\footnotesize 1.29 (-)} & {\footnotesize -80.67\%} & {\footnotesize 267.63\%} & {\footnotesize 317.01\%} & {\footnotesize 326.45\%} & {\footnotesize 399.47\%} & {\footnotesize 368.58\%}\tabularnewline
\hline
\hline
{\footnotesize 5-percentile UPTs (overall)} & {\footnotesize $\begin{array}{c}
\textrm{Sch.~A (Mbps)}
\end{array}$ } & {\footnotesize $\begin{array}{c}
\textrm{Sch.~B}
\end{array}$} & {\footnotesize $\begin{array}{c}
\textrm{Sch.~C}
\end{array}$ } & {\footnotesize $\begin{array}{c}
\textrm{Sch.~D}
\end{array}$} & {\footnotesize $\begin{array}{c}
\textrm{Sch.~E}
\end{array}$ } & {\footnotesize $\begin{array}{c}
\textrm{Sch.~F}
\end{array}$ } & {\footnotesize $\begin{array}{c}
\textrm{Sch.~F(b)}
\end{array}$}\tabularnewline
\hline
\hline
{\footnotesize DL $\lambda^{\textrm{DL}}\left(u(q)\right)=0.1$} & {\footnotesize 6.13 (-)} & {\footnotesize 7.95\%} & {\footnotesize 4.34\%} & {\footnotesize 37.84\%} & {\footnotesize 37.84\%} & {\footnotesize 38.72\%} & {\footnotesize 38.72\%}\tabularnewline
\hline
{\footnotesize DL $\lambda^{\textrm{DL}}\left(u(q)\right)=0.3$} & {\footnotesize 0.16 (-)} & {\footnotesize 21.51\%} & {\footnotesize 439.40\%} & {\footnotesize 492.22\%} & {\footnotesize 493.72\%} & {\footnotesize 494.22\%} & {\footnotesize 493.72\%}\tabularnewline
\hline
{\footnotesize DL $\lambda^{\textrm{DL}}\left(u(q)\right)=0.5$} & {\footnotesize 0.09 (-)} & {\footnotesize 3.71\%} & {\footnotesize 153.15\%} & {\footnotesize 172.46\%} & {\footnotesize 170.47\%} & {\footnotesize 171.55\%} & {\footnotesize 170.28\%}\tabularnewline
\hline
{\footnotesize UL $\lambda^{\textrm{UL}}\left(u(q)\right)=0.1/2$} & {\footnotesize 3.43 (-)} & {\footnotesize -61.65\%} & {\footnotesize 51.94\%} & {\footnotesize -25.50\%} & {\footnotesize 78.90\%} & {\footnotesize 79.34\%} & {\footnotesize 79.34\%}\tabularnewline
\hline
{\footnotesize UL $\lambda^{\textrm{UL}}\left(u(q)\right)=0.3/2$} & {\footnotesize 0.30 (-)} & {\footnotesize -84.21\%} & {\footnotesize 380.98\%} & {\footnotesize -74.29\%} & {\footnotesize 413.76\%} & {\footnotesize 417.71\%} & {\footnotesize 412.59\%}\tabularnewline
\hline
{\footnotesize UL $\lambda^{\textrm{UL}}\left(u(q)\right)=0.5/2$} & {\footnotesize 0.13 (-)} & {\footnotesize -72.09\%} & {\footnotesize 165.06\%} & {\footnotesize -57.66\%} & {\footnotesize 193.40\%} & {\footnotesize 210.47\%} & {\footnotesize 205.88\%}\tabularnewline
\hline
\hline
\hline
{\footnotesize 95-percentile UPTs (macrocells)} & {\footnotesize $\begin{array}{c}
\textrm{Sch.~A (Mbps)}
\end{array}$ } & {\footnotesize $\begin{array}{c}
\textrm{Sch.~B}
\end{array}$} & {\footnotesize $\begin{array}{c}
\textrm{Sch.~C}
\end{array}$ } & {\footnotesize $\begin{array}{c}
\textrm{Sch.~D}
\end{array}$} & {\footnotesize $\begin{array}{c}
\textrm{Sch.~E}
\end{array}$ } & {\footnotesize $\begin{array}{c}
\textrm{Sch.~F}
\end{array}$ } & {\footnotesize $\begin{array}{c}
\textrm{Sch.~F(b)}
\end{array}$}\tabularnewline
\hline
\hline
{\footnotesize DL $\lambda^{\textrm{DL}}\left(u(q)\right)=0.1$} & {\footnotesize 47.20 (-)} & {\footnotesize 0.47\%} & {\footnotesize -10.79\%} & {\footnotesize -10.79\%} & {\footnotesize -11.58\%} & {\footnotesize -10.79\%} & {\footnotesize -10.79\%}\tabularnewline
\hline
{\footnotesize DL $\lambda^{\textrm{DL}}\left(u(q)\right)=0.3$} & {\footnotesize 6.04 (-)} & {\footnotesize 18.61\%} & {\footnotesize 209.90\%} & {\footnotesize 215.82\%} & {\footnotesize 215.82\%} & {\footnotesize 215.97\%} & {\footnotesize 212.83\%}\tabularnewline
\hline
{\footnotesize DL $\lambda^{\textrm{DL}}\left(u(q)\right)=0.5$} & {\footnotesize 2.92 (-)} & {\footnotesize 9.59\%} & {\footnotesize 159.20\%} & {\footnotesize 174.26\%} & {\footnotesize 170.47\%} & {\footnotesize 174.26\%} & {\footnotesize 174.26\%}\tabularnewline
\hline
{\footnotesize UL $\lambda^{\textrm{UL}}\left(u(q)\right)=0.1/2$} & {\footnotesize 13.08 (-)} & {\footnotesize -33.71\%} & {\footnotesize 24.69\%} & {\footnotesize -20.20\%} & {\footnotesize 26.38\%} & {\footnotesize 29.00\%} & {\footnotesize 27.33\%}\tabularnewline
\hline
{\footnotesize UL $\lambda^{\textrm{UL}}\left(u(q)\right)=0.3/2$} & {\footnotesize 5.12 (-)} & {\footnotesize -87.22\%} & {\footnotesize 70.14\%} & {\footnotesize -76.66\%} & {\footnotesize 89.80\%} & {\footnotesize 89.80\%} & {\footnotesize 89.71\%}\tabularnewline
\hline
{\footnotesize UL $\lambda^{\textrm{UL}}\left(u(q)\right)=0.5/2$} & {\footnotesize 2.08 (-)} & {\footnotesize -86.27\%} & {\footnotesize 152.76\%} & {\footnotesize -79.34\%} & {\footnotesize 184.90\%} & {\footnotesize 185.24\%} & {\footnotesize 177.72\%}\tabularnewline
\hline
\hline
{\footnotesize 50-percentile UPTs (macrocells)} & {\footnotesize $\begin{array}{c}
\textrm{Sch.~A (Mbps)}
\end{array}$ } & {\footnotesize $\begin{array}{c}
\textrm{Sch.~B}
\end{array}$} & {\footnotesize $\begin{array}{c}
\textrm{Sch.~C}
\end{array}$ } & {\footnotesize $\begin{array}{c}
\textrm{Sch.~D}
\end{array}$} & {\footnotesize $\begin{array}{c}
\textrm{Sch.~E}
\end{array}$ } & {\footnotesize $\begin{array}{c}
\textrm{Sch.~F}
\end{array}$ } & {\footnotesize $\begin{array}{c}
\textrm{Sch.~F(b)}
\end{array}$}\tabularnewline
\hline
\hline
{\footnotesize DL $\lambda^{\textrm{DL}}\left(u(q)\right)=0.1$} & {\footnotesize 17.47 (-)} & {\footnotesize 0.00\%} & {\footnotesize 8.53\%} & {\footnotesize 9.05\%} & {\footnotesize 9.05\%} & {\footnotesize 9.57\%} & {\footnotesize 9.31\%}\tabularnewline
\hline
{\footnotesize DL $\lambda^{\textrm{DL}}\left(u(q)\right)=0.3$} & {\footnotesize 0.89 (-)} & {\footnotesize 4.73\%} & {\footnotesize 295.75\%} & {\footnotesize 308.74\%} & {\footnotesize 309.48\%} & {\footnotesize 309.11\%} & {\footnotesize 306.51\%}\tabularnewline
\hline
{\footnotesize DL $\lambda^{\textrm{DL}}\left(u(q)\right)=0.5$} & {\footnotesize 0.29 (-)} & {\footnotesize 5.66\%} & {\footnotesize 265.23\%} & {\footnotesize 282.56\%} & {\footnotesize 282.06\%} & {\footnotesize 285.02\%} & {\footnotesize 283.83\%}\tabularnewline
\hline
{\footnotesize UL $\lambda^{\textrm{UL}}\left(u(q)\right)=0.1/2$} & {\footnotesize 8.02 (-)} & {\footnotesize -51.97\%} & {\footnotesize 23.51\%} & {\footnotesize -36.07\%} & {\footnotesize 26.01\%} & {\footnotesize 26.33\%} & {\footnotesize 25.69\%}\tabularnewline
\hline
{\footnotesize UL $\lambda^{\textrm{UL}}\left(u(q)\right)=0.3/2$} & {\footnotesize 1.03 (-)} & {\footnotesize -88.27\%} & {\footnotesize 247.12\%} & {\footnotesize -79.87\%} & {\footnotesize 272.84\%} & {\footnotesize 280.16\%} & {\footnotesize 275.01\%}\tabularnewline
\hline
{\footnotesize UL $\lambda^{\textrm{UL}}\left(u(q)\right)=0.5/2$} & {\footnotesize 0.44 (-)} & {\footnotesize -80.54\%} & {\footnotesize 171.67\%} & {\footnotesize -73.01\%} & {\footnotesize 209.91\%} & {\footnotesize 220.84\%} & {\footnotesize 214.40\%}\tabularnewline
\hline
\hline
{\footnotesize 5-percentile UPTs (macrocells)} & {\footnotesize $\begin{array}{c}
\textrm{Sch.~A (Mbps)}
\end{array}$ } & {\footnotesize $\begin{array}{c}
\textrm{Sch.~B}
\end{array}$} & {\footnotesize $\begin{array}{c}
\textrm{Sch.~C}
\end{array}$ } & {\footnotesize $\begin{array}{c}
\textrm{Sch.~D}
\end{array}$} & {\footnotesize $\begin{array}{c}
\textrm{Sch.~E}
\end{array}$ } & {\footnotesize $\begin{array}{c}
\textrm{Sch.~F}
\end{array}$ } & {\footnotesize $\begin{array}{c}
\textrm{Sch.~F(b)}
\end{array}$}\tabularnewline
\hline
\hline
{\footnotesize DL $\lambda^{\textrm{DL}}\left(u(q)\right)=0.1$} & {\footnotesize 5.72 (-)} & {\footnotesize -4.45\%} & {\footnotesize 13.88\%} & {\footnotesize 20.41\%} & {\footnotesize 20.24\%} & {\footnotesize 21.01\%} & {\footnotesize 21.07\%}\tabularnewline
\hline
{\footnotesize DL $\lambda^{\textrm{DL}}\left(u(q)\right)=0.3$} & {\footnotesize 0.12 (-)} & {\footnotesize 14.05\%} & {\footnotesize 298.88\%} & {\footnotesize 305.64\%} & {\footnotesize 307.90\%} & {\footnotesize 309.19\%} & {\footnotesize 308.71\%}\tabularnewline
\hline
{\footnotesize DL $\lambda^{\textrm{DL}}\left(u(q)\right)=0.5$} & {\footnotesize 0.08 (-)} & {\footnotesize 0.31\%} & {\footnotesize 91.85\%} & {\footnotesize 95.06\%} & {\footnotesize 94.94\%} & {\footnotesize 96.29\%} & {\footnotesize 95.36\%}\tabularnewline
\hline
{\footnotesize UL $\lambda^{\textrm{UL}}\left(u(q)\right)=0.1/2$} & {\footnotesize 2.87 (-)} & {\footnotesize -65.84\%} & {\footnotesize 53.68\%} & {\footnotesize -41.92\%} & {\footnotesize 62.30\%} & {\footnotesize 62.99\%} & {\footnotesize 61.79\%}\tabularnewline
\hline
{\footnotesize UL $\lambda^{\textrm{UL}}\left(u(q)\right)=0.3/2$} & {\footnotesize 0.23 (-)} & {\footnotesize -82.38\%} & {\footnotesize 249.81\%} & {\footnotesize -76.62\%} & {\footnotesize 268.03\%} & {\footnotesize 274.18\%} & {\footnotesize 268.29\%}\tabularnewline
\hline
{\footnotesize UL $\lambda^{\textrm{UL}}\left(u(q)\right)=0.5/2$} & {\footnotesize 0.11 (-)} & {\footnotesize -74.12\%} & {\footnotesize 58.54\%} & {\footnotesize -68.27\%} & {\footnotesize 73.05\%} & {\footnotesize 82.26\%} & {\footnotesize 76.25\%}\tabularnewline
\hline
\hline
\hline
{\footnotesize 95-percentile UPTs (small cells)} & {\footnotesize $\begin{array}{c}
\textrm{Sch.~A (Mbps)}
\end{array}$ } & {\footnotesize $\begin{array}{c}
\textrm{Sch.~B}
\end{array}$} & {\footnotesize $\begin{array}{c}
\textrm{Sch.~C}
\end{array}$ } & {\footnotesize $\begin{array}{c}
\textrm{Sch.~D}
\end{array}$} & {\footnotesize $\begin{array}{c}
\textrm{Sch.~E}
\end{array}$ } & {\footnotesize $\begin{array}{c}
\textrm{Sch.~F}
\end{array}$ } & {\footnotesize $\begin{array}{c}
\textrm{Sch.~F(b)}
\end{array}$}\tabularnewline
\hline
\hline
{\footnotesize DL $\lambda^{\textrm{DL}}\left(u(q)\right)=0.1$} & {\footnotesize 36.94 (-)} & {\footnotesize 29.39\%} & {\footnotesize 38.84\%} & {\footnotesize 74.67\%} & {\footnotesize 74.67\%} & {\footnotesize 74.67\%} & {\footnotesize 74.67\%}\tabularnewline
\hline
{\footnotesize DL $\lambda^{\textrm{DL}}\left(u(q)\right)=0.3$} & {\footnotesize 20.00 (-)} & {\footnotesize 53.61\%} & {\footnotesize 66.67\%} & {\footnotesize 122.22\%} & {\footnotesize 117.39\%} & {\footnotesize 122.22\%} & {\footnotesize 117.39\%}\tabularnewline
\hline
{\footnotesize DL $\lambda^{\textrm{DL}}\left(u(q)\right)=0.5$} & {\footnotesize 15.29 (-)} & {\footnotesize 51.56\%} & {\footnotesize 61.63\%} & {\footnotesize 112.68\%} & {\footnotesize 111.22\%} & {\footnotesize 115.14\%} & {\footnotesize 114.50\%}\tabularnewline
\hline
{\footnotesize UL $\lambda^{\textrm{UL}}\left(u(q)\right)=0.1/2$} & {\footnotesize 13.85 (-)} & {\footnotesize -23.40\%} & {\footnotesize 13.81\%} & {\footnotesize 80.56\%} & {\footnotesize 80.56\%} & {\footnotesize 85.19\%} & {\footnotesize 82.85\%}\tabularnewline
\hline
{\footnotesize UL $\lambda^{\textrm{UL}}\left(u(q)\right)=0.3/2$} & {\footnotesize 12.83 (-)} & {\footnotesize -31.52\%} & {\footnotesize 15.87\%} & {\footnotesize 70.37\%} & {\footnotesize 68.57\%} & {\footnotesize 91.22\%} & {\footnotesize 86.70\%}\tabularnewline
\hline
{\footnotesize UL $\lambda^{\textrm{UL}}\left(u(q)\right)=0.5/2$} & {\footnotesize 12.38 (-)} & {\footnotesize -32.14\%} & {\footnotesize 8.35\%} & {\footnotesize 50.93\%} & {\footnotesize 49.09\%} & {\footnotesize 72.73\%} & {\footnotesize 61.14\%}\tabularnewline
\hline
\hline
{\footnotesize 50-percentile UPTs (small cells)} & {\footnotesize $\begin{array}{c}
\textrm{Sch.~A (Mbps)}
\end{array}$ } & {\footnotesize $\begin{array}{c}
\textrm{Sch.~B}
\end{array}$} & {\footnotesize $\begin{array}{c}
\textrm{Sch.~C}
\end{array}$ } & {\footnotesize $\begin{array}{c}
\textrm{Sch.~D}
\end{array}$} & {\footnotesize $\begin{array}{c}
\textrm{Sch.~E}
\end{array}$ } & {\footnotesize $\begin{array}{c}
\textrm{Sch.~F}
\end{array}$ } & {\footnotesize $\begin{array}{c}
\textrm{Sch.~F(b)}
\end{array}$}\tabularnewline
\hline
\hline
{\footnotesize DL $\lambda^{\textrm{DL}}\left(u(q)\right)=0.1$} & {\footnotesize 17.70 (-)} & {\footnotesize 45.81\%} & {\footnotesize 25.56\%} & {\footnotesize 60.28\%} & {\footnotesize 60.28\%} & {\footnotesize 60.28\%} & {\footnotesize 59.15\%}\tabularnewline
\hline
{\footnotesize DL $\lambda^{\textrm{DL}}\left(u(q)\right)=0.3$} & {\footnotesize 6.62 (-)} & {\footnotesize 64.27\%} & {\footnotesize 86.00\%} & {\footnotesize 158.89\%} & {\footnotesize 157.23\%} & {\footnotesize 160.56\%} & {\footnotesize 157.23\%}\tabularnewline
\hline
{\footnotesize DL $\lambda^{\textrm{DL}}\left(u(q)\right)=0.5$} & {\footnotesize 3.17 (-)} & {\footnotesize 63.06\%} & {\footnotesize 111.22\%} & {\footnotesize 195.31\%} & {\footnotesize 187.80\%} & {\footnotesize 201.15\%} & {\footnotesize 195.32\%}\tabularnewline
\hline
{\footnotesize UL $\lambda^{\textrm{UL}}\left(u(q)\right)=0.1/2$} & {\footnotesize 11.49 (-)} & {\footnotesize -23.01\%} & {\footnotesize -0.57\%} & {\footnotesize 59.63\%} & {\footnotesize 61.11\%} & {\footnotesize 65.71\%} & {\footnotesize 60.37\%}\tabularnewline
\hline
{\footnotesize UL $\lambda^{\textrm{UL}}\left(u(q)\right)=0.3/2$} & {\footnotesize 9.98 (-)} & {\footnotesize -40.19\%} & {\footnotesize -7.39\%} & {\footnotesize 29.56\%} & {\footnotesize 30.62\%} & {\footnotesize 43.47\%} & {\footnotesize 35.93\%}\tabularnewline
\hline
{\footnotesize UL $\lambda^{\textrm{UL}}\left(u(q)\right)=0.5/2$} & {\footnotesize 7.84 (-)} & {\footnotesize -54.34\%} & {\footnotesize -16.87\%} & {\footnotesize 4.81\%} & {\footnotesize 4.38\%} & {\footnotesize 20.39\%} & {\footnotesize 12.55\%}\tabularnewline
\hline
\hline
{\footnotesize 5-percentile UPTs (small cells)} & {\footnotesize $\begin{array}{c}
\textrm{Sch.~A (Mbps)}
\end{array}$ } & {\footnotesize $\begin{array}{c}
\textrm{Sch.~B}
\end{array}$} & {\footnotesize $\begin{array}{c}
\textrm{Sch.~C}
\end{array}$ } & {\footnotesize $\begin{array}{c}
\textrm{Sch.~D}
\end{array}$} & {\footnotesize $\begin{array}{c}
\textrm{Sch.~E}
\end{array}$ } & {\footnotesize $\begin{array}{c}
\textrm{Sch.~F}
\end{array}$ } & {\footnotesize $\begin{array}{c}
\textrm{Sch.~F(b)}
\end{array}$}\tabularnewline
\hline
\hline
{\footnotesize DL $\lambda^{\textrm{DL}}\left(u(q)\right)=0.1$} & {\footnotesize 7.28 (-)} & {\footnotesize 58.61\%} & {\footnotesize -12.36\%} & {\footnotesize 44.60\%} & {\footnotesize 45.31\%} & {\footnotesize 45.49\%} & {\footnotesize 44.28\%}\tabularnewline
\hline
{\footnotesize DL $\lambda^{\textrm{DL}}\left(u(q)\right)=0.3$} & {\footnotesize 1.10 (-)} & {\footnotesize 56.61\%} & {\footnotesize 46.74\%} & {\footnotesize 208.88\%} & {\footnotesize 196.80\%} & {\footnotesize 209.30\%} & {\footnotesize 189.65\%}\tabularnewline
\hline
{\footnotesize DL $\lambda^{\textrm{DL}}\left(u(q)\right)=0.5$} & {\footnotesize 0.49 (-)} & {\footnotesize 53.18\%} & {\footnotesize 66.79\%} & {\footnotesize 187.80\%} & {\footnotesize 176.36\%} & {\footnotesize 222.03\%} & {\footnotesize 215.88\%}\tabularnewline
\hline
{\footnotesize UL $\lambda^{\textrm{UL}}\left(u(q)\right)=0.1/2$} & {\footnotesize 6.76 (-)} & {\footnotesize -26.98\%} & {\footnotesize -11.94\%} & {\footnotesize 44.27\%} & {\footnotesize 45.41\%} & {\footnotesize 48.40\%} & {\footnotesize 43.71\%}\tabularnewline
\hline
{\footnotesize UL $\lambda^{\textrm{UL}}\left(u(q)\right)=0.3/2$} & {\footnotesize 3.78 (-)} & {\footnotesize -38.84\%} & {\footnotesize -6.66\%} & {\footnotesize 37.14\%} & {\footnotesize 38.71\%} & {\footnotesize 51.25\%} & {\footnotesize 34.63\%}\tabularnewline
\hline
{\footnotesize UL $\lambda^{\textrm{UL}}\left(u(q)\right)=0.5/2$} & {\footnotesize 2.05 (-)} & {\footnotesize -68.04\%} & {\footnotesize -11.37\%} & {\footnotesize -3.71\%} & {\footnotesize -4.40\%} & {\footnotesize 31.14\%} & {\footnotesize 26.56\%}\tabularnewline
\hline
\end{tabular}
}
\vspace{-0.6cm}
\end{table*}
\end{center}

%[Ming]: Adding another table with absolute results would consume a lot of space. Hence, I propose that we only add the absolute results for Scheme~a. What do you think of it?
%[David]: agreed.

%Gains and loses of straightforward dynamic TDD
As it can be seen from Table~\ref{tab:Het_relative_gain_results},
it is easy to conclude that
the straightforward dynamic TDD in the small cell tier (Scheme~B) leads to substantial performance degradation in the UL of a HetNet,
particularly for the macrocell tier,
i.e., macrocell UL UPTs degradation of up to 88.27\,\%.
This is due to the significant inter-tier DL-to-UL interference,
which indicates the great difficulties in introducing dynamic TDD into HetNets,
if inter-link interference is not properly managed.
Similar observations were drawn for HomSCNs in Section~\ref{sec:homoResults}.
In contrast,
Scheme~B leads to performance gains in the DL of a HetNet,
especially for the small cell tier,
because the scheduler favours the UL in the small cell tier to combat the above mentioned strong inter-tier DL-to-UL interference,
and thus the interference experienced by the DL transmissions in the small cell tier is significantly reduced.
However, the observed gains in the DL UPTs do not justify the use of Scheme~B in a HetNet,
because the UL UPT reductions are enormous.
%It is important to note that such UL deficiency,
%more subframes are diverted to the UL,
%leads to a DL resource shortage.
%As will be discussed in the sequel,
%the CRE and ABS operations are much more efficient than the straightforward dynamic TDD scheme in enhancing the DL UPTs,
%while bringing considerable improvement to the UL UPTs too.
%(Note by Ming: we have a different observation this time because the newly adopted ideal LA algorithm saves the UL performance from the fatal failure,
%and thus the DL resource shortage is not so severe this time.)
%[David]: Should be comment on link adaptation?
%[Ming]: I comment on the ideal genie-aided LA mechanism at the end of Section V. Please have a look. Also I have added some related remarked in Section I.

%Gains of REB and ABS
Let us now compare the baseline scheme (Scheme~A) with the static TDD scheme with CRE and ABS operations (Scheme~C).
When the traffic load is low,
e.g., $\lambda^{\textrm{DL}}(u(q))=0.1$,
the performance gains of Scheme~C are low.
This is because interference is not a severe problem,
and thus the gain of eICIC is small  or even negative.
The 95-percentile macrocell DL UPT suffers from a loss of 10.79\,\% because 2 subframes have been converted from DL to ABS,
resulting in a moderate resource shortage.
In contrast,
when the traffic load is medium to high,
e.g., $\lambda^{\textrm{DL}}\left(u(q)\right)\geq0.3$,
in other words, when the interference is high,
it can be seen that the performance gains of Scheme~C are significant in almost all categories.
The only exception is that the 50-percentile and the 5-percentile UL UPTs of small cell UEs suffer from a slight performance loss of
6.66\,\%$\sim$ 16.87\,\%.
%about 6$\sim$16\,\%.
This negative impact is mostly caused due to the larger number of small cell UEs to share the small cell resources,
as a result of macrocell offloading through range expansion.
Having said that, it is important to notice that  the 50-percentile and the 5-percentile UL UPTs  of all UEs together still rise by
121.28\,\%$\sim$380.98\,\%,
%about 120$\sim$380\,\%,
indicating that CRE and eICIC benefit UL performance in general.
This is because the RAN has been brought closer to small cell ER UEs,
thus greatly improving the qualities of ULs.
%Regarding other performance categories,
%when the traffic load is medium, e.g., $\lambda^{\textrm{DL}}\left(u(q)\right)=0.3$,
%the 50-percentile macrocell DL and UL UPTs are boosted by about 300\,\% and 250\,\%, respectively,
%thanks to the traffic offload from the macrocell tier to the small cell tier by means of small cell range expansion.
%Moreover, in the small cell tier,
%more than 60\,\% gains in terms of 95-percentile DL UPT are achieved mainly because from time to time non-ER UEs can access the good-quality subframes aligned with macrocell ABSs
%when such subframes are not used by ER UEs.
%The seemingly performance losses in terms of the 5-percentile small cell DL UPT and the 5-percentile small cell UL UPT are also caused by the poor performance of small cell ER UEs off-loaded from the macrocell tier,
%as  has been explained early.
%[David]: I don't think we need to explain in detail the advantages of range expansion and ABS, since the paper is not about it.
%[Ming]: Agreed.

Regarding Scheme~F,
when the traffic load is low,
e.g., $\lambda^{\textrm{DL}}\left(u(q)\right)=0.1$,
the 95-percentile small cell DL UPT and the 95-percentile small cell UL UPT are basically contributed by cell-interior UEs.
This is because these UEs suffer from low  inter-cell interference and the IC function is engaged to mitigate inter-link interference.
Besides, when the traffic load is low,
the coupling of DL scheduling and UL scheduling is quite weak.
Hence, the UPT gains are mainly determined by the amount of additional transmission subframes in the DL or in the UL.
Considering that in dynamic TDD the numbers of the available DL and UL subframes per 10 subframes respectively increase from 7 to 9 and from 3 to 5,
the UPT gains of Scheme~F compared with Scheme~C in terms of the 95-percentile small cell DL UPT and the 95-percentile small cell UL UPT should be around 2/7 and 3/5, respectively.
The corresponding numerical results Table~\ref{tab:Het_relative_gain_results} confirm this observation,
indicating 95-percentile small cell DL UPT and the 95-percentile small cell UL UPT gains around 26\,\%((1.7467-1.3884)/1.3884) and 63\,\%((1.8519-1.1381)/1.1381), respectively.
Note that such insightful observation cannot be obtained in our previous work~\cite{HetNetdynTDD_ICC} because of the non-ideal link adaptor.

%DL gains of proposed schemes
As can be further observed from Table~\ref{tab:Het_relative_gain_results},
compared with Scheme~C,
the proposed schemes (Schemes~D,~E and~F) achieve superior performances in all DL UPT categories.
The additional gains on top of those of Scheme~C over Scheme~A are particularly significant for the small cell tier.
To be more specific,
additional gains of
35.83\,\%(74.67\,\%-38.84\,\%)$\sim$55.55\,\%(122.22\,\%-66.67\,\%), 34.72\,\%(60.28\,\%-25.56\,\%)$\sim$89.93\,\%(201.15\,\%-111.22\,\%), and 56.96\,\%(44.60\,\%-(-12.36\,\%))$\sim$162.56\,\%(209.30\,\%-46.74\,\%)
%about 40$\sim$60\,\%, 45$\sim$90\,\%, 60$\sim$160\,\%
can be observed in terms of 95-, 50- and 5-percentile small cell DL UPTs, respectively.
The reason for these extra gains is that dynamic TDD is able to divert idle UL subframes for DL usage,
thus boosting DL capacity.
In the proposed scheduling policy,
an ER UE may occupy as many as 5 dynamic TDD subframes for its DL transmission,
thus greatly improving the 5-percentile small cell DL UPT.

%UL gains of proposed schemes
As for the UL UPT,
gains or losses maybe observed depending on the tier and the used scheme.
When the traffic load is low to medium,
e.g., $\lambda^{\textrm{DL}}(u(q))\leq$0.3,
the small cell UL UPT performance of the proposed schemes (Schemes~D,~E and~F) improves.
In more detail, extra performance gains of
52.70\,\%(68.57\,\%-15.87\,\%)$\sim$75.35\,\%(91.22\,\%-15.87\,\%), 36.95\,\%(29.56\,\%-(-7.39\,\%))$\sim$66.28\,\%(65.71\,\%-(-0.57\,\%)) and 43.80\,\%(37.14\,\%-(-6.66\,\%))$\sim$60.34\,\%(48.40\,\%-(-11.94\,\%))
%55$\sim$75\,\%, 40$\sim$65\,\% and 45$\sim$60\,\%
are observed for the proposed schemes on top of those for Scheme~C
in terms of 95-, 50- and 5-percentile small cell UL UPTs, respectively.
The story is different for the macrocell tier,
when using Scheme~D.
In this case, macrocell UL UPTs suffer from a severe performance degradation as high as
79.87\,\%,
%70$\sim$80\,\%,
indicating that the inter-tier inter-link interference from the small cell DL is overwhelming for the macrocell UL.
Thus, the inter-tier small cell DL to macrocell UL IC is necessary for the macrocell UL to efficiently function,
if small cell dynamic TDD operation is introduced into HetNets.
When the traffic load is relatively high,
e.g., $\lambda^{\textrm{DL}}(u(q))=0.5$,
the small cell UL UPT performance of the proposed schemes (Schemes~D,~E)  improves but not as much as with low to medium traffic loads.
In more detail, extra performance gains of
40.74\,\%(49.09\,\%-8.35\,\%), 21.68\,\%(4.81\,\%-(-16.87\,\%)) and 7.66\,\%(-3.71\,\%-(-11.37\,\%))
%40\,\%, 20\,\% and 7$\sim$8\,\%
are observed for the proposed schemes (Schemes~D and~E) on top of those for Scheme~C
in terms of 95-, 50- and 5-percentile UL UPTs, respectively.
This shows  that inter-tier IC is helpless in dealing with inter-link interference inside the small cell tier.
In contrast,
the proposed Scheme~F with the required double IC,
i.e., inter-tier small cell DL to macrocell UL IC and DL-to-UL IC in the small cell tier,
considerably outperforms Scheme~C,
providing additional gains of
64.38\,\%(72.73\,\%-8.35\,\%), 37.26\,\%(20.39\,\%-(-16.87\,\%) and 42.51\,\%(31.14\,\%-(-11.37\,\%))
%65\,\%, 35\,\% and 40\,\%
in terms of 95-, 50- and 5-percentile UL UPTs, respectively.
Double IC is thus necessary to aid the UL
%especially for cell-edge UEs,
at high traffic loads.

It is important to note that the proposed partial IC scheme, i.e., Scheme~F(b),
turns out to be very efficient,
resulting in small performance losses and low complexity compared with Scheme~F (full IC).
Therefore, we conclude that for the used of dynamic TDD in HetNets,
Scheme~F(b) is a good choice,
which strikes a beneficial balance between performance and complexity.
Its nearly perfect score sheet is due to two reasons,
i.e., \emph{(i)} the CRE, ABS and small cell DL to macrocell UL IC operations handle the inter-tier interference that paves the way for efficient dynamic TDD transmissions in the small cell tier,
and \emph{(ii)} the adaptive dynamic TDD transmission, the IC operation to mitigate the small cell DL to small cell UL interference together with the proposed scheduling policy in the small cell tier make the best of the transmission opportunities created by macrocell ABS and UL subframes.
The only downside of Scheme~F(b) is that the 95-percentile macrocell DL UPT suffers from a loss of 10.79\,\%
%about 10\,\%
when $\lambda^{\textrm{DL}}(u(q))=0.1$.
As explained earlier for Scheme~D,
this is because macrocells experience resource shortage when the traffic load is low due to the ABS operation of muting 2 subframes per 10 subframes.

\section{Comparison of Dynamic TDD Operations in HomSCNs and HetNets}
\label{sec:dynTDD_comp_homo_het}

Our study is coherent for both HomSCNs and HetNets,
because
\emph{(i)} the optimization objectives are same for both network scenarios,
\emph{(ii)} the additional complication of scheduler in HetNets compared with that in HomSCNs is removed by the ideal genie-aided LA mechanism,
and \emph{(iii)} LTE-compliant DL/UL MIMO operations are considered for both network scenarios.
Therefore, the performance results of HomSCNs and those of HetNets can be compared head to head,
and thus we can draw some useful insights on the application of dynamic TDD in future networks as follows.

\begin{remark}
A higher flexibility of TDD configurations promises higher potential performance gains of dynamic TDD.
\end{remark}

From Table~\ref{tab:Homo_relative_gain_results_basicILIM} and Table~\ref{tab:Het_relative_gain_results},
it can be observed that the performance gain of dynamic TDD is smaller in HetNets than that in HomSCNs,
especially in the UL.
This is mainly because only limited flexibility of dynamic TDD can be achieved in HetNets due to the existence of ABSs and the restrictions it imposes on dynamic TDD transmissions.
In more detail, in the HomSCNs,
all subframes can be dynamic TDD subframes,
and the DL-to-UL subframe ratio ranges from 2:3 (LTE Release~12) or 1:9 (LTE future releases) to 9:1,
as discussed in Section~\ref{sec:homoResults}.
In contrast, in the HetNet small cell tier,
the DL-to-UL subframe ratio ranges from 5:5 to 9:1,
since not all subframes can be dynamic TDD subframes,
as discussed in Section~\ref{sec:hetnetResults}.
Hence, compared with dynamic TDD in HetNets,
it's counterpart in HomSCNs benefits form a much wider range of DL-to-UL subframe ratios,
leading to larger performance gains due to the traffic-adaptive scheduling.

\begin{remark}
Interference mitigation is more crucial for the successful dynamic TDD operation in HetNets than in HomSCNs.
\end{remark}

As can be seen from Table~\ref{tab:Homo_relative_gain_results_basicILIM} and Table~\ref{tab:Het_relative_gain_results},
we can conclude that interference mitigation is more crucial for the successful dynamic TDD operation in HetNets than that in HomSCNs.
In more detail, the straightforward dynamic TDD operation is able to stand on its own with positive performance gains in HomSCNs
(see Scheme~1 vs. Scheme~3 in Table~\ref{tab:Homo_relative_gain_results_basicILIM}),
while the straightforward dynamic TDD operation suffers from huge performance losses in HetNets (see Scheme~A vs. Scheme~B or Scheme~C vs. Scheme~D in Table~\ref{tab:Het_relative_gain_results}),
particularly in terms of the macrocell UL UPTs due to the devastating interference from small cell DL to macrocell UL.
Hence, proper interference mitigation must be in place to handle the inter-tier inter-link interference for dynamic TDD in HetNets.

\begin{remark}
Proper LA algorithms are essential for both HomSCNs and HetNets to reap the performance gains offered by dynamic TDD.
\end{remark}

Comparing the results of a given dynamic TDD scheme in this study and those in our previous work~\cite{HomodynTDD_ICC} and~\cite{HetNetdynTDD_ICC},
we can find that some results in~\cite{HomodynTDD_ICC} and~\cite{HetNetdynTDD_ICC} were lack of insights and seemed counter-intuitively small because a simple LA mechanism was assumed in our previous work, and hence the potential gains of dynamic TDD were not fully reaped.
Such examples include Scheme~3 for the HomSCNs as discussed in~\ref{sec:homoResults},
Scheme~B for the HetNets as discussed in~\ref{sec:hetnetResults}, etc.
Therefore, it is very important for dynamic TDD networks,
both HomSCNs and HetNets, to have proper LA algorithms to predict the drastic interference fluctuation due to the dynamic and non-uniform TDD configurations in neighbouring cells.

\section{Conclusion}
\label{sec:conclusion}

%[David]: To be completed.
%[Ming]: To start with, the conclusions of our previous papers have been shown in the following.
%[Ming]: I have completed the Conclusion.
In this paper, using a unified framework,
we present new results on dynamic TDD transmissions in both HomSCNs and HetNets,
%From our study on dynamic TDD in HomSCNs,
and we draw the following conclusion,
%variou ILIM schemes as well as their combinations have been investigated, which lead to the following conclusion,
\begin{itemize}
  \item
  The dynamic TDD with (partial) IC is shown to provide large gains, especially in terms of the 5-percentile UL UPT,
  when the traffic load is medium to relatively high.
  %where the DL-to-UL interference occurs frequently.
  \item
  The combination of CC and IC, with or without ULPB, is not an efficient strategy because the CC scheme and IC scheme are somehow redundant.
  \item
  The combination of CC and ULPB is recommended for low-complexity implementation,
  while that of ULPB and IC can bring much more performance gains at the expense of higher complexity.
\end{itemize}

In our study on dynamic TDD in HetNets,
%based on a novel scheduling policy in small cells,
%we propose an algorithm to jointly determine UE cell association and macrocell ABS duty cycle,
%together with another algorithm to split DL/UL subframes in dynamic TDD small cells.
%From simulation results,
%we can conclude that significant performance gains can be achieved by introducing dynamic TDD transmissions into HetNets.
we show that in order to make dynamic TDD operate properly in HetNets,
\begin{itemize}
  \item
  Small cell DL to macrocell UL IC is indispensable for the macrocells to achieve reasonable UL UPTs.
  \item
  Another DL-to-UL IC in the small cell tier is required to mitigate the inter-link interference among small cells,
  especially when the traffic load is medium or high.
  \item
  The proposed BOIC scheme results in small performance losses and low complexity compared with the full IC scheme,
  making it a good choice for practical use.
\end{itemize}

%In summary, we conclude that
%\begin{itemize}
%  \item A higher flexibility of TDD configurations promises higher potential performance gains of dynamic TDD.
%  \item ILIM is crucial for the successful operation of dynamic TDD in both HomSCNs and HetNets.
%  \item Proper LA algorithms are indispensable for the network to reap the performance gains offered by dynamic TDD.
%\end{itemize}

To improve the feasibility and the generality of the proposed algorithms,
as future work,
we will consider more practical assumptions in our study such as errors in buffer size,
investigate practical LA algorithms and more practical non-IC receivers,
as well as use theories such as machine learning techniques, game theory, distributed optimization, etc.,
to design low-complexity algorithms, particularly for dynamic TDD in HetNets.
%dynamic CC schemes considering joint optimization of DL/UL scheduling among multiple small cells for dynamic TDD networks. Also, we will design

\section*{Acknowlegements}

The authors would like to thank all the anonymous reviewers for their helpful comments and constructive suggestions to improve early drafts of this paper.

%[Ming]: I have double checked the reference and downsize the list to 22 references.
%\vspace{1cm}


\begin{thebibliography}{1}
\bibitem{Cisco2012}
Cisco, Cisco Visual Networking Index: Global Mobile Data Traffic Fore-cast Update, 2011$-$2016. Feb. 2012.

\bibitem{Book_LTE-A}
S. Sesia, I. Toufik, M. Baker, LTE: The UMTS Long Term Evolution (2nd Edition). John Wiley and Sons, USA, 2011.

\bibitem{Tutor_5G}
D. L\'opez-P\'erez, M. Ding, H. Claussen, and A. Jafari, "Towards 1 Gbps/UE in Cellular Systems: Understanding Ultra-Dense Small Cell Deployments," \emph{IEEE Communications Surveys and Tutorials}, vol. 17, no. 4, pp. 2078-2101, Jun. 2015.

\bibitem{RAN1conf77}
ETSI MCC, Draft Report of 3GPP TSG RAN meeting \#66. Dec. 2014.

\bibitem{reviewer_add1}
H. Zhang, C. Jiang, J. Cheng, V. C. M. Leung, "Cooperative interference mitigation and handover management for heterogeneous cloud small cell networks," \emph{IEEE Mag. on Wireless Commun.}, vol. 22, no. 3, pp. 92-99, Jun. 2015.

\bibitem{LAA_HetNet2015}
H. Zhang, X. Chu, W. Guo, S. Wang, "Coexistence of Wi-Fi and heterogeneous small cell networks sharing unlicensed spectrum," \emph{IEEE Comm. Mag.}, vol. 53, no. 3, pp. 158-164, Mar. 2015.

\bibitem{LAA_HomoNet2015}
Z. Luo, M. Ding, H. Luo, "CC On/Off Scheduling Using Learning-Based Prediction for LTE in the Unlicensed Spectrum," \emph{IEEE Communications Letters}, vol. 19, no. 12, pp. 2158-2161, Dec. 2015.

\bibitem{D2D2014}
Y. Qin, M. Ding, M. Zhang, H. Yu, H. Luo, "Relaying Robust Beamforming for Device-to-Device Communication with Channel Uncertainty," \emph{IEEE Communications Letters}, vol. 18, no. 10, pp. 1859-1862, Oct. 2014.

\bibitem{D2D2015}
Y. Niu, C. Gao, Y. Li, L. Su, D. Jin, A. V. Vasilakos, "Exploiting Device-to-Device Communications in Joint Scheduling of Access and Backhaul for mmWave Small Cells," IEEE Journal on Selected Areas in Communications, vol. 33, no. 10, pp. 2052-2069, Oct. 2015.

\bibitem{reviewer_add2}
H. Zhang, C. Jiang, N. C. Beaulieu, X. Chu, X. Wang, T. Q. S. Quek, "Resource allocation for cognitive small cell networks: A cooperative bargaining game theoretic approach," \emph{IEEE Trans. on Wireless Commun.}, vol. 14, no. 6, pp. 3481-3493, Jun. 2015.

%\bibitem{reviewer_add3}
%H. Zhang, C. Jiang, R. Q. Hu, Y. Qian, "Self-organization in disaster resilient heterogeneous small cell networks," to appear in \emph{IEEE Networks}, arXiv:1505.03209 [cs.NI], May 2015.

\bibitem{reviewer_add4}
H. Zhang, C. Jiang, X. Mao, H. Chen, "Interference-limited resource optimization in cognitive femtocells with fairness and imperfect spectrum sensing," to appear in \emph{IEEE Trans. on Vehicular Technology}.

\bibitem{reviewer_add5}
H. Zhang, C. Jiang, N. C. Beaulieu, X. Chu, X. Wen, M. Tao, "Resource allocation in spectrum-sharing OFDMA femtocells with heterogeneous services," \emph{IEEE Trans. on Commun.}, vol. 62, no. 7, pp. 2366-2377, July 2014.

\bibitem{Lopez2013b}
D. L\'opez-P\'erez, X. Chu, A.V. Vasilakos, H. Claussen, "On Distributed and Coordinated Resource Allocation for Interference Mitigation in Self-Organizing LTE Networks," \emph{IEEE/ACM Transactions on Networking}, vol. 21, no. 4, pp. 1145-1158, Aug. 2013.

\bibitem{Lopez2014}
D. L\'opez-P\'erez, X. Chu, A.V. Vasilakos, H. Claussen, "Power Minimization Based Resource Allocation for Interference Mitigation in OFDMA Femtocell Networks," IEEE Journal on Selected Areas in Communications, vol. 32, no. 2, pp. 333-344, Feb. 2014.

%\bibitem{DoCoMo2012}
%NTT DoCoMo, RWS-120010: Requirements, candidate solutions \& technology roadmap for LTE Rel-12 Onward. In: 3GPP workshop on Release 12 onward, Ljubljana, Slovenia, June 2012.
%
%\bibitem{SK2012}
%SK Telecom, RWS-120020: efficient spectrum resource usage for next-generation network. In: 3GPP workshop on Release 12 Onward, Ljubljana, Slovenia, June 2012.

\bibitem{Andrews2013}
J. G. Andrews, "Seven ways that HetNets are a cellular paradigm shift," \emph {IEEE Comm. Mag.}, vol. 51, no. 3, pp.136-144, Mar. 2013.

\bibitem{Lopez2011}
D. L\'opez-P\'erez, I. Guvenc, Guillaume de la Roche, M. Kountouris, T. Q. S. Quek and J. Zhang, "Enhanced Inter-cell Interference Coordination Challenges in Heterogeneous Networks," \emph{IEEE Wireless Comm. Mag.}, vol. 18, no. 3, pp. 22-30, Jun., 2011.

\bibitem{Lopez2012}
D. L\'opez-P\'erez, C. Xiaoli, I. Guvenc, "On the expanded region of picocells in heterogeneous networks," \emph{IEEE Selected Topics in Signal Processing}, vol. 6, no. 3, pp. 281-294, Jun. 2012.

\bibitem{SmallCell1}
M. Ding, D. L\'opez-P\'erez, G. Mao, P. Wang, and Z. Lin, "Will the Area Spectral Efficiency Monotonically Grow as Small Cells Go Dense?,"1¤7to appear in IEEE Globecom 2015, arXiv: 1505.01920 [cs.NI], Dec. 2015.

\bibitem{SmallCell2}
M. Ding, P. Wang, D. L\'opez-P\'erez, G. Mao, and Z. Lin, "Performance Impact of LoS and NLoS Transmissions in Dense Cellular Networks,"1¤7to appear in  \emph{IEEE Transactions on Wireless Communications}, arXiv:1503.04251 [cs.IT], Dec. 2015.

\bibitem{SmallCell3}
M. Ding, D. L\'opez-P\'erez, G. Mao, and Z. Lin, "DNA-GA: A New Approach of Network Performance Analysis,"1¤7submitted to IEEE ICC 2016, arXiv:1512.05429 [cs.IT], Dec. 2015.

%\bibitem{Pedersen2013}
%K. I. Pedersen, W. Yuanye, S. Strzyz, F. Frederiksen, "Enhanced inter-cell interference coordination in co-channel multi-layer LTE-Advanced networks," \emph{IEEE Wireless communications}, vol. 20, no. 3, pp. 120-127, Jun. 2013.
%
%\bibitem{Soret2012}
%B. Soret and K. I. Pedersen, "Macro transmission power reduced for HetNet co-channel deployments," \emph{IEEE Globecom},
%Anaheim, CA, USA, pp. 4126-4130, Dec. 2012.

%\bibitem{Larson2013}
%E. G. Larsson, F. Tufvesson, O. Edfors and T. L. Marzetta, "Massive MIMO for Next Generation Wireless Systems," \emph{IEEE Comm. Mag.}, vol. 52, no. 2, pp. 186-195, Feb. 2014.

\bibitem{TS36.213}
3GPP, "TS 36.213 (V11.3.0): Physical layer procedures (Release 11)," Jun. 2013.

\bibitem{FullDup2014}
C. I, C. Rowell, S. Han, Z. Xu, G. Li, Z. Pan, "Toward green and soft: a 5G perspective," \emph{IEEE Comm. Mag.}, vol. 52, no. 2, pp. 66-73, Feb. 2014.

\bibitem{Shen2012}
Z. Shen, A. Khoryaev, E. Eriksson, X. Pan, "Dynamic uplink-downlink configuration and interference management in TD-LTE," \emph{IEEE Comm. Mag.}, vol. 50, no. 11, pp. 51-59, Nov. 2012.

\bibitem{HomodynTDD_ICC}
M. Ding, D. L\'opez-P\'erez, A. V. Vasilakos,
and W. Chen, \textquotedblleft{}Dynamic TDD transmissions in homogeneous
small cell networks,\textquotedblright{} \emph{IEEE ICC 2014}, pp. 616-621, Jun. 2014.

%\bibitem{HomodynTDD_GC}
%M. Ding, D. L\'opez-P\'erez, A. V. Vasilakos,
%and W. Chen, \textquotedblleft{}Analysis on the SINR Performance of Dynamic TDD in Homogeneous Small
%Cell Networks,\textquotedblright{} submitted to \emph{IEEE Globecom 2014}.

\bibitem{SG_dynTDD_ICC}
S. Song, Y. Chang, H. Xu, D. Zheng, and D. Yang, \textquotedblleft{}Energy efficiency model based on stochastic geometry in dynamic TDD cellular networks,\textquotedblright{} \emph{IEEE ICC 2014}, pp. 889-894, Jun. 2014.

\bibitem{TR36.828}
3GPP, "TR 36.828 (V11.0.0): Further enhancements to LTE Time Division Duplex (TDD) for Downlink-Uplink (DL-UL) interference management and traffic adaptation (Release 11)," Jun. 2012.

\bibitem{HetNetdynTDD_ICC}
M. Ding, D. L\'opez-P\'erez, R. Xue, A. V. Vasilakos,
and W. Chen, \textquotedblleft{}Small Cell Dynamic TDD Transmissions in Heterogeneous Networks,\textquotedblright{} \emph{IEEE ICC 2014}, pp. 4881-4887, Jun. 2014.

\bibitem{SHARP_DLPC_dynTDD}
Sharp, \textquotedblleft{}R1-132350: DL power control based interference mitigation for eIMTA,\textquotedblright{}
3GPP RAN1 Meeting \#73, Fukuoka, Japan, May, 2013.

\bibitem{SHARP_ULPC_dynTDD}
Sharp, \textquotedblleft{}R1-132351: UL power control based interference mitigation for eIMTA,\textquotedblright{}
3GPP RAN1 Meeting \#73, Fukuoka, Japan, May, 2013.

\bibitem{Machine_Learning1}
A.	Galindo-Serrano, L. Giupponi, \textquotedblleft{}Distributed Q-Learning for Aggregated Interference Control in Cognitive Radio Networks,\textquotedblright{} \emph{IEEE Transactions on Vehicular Technology}, vol. 59, no. 4, pp. 1823-1834, May 2010.

\bibitem{Machine_Learning2}
L. Giupponi, A. Galindo-Serrano, P. Blasco, M. Dohler, \textquotedblleft{}Docitive networks: an emerging paradigm for dynamic spectrum management [Dynamic Spectrum Management],\textquotedblright{} \emph{IEEE Transactions on Wireless Communications}, vol. 17, no. 4, pp. 47-54, Aug. 2010.

\bibitem{R1-132304}
Nokia Siemens Networks, Nokia, "R1-132304: Enhanced fast ABS adaptation for Rel-12 small cell scenario 1," 3GPP RAN1 Meeting \#73, Fukuoka, Japan, May, 2013.

\bibitem{Lopez2013}
D. L\'opez-P\'erez and H.Claussen, "Duty cycles and load balancing in HetNets with eICIC almost blank subframes," \emph{IEEE PIMRC}, London, UK, pp. 1-6, Sep. 2013.

\bibitem{simulator}
http://wnt.sjtu.edu.cn/flint/html/index.html

\bibitem{TR36.814}
3GPP, "TR 36.814 (V9.0.0): Further advancements for E-UTRA physical layer aspects (Release 9)," Mar. 2010.

%\bibitem{TR36.872}
%3GPP, "TR 36.872 (V1.0.1): Small cell enhancements for E-UTRA and E-UTRAN - Physical layer aspects (Release 12)," Aug. 2013.

\bibitem{EPAchannel}
3GPP, "TS 36.104 (V11.4.0): Base Station (BS) radio transmission and reception," Mar. 2013.

\end{thebibliography}
\end{document}